\documentclass[11pt,a4paper,usenames,dvipsnames]{article}
\RequirePackage{pdf14}
\usepackage{jheppub}

\usepackage{amsthm}

\usepackage[utf8]{inputenc}

\usepackage{graphicx,tikz}
\usepackage{amsfonts,amsmath,amssymb}
\usepackage{dsfont}
\usepackage{mathtools}
\usepackage{makecell}

\usetikzlibrary{positioning}

\usepackage{comment} 
\usepackage{ulem}

\newtheorem*{nono-question}{Question}

\newcommand{\Rrep}{\mathcal{R}}
\newcommand{\USp}{\text{USp}}
\newcommand{\SO}{\text{SO}}
\newcommand{\SU}{\text{SU}}
\newcommand{\U}{\text{U}}
\newcommand{\pchi}{\varkappa}

\makeatletter
\gdef\@fpheader{\break}
\makeatother

\title{Higgsless Lagrangian SCFTs and Strongly Finite VOAs}

\author[]{Siqi Chen,\!}
\author[]{Anirudh Deb\!} 
\author[]{and Leonardo Rastelli}

\emailAdd{siqi.chen.1@stonybrook.edu}
\emailAdd{anirudh.deb@stonybrook.edu}
\emailAdd{leonardo.rastelli@stonybrook.edu}

\affiliation[]{C. N. Yang Institute for Theoretical Physics, Stony Brook University, Stony Brook, NY 11794-3840,
USA}

\abstract{
Vertex operator algebras (VOAs) are well studied in both mathematics and physics. The best understood class is that of {\it strongly rational} VOAs, whose representation category is maximally well behaved: indeed, it is a modular tensor category.
At the next level of complexity are  {\it strongly finite} but non-rational VOAs. Their representation category is not semisimple (it is ``logarithmic''), but maintains   nice structural properties.
 Only a few families of examples in this class are known, a fact that may have hindered the development of a comprehensive mathematical theory.
The SCFT/VOA correspondence provides a natural way to generate more examples:
strongly finite but non-rational VOAs are expected to arise from four-dimensional ${\cal N}=2$ Lagrangian superconformal field theories (SCFTs) that do {\it not} admit a Higgs branch moduli space of vacua. We
tackle the  combinatorial task of
classifying {\it all} such ``Higgsless'' Lagrangian SCFTs. To our surprise, this
set  turns out to be rather sparse. Free vector multiplets and their discrete gaugings are immediate examples. The interacting Higgsless theories comprise
  one infinite sequence of SO/USp
quivers and three sporadic examples.
We construct and study
the novel VOAs associated to two of the sporadic examples, and confirm that they are indeed strongly finite and logarithmic.

}

\begin{document} 
	\maketitle

	\section{Introduction}

Moduli spaces of vacua of supersymmetric field theories are a subject of great
interest in physical mathematics.
In this paper we leverage the special role that the Higgs branch of vacua  plays in the SCFT/VOA correspondence~\cite{Beem:2013sza} to
generate new examples of strongly finite but non-rational (logarithmic) vertex operator algebras (VOAs).

Let us review the general framework, starting on the physics side. We focus on 
${\cal N}=2$ four-dimensional superconformal field theories, henceforth simply SCFTs (the qualifiers ``four-dimensional'' and ``${\cal N}=2$'' being understood).
The two canonical branches of vacua of  a SCFT are the Higgs branch (where the $\SU(2)_R$ subgroup of the R-symmetry group is broken, while $\U(1)_r$ is unbroken) and the Coulomb branch (where conversely $\U(1)_r$ is broken and $\SU(2)_R$ unbroken). It is believed that any interacting SCFT has a nontrivial Coulomb branch. On the other hand, while the Higgs branch ${\cal M}_H[{\cal T}]$ 
of a generic SCFT ${\cal T}$ is nontrivial, there are interesting examples of ``Higgsless'' SCFTs, notably certain infinite classes of Argyres-Douglas theories. The  coordinate ring  of ${\cal M}_H[{\cal T}]$
(in physics parlance,  the Higgs branch chiral ring) is embedded in a much richer algebraic invariant of the SCFT, its vertex operator algebra $\mathcal{V}[{\cal T}]$.  A central conjecture~\cite{Beem:2017ooy} states that the
Higgs branch of ${\cal T}$ (as a holomorphic symplectic variety) can be canonically recovered from the vertex algebra,
being isomorphic with the {\it associated variety}~\cite{Arakawa:2010dtu} of the corresponding VOA,
\begin{equation} \label{higgsconj}
{\cal M}_H[{\cal T}] = \mathcal{X}_{\mathcal{V}[{\cal T}]}\,.
\end{equation}
We'll refer to (\ref{higgsconj}) as the {\it Higgs branch conjecture}.\footnote{
The relation of the VOA with the physics of the Higgs branch (when one exists) runs even deeper. For several classes of SCFTs one can leverage the effective field theory  on the Higgs branch to motivate very interesting free field realizations of the associated VOAs~\cite{Bonetti:2018fqz,Beem:2019tfp,Beem:2019snk,Beem:2021jnm,Beem:2024fom}.}

The mathematical significance of the associated variety is that it controls the representation theory of its vertex algebra. Informally,
a smaller
$\mathcal{X}_\mathcal{V}$ forces a better representation theory.
 In the extreme case that $\mathcal{X}_\mathcal{V}$  is trivial (i.e.~a point), the vertex algebra is called {\it lisse} or {\it $C_2$-cofinite}, and under certain additional technical conditions\footnote{The technical conditions are that the VOA must be simple (which holds if the SCFT ${\cal T}$ itself is simple, i.e.~not a tensor product theory), of CFT type (there is a unique $SL(2)$-invariant vacuum, which is the state of minimum $L_0$) and self-contragredient (equivalent to the condition that the $L_0 =1$ eigenstates are annihilated by $L_1$).} that are automatically true for VOAs that arise from SCFTs, {\it strongly finite}. Strongly finite VOAs
are not necessarily rational --  in general, they may admit indecomposable modules where $L_0$ acts with a Jordan block structure. If one further assumes semisimplicity, we are in the {\it strongly rational} case \cite{McRae:2021yyb}, which is the familiar maximally nice story:
the representation category is a modular tensor  category, 
the characters of the simple modules furnish a finite dimensional representation of $SL(2, \mathbb{Z})$, and the fusion coefficients
are given by the Verlinde formula in terms of the modular $S$
matrix.

 Strongly finite but non-rational VOAs sit at the next level of complexity.
 Their representation category, which now also includes 
indecomposable (``logarithmic'') modules, is still relatively well behaved -- for example,
there are still finitely many {\it simple} modules, which are all ordinary.\footnote{An ordinary module is what it sounds to a physicist's ear: it is graded by $L_0$, with $L_0$ bounded from below and each graded component finite-dimensional.} Modularity is more involved: the characters of the simple modules, which are power series in the nome $q$, do not close under $SL(2, \mathbb{Z})$. To restore modularity one must include (a finite number of) {\it pseudocharacters}, which contain  powers of $\log \, q$ in their $q$-series expansions \cite{Miyamoto:2002ar}.
Logarithmic strongly-finite VOAs are a subject of
active mathematical research, but a complete theory is still lacking, see~\cite{Creutzig_2017} for an insightful discussion. Progress has perhaps been hindered by the relative scarcity of examples compared to the rational case.

In view of the Higgs branch conjecture (\ref{higgsconj}), we can expect the SCFT/VOA correspondence to generate more examples of strongly-finite vertex algebras, by exploring the landscape of Higgsless SCFTs. Curiously, it turns out that all known interacting examples of Higgsless SCFTs (the Argyres-Douglas theories mentioned above) give rise to strongly {\it rational} VOAs, indeed one recovers infinite families of rational W-algebras~\cite{Xie:2019vzr,Xie:2016evu,Xie:2019yds,Xie:2019zlb}. On the other hand, discrete gaugings of free vector multiplet SCFTs, while trivial from a physics perspective, correspond to textbook examples of non-rational strongly finite vertex algebras \cite{Creutzig:2014xea}. It is then natural to consider the obvious generalization  of free vector multiplets, namely ${\cal N}=2$ conformal {\it Lagrangian} gauge theories with trivial Higgs branch. To decide whether the associated vertex algebras are rational or logarithmic, we can look at the modular transformation properties of the vacuum character,
which coincides with the Schur index of the parent SCFT. The Schur index of a Lagrangian theory is computed by a certain matrix integral, whose modular properties have been studied in~\cite{ArabiArdehali:2023bpq}:
the generic expectation is that the S-transformed index contains powers of $\log \, q$, the tell-tale sign of non-rationality.
Heuristically, the logarithms can be traced to the vector multiplet sector: the gauginos contribute symplectic-fermion-like degrees of freedom to the VOA, and symplectic fermions are the archetypal logarithmic system. Perhaps this is why the Lagrangian Higgsless theories land on the logarithmic side of the strongly finite world, while the (non-Lagrangian) Argyres-Douglas examples are rational.
All in all, we are led to the following

\medskip
\noindent
{\it \bf Conjecture:} {\it For any four-dimensional ${\cal N}=2$ superconformal  Lagrangian gauge theory~${\cal T}$ with trivial Higgs branch, the vertex algebra $\mathbb{V}[{\cal T}]$ is strongly finite and non-rational.}

\medskip
\noindent
We emphasize that this is a well-posed mathematical conjecture, as 
the VOAs
associated to Lagrangian SCFTs are defined by a precise
cohomological construction~\cite{Beem:2013sza}. One is instructed to perform a certain BRST reduction of the free field VOA comprising symplectic bosons and fermions, corresponding respectively to four-dimensional (half-)hyper multiplets and vector multiplets, see e.g.~\cite{Arakawa:2018egx, Beem:2025guj} for careful mathematical presentations.

An ${\cal N}=2$ Lagrangian gauge theory is specified by the pair $(G, \rho)$, where 
$G$ is a semisimple gauge group, and $\rho$ the pseudoreal
$G$-representation under which the half-hypermultiplets transform. To restrict to 
{\it conformal} gauge theories, it is sufficient
to impose vanishing of the one-loop beta function for each 
of the simple factors of $G$. This intricate combinatorial
 problem was solved in~\cite{Bhardwaj:2013qia}; the resulting list of theories is rather baroque, comprising  infinite families of generalized quivers and several sporadic cases. As the Higgs branch 
 of a Lagrangian gauge theory 
 is given
 by the hyperk\"ahler quotient $\mathcal{M}_H= \mathbb{C}^{{\rm dim}(\rho) } /\!\!/\!\!/G$,  the task of restricting to Higgsless
 theories is conceptually straightforward, but computationally challenging. A direct calculation of the requisite invariant (the Higgs branch Hilbert series) gets quickly out of hand as the rank of the gauge group increases. 
  In practice,
 we winnow the list by imposing 
 the necessary condition that the theory should not admit any continuous global symmetry (as flavor symmetry moment maps are generators of the Higgs chiral ring), and then proceed with a case by case analysis. 
Many of the candidates are ruled out because we are able to  explicitly construct non-trivial  Higgs chiral operators.  All in all, we are left with three sporadic cases  and with one infinite sequence of SO/USp quivers, which we are able to argue  have trivial Higgs branch. This is a curious result.
The list of Higgsless theories is rather sparse and does not conform to any obvious pattern, though the latter could perhaps also be said for the complete list~\cite{Bhardwaj:2013qia} of ${\cal N}=2$ Lagrangian theories.

The three Higgsless sporadic examples are (i) the $\USp(4)$ gauge theory with half-hypers in the ${\bf 16}$ representation, (ii) the $\SU(3)\times \SU(2)$ gauge theory with half-hypers in  the $\bf 8\times 2$ representation, 
and (iii) the $\mathbf{\frac{1}{2}asym3}\,$--$\,\USp(8)\,$--$\,\SO(6)$ quiver.
We bootstrap the  VOAs of the first two sporadic cases, by identifying their strong generators, making an ansatz for their  singular OPEs and imposing associativity, which holds thanks to the existence of  nontrivial null relations. We confirm by direct calculation that the associated varieties of these vertex algebras are trivial, as expected from the Higgs branch conjecture. In both cases, we have found closed form expressions for the vacuum characters, and find that logarithmic pseudocharacters appear in their modular orbits. 
  To the best of our knowledge, these two vertex algebras are novel examples of strongly finite but non-rational VOAs.  The construction and study of the VOAs associated to the  third sporadic case and to the infinite sequence of  Higgsless quivers is left for future work.

	\medskip
    \noindent
The remainder of the paper is organized as follows. In Section \ref{sec:higgslessc2cofinite} we  review basic facts about the SCFT/VOA
    correspondence, with an emphasis on 
    the Higgs branch conjecture. We also review various diagnostics for the absence of a Higgs branch. In Section \ref{sec:classrec} we tackle the classification problem of Higgsless Lagrangian SCFTs. In Section \ref{sec:examples}, after briefly discussing the examples of free vector multiplets and their discrete gaugings, we analyze in detail  two of the Higgsless sporadic examples identified by the classification. We conclude in Section~\ref{sec:discussion} with a brief discussion of open questions. Several appendices supplement the main text with additional technical material.

\section{Higgsless theories and strongly finite VOAs}
\label{sec:higgslessc2cofinite}

This section reviews the conceptual framework on which the rest of the paper rests. We begin with a lightning review of the SCFT/VOA correspondence, emphasizing the special role played by the Higgs chiral ring within the larger algebra of Schur operators.
We then
recall the relation between the Higgs branch as a variety and the Higgs chiral ring,  introduce the notion of  associated variety of a VOA and state the Higgs branch conjecture. The main observation is that Higgslessness of the SCFT is a {\it necessary} condition for $C_2$-cofiniteness of the associated VOA, independently of the full Higgs branch conjecture; this provides the unconditional logical motivation for the rest of the paper. We then pivot to a discussion of 
the modularity-based criteria for non-rationality. Finally we collect the practical diagnostics, namely the necessary conditions, Higgs branch Hilbert series, Hall-Littlewood index, that are used in the case-by-case analysis of Section~\ref{sec:classrec}.

\subsection{The SCFT/VOA correspondence}
\label{subsec:scftvoa}

To each four-dimensional $\mathcal{N}=2$ superconformal field theory $\mathcal{T}$ one associates a two-dimensional vertex operator algebra $\mathcal{V}[\mathcal{T}]$, obtained by passing to the cohomology of a particular nilpotent supercharge of $\mathcal{T}$~\cite{Beem:2013sza}. The local operators that survive in cohomology are called {\it Schur operators}, and their $\mathcal{N}=2$ superconformal quantum numbers $(E, j_1, j_2, R, r)$ satisfy
\begin{equation}
E - (j_1+j_2) - 2R = 0\,, \qquad r + j_1 - j_2 = 0\,,
\end{equation}
where $(E, j_1, j_2, R, r)$ are the conformal dimension, the two Lorentz Cartan eigenvalues, and the $SU(2)_R \times U(1)_r$ R-charges. A Schur operator descends to a 2d chiral operator of weight
\begin{equation}\label{eq:hERrel}
h \;=\; \frac{E+j_1+j_2}{2} \;=\; E - R\,.
\end{equation}
The stress-tensor multiplet of $\mathcal{T}$ contains the $\SU(2)_R$ current, whose Schur cohomology class is the 2d stress tensor $T(z)$ with central charge
\begin{equation}\label{eq:c2d4d}
c_{\rm 2d} \;=\; -12\, c_{\rm 4d}\,.
\end{equation}
Schur operators come from four types of $\mathcal{N}=2$ superconformal multiplets, listed in Table~\ref{tab:schurmultipletslagletters} in the notation of~\cite{Dolan:2002zh}: $\hat{\mathcal{B}}_R$, $\mathcal{D}_{R(0,j_2)}$, $\bar{\mathcal{D}}_{R(j_1,0)}$, and $\hat{\mathcal{C}}_{R(j_1,j_2)}$. Of these, $\hat{\mathcal{B}}_R$ is the multiplet that plays the central role in this paper: its bottom component is a {\it Higgs chiral operator}, satisfying $E = 2 R$.
 Higgs chiral operators are the $j_1=j_2=0$ subset of the Schur sector, and the Higgs chiral ring $\mathcal{R}_H$ is a subring of the larger Schur algebra.

The Schur limit of the superconformal index \cite{Gadde:2011ik,Gadde:2011uv}, which counts Schur operators with a sign, 
\begin{equation}\label{SchurIndex}
\mathcal{I}_{\rm Schur}(q)  \;=\;  \mathrm{Tr}\,(-1)^F\,q^{E-R}  = \mathrm{Tr}\,(-1)^F\,q^{h} \, ,
\end{equation}
coincides up to a vacuum-energy prefactor with the graded vacuum character of the VOA, 
\begin{equation}\label{SchurVac}
\chi_{\rm vac}(q)    \;=\;  \mathrm{Tr}\,(-1)^F\,q^{- c_{\rm 2d}/24 + h}= q^{- c_{\rm 2d}/24} \, \mathcal{I}_{\rm Schur}(q) \,.
\end{equation}

\begin{table}[t]
\renewcommand{\arraystretch}{1.2}
\centering
\begin{tabular}{|c|c|c|c|}
\hline
Multiplet  & $h$ & $r$ & Lagrangian letters \\
\hline\hline
$\hat{\mathcal{B}}_R$  & $R$ & $0$ & $q$ \\
\hline
$\mathcal{D}_{R(0,j_2)}$  & $R+j_2+1$ & $j_2+\tfrac12$ & $q$, $\tilde{\lambda}^1_{\dot +}$ \\
\hline
$\bar{\mathcal{D}}_{R(j_1,0)}$ & $R+j_1+1$ & $-j_1-\tfrac12$ & $q$, $\lambda^1_{+}$ \\
\hline
$\hat{\mathcal{C}}_{R(j_1,j_2)}$ & $R+j_1+j_2+2$ & $j_2-j_1$ & $D^n_{+\dot +}q$, $D^n_{+\dot +}\lambda^1_{+}$, $D^n_{+\dot +}\tilde{\lambda}^1_{\dot +}$ \\
\hline
\end{tabular}
\caption{The four 4d $\mathcal{N}=2$ superconformal multiplet types whose bottom or descendant Schur operator survives the chiral cohomology. The last column lists the Lagrangian letters that build the Schur operator: $q$ denotes a half-hypermultiplet scalar, $\lambda^1_+$ and $\tilde{\lambda}^1_{\dot +}$ are gauginos, and $D_{+\dot +}$ is the covariant derivative ($n\geq 0$). 
}
\label{tab:schurmultipletslagletters}
\end{table}

\subsection{The Higgs branch conjecture}

The {\it Higgs branch} $\mathcal{M}_H$ of a four-dimensional $\mathcal{N}=2$ SCFT is the component of the moduli space of vacua on which the $\SU(2)_R$ symmetry is spontaneously broken and the $\U(1)_r$ symmetry is preserved. The Higgs branch is a holomorphic symplectic cone with $\mathbb{C}^*$ action generated by $\U(1)_R$.

In any $\mathcal{N}=2$ theory, the geometry of the Higgs branch and the Higgs chiral ring $\mathcal{R}_H$ (which comprises, as we have mentioned, local operators satisfying $E = 2R$)
are expected to be related by
\begin{equation}\label{eq:HBgeom}
\mathbb{C}[\mathcal{M}_H] \;=\; \left(\mathcal{R}_H\right)_{\rm red}\,,
\end{equation}
that is, the coordinate ring of $\mathcal{M}_H$ as a variety equals the {\it reduced} part of the chiral ring (obtained by quotienting out the nilradical). For Lagrangian theories this relation can be established directly along the lines of the analysis of moduli spaces of supersymmetric gauge theories in~\cite{Luty:1995sd}, adapted to the $\mathcal{N}=2$ case: the Higgs branch is realized as the hyperk\"ahler quotient (which we review in Section~\ref{subsec:diagnostics}), and $\mathcal{R}_H$ is the gauge-invariant ring of hypermultiplet scalars modulo the F-term relations. For non-Lagrangian theories~(\ref{eq:HBgeom}) is still expected to hold but is generally proved on a case-by-case basis.

A stronger statement is conjectured to hold for {\it conformal} $\mathcal{N}=2$ theories: $\mathcal{R}_H$ itself contains no nilpotent elements. Combined with~(\ref{eq:HBgeom}), this gives the {\it Higgs branch geometrization conjecture}~\cite{Beem:2014zpa}: 
\begin{equation}\label{eq:HBgeo}
\mathbb{C}[\mathcal{M}_H] \;=\; \mathcal{R}_H\,.
\end{equation}

\label{subsec:assvar_HBconj}

On the other hand, on the VOA side, one has the following  natural geometric construction.
Given a vertex operator algebra $\mathcal{V}$, consider the subspace $C_2(\mathcal{V}) \subset \mathcal{V}$ spanned by states of the form $u^i_{-h_i - 1}\phi$ for $u^i, \phi \in \mathcal{V}$ and $u^i$ of conformal weight $h_i$~\cite{zhu1990vertex}.\footnote{This is the convention of the physics literature, where the mode expansion of a weight-$h_i$ field is $u(z) = \sum_n u_{-h_i -n}\,z^n$.} In words, $C_2(\mathcal{V})$ is the subspace of $\mathcal{V}$ consisting of states with at least one $z$-derivative; the quotient
\begin{equation}
\mathcal{R}_\mathcal{V} \;:=\; \mathcal{V}/C_2(\mathcal{V})
\end{equation}
inherits a commutative associative product (from the singular part of the OPE evaluated at coincident points) and a Poisson bracket, and is known as {\it Zhu's $C_2$ algebra}. The geometric object dual to $\mathcal{R}_\mathcal{V}$ is the {\it associated variety}
\begin{equation}
\mathcal{X}_\mathcal{V} \;:=\; \mathrm{Specm}\bigl(\mathcal{R}_\mathcal{V}\bigr)\,,
\end{equation}
the maximal spectrum of $\mathcal{R}_\mathcal{V}$. It carries a Poisson structure inherited from $\mathcal{R}_\mathcal{V}$, and a $\mathbb{C}^*$ action coming from the conformal grading.

The {\it Higgs branch conjecture}~\cite{Beem:2017ooy} asserts that for the VOA $\mathcal{V}[\mathcal{T}]$ associated to a four-dimensional $\mathcal{N}=2$ SCFT $\mathcal{T}$, the associated variety coincides with the Higgs branch as a Poisson variety with $\mathbb{C}^*$ action:
\begin{equation}\label{eq:HBconj}
\mathcal{X}_\mathcal{V} \;=\; \mathcal{M}_H\,.
\end{equation}
Equivalently, at the level of rings, $\bigl(\mathcal{R}_\mathcal{V}\bigr)_{\rm red} = \mathbb{C}[\mathcal{M}_H]$, which under~(\ref{eq:HBgeo}) becomes $\bigl(\mathcal{R}_\mathcal{V}\bigr)_{\rm red} = \mathcal{R}_H$.

\subsection{Higgslessness as a necessary condition for \texorpdfstring{$C_2$}{C\_2}-cofiniteness}
\label{subsec:higgslessneccond}

A VOA $\mathcal{V}$ is called {\it $C_2$-cofinite} (or {\it lisse}) if $\mathcal{R}_\mathcal{V}$ is finite-dimensional, or equivalently, if the associated variety $\mathcal{X}_\mathcal{V}$ consists of a single point. 
A simple, $C_2$-cofinite VOA which is also of CFT type and self-contragredient (these two latter conditions being automatic for VOAs that descend from 4d SCFTs) is called {\it strongly finite}.

A point that we wish to emphasize at the outset is that one direction of the link between Higgsless SCFTs and $C_2$-cofinite VOAs follows from the structure of the SCFT/VOA correspondence alone, {\it without} invoking the Higgs branch conjecture~(\ref{eq:HBconj}).
Consider now the $R$-filtration on $\mathcal{V}$~\cite{Beem:2017ooy}. This filtration is a structural fact about any VOA that descends from a four-dimensional parent: the underlying vector space of $\mathcal{V}$ is the space of Schur operators, which is graded by the $\SU(2)_R$ weight $R$, and the $\SU(2)_R$ selection rules of the four-dimensional OPE imply that the VOA operations can only preserve or lower the total $R$-weight, so that $\mathcal{V}$ automatically inherits a filtration (though not a grading) by $R$. Higgs chiral operators are the Schur operators of maximal $\SU(2)_R$ weight at fixed chiral dimension, and in the associated graded of the $R$-filtration the singular OPE of two $\hat{\mathcal{B}}$-type operators reproduces, in its top $R$-component, the Higgs chiral ring product -- again a consequence of the four-dimensional selection rules. Setting to zero the (classes of the) strong generators that do {\it not} descend from $\hat{\mathcal{B}}$ multiplets is then a well-defined ring homomorphism, i.e.~one obtains a surjection
$
\mathcal{R}_\mathcal{V}\;\twoheadrightarrow\;\mathcal{R}_H\,.
$
Passing to reduced rings and using~(\ref{eq:HBgeom}), this yields a surjection $\bigl(\mathcal{R}_\mathcal{V}\bigr)_{\rm red}\twoheadrightarrow \bigl(\mathcal{R}_H\bigr)_{\rm red} = \mathbb{C}[\mathcal{M}_H]$, and dually a closed embedding of varieties
$
\mathcal{M}_H \;\subseteq\; \mathcal{X}_\mathcal{V}\,.
$
The Higgs branch conjecture~(\ref{eq:HBconj}) upgrades this inclusion to an equality. The inclusion alone, however, is enough for the following:

\medskip
\noindent
{\bf Observation:}~{\it If the VOA $\mathcal{V}[\mathcal{T}]$ is $C_2$-cofinite, then the SCFT $\mathcal{T}$ is Higgsless.}

\medskip
\noindent
Indeed, if $\mathcal{X}_\mathcal{V}$ is a point, then so is $\mathcal{M}_H$. The converse, that Higgslessness of $\mathcal{T}$ implies $C_2$-cofiniteness of $\mathcal{V}[\mathcal{T}]$, is the genuinely conjectural part, and follows from the full Higgs branch conjecture~(\ref{eq:HBconj}).
This observation justifies our strategy: to find candidate strongly finite VOAs we look for Higgsless 4d $\mathcal{N}=2$ SCFTs, since these are the only theories that could produce them. Whether each candidate actually gives a $C_2$-cofinite VOA must then be verified case by case by exhibiting the strong generators and showing that they are nilpotent in $\mathcal{R}_\mathcal{V}$. This case-by-case verification, which we carry out in Section~\ref{sec:examples} for the two Lagrangian sporadic cases, provides at the same time non-trivial evidence for the Higgs branch conjecture itself.

The simplest examples of strongly finite VOAs arising from 4d are the Virasoro VOAs $\mathrm{Vir}_{(2,2n+3)}$ associated with the Argyres-Douglas theories $(A_1,A_{2n})$~\cite{Argyres:1995jj,Xie:2012hs}: these are the chiral algebras of the $(2,2n+3)$ minimal models, generated by the 2d stress tensor alone, with $c_{\rm 2d} = -2n(6n+5)/(2n+3)$. A larger class of $C_2$-cofinite VOAs associated to non-Lagrangian 4d SCFTs was studied in~\cite{Xie:2019vzr}; these are all {\it rational} $C_2$-cofinite VOAs. The non-rational $C_2$-cofinite VOAs that interest us~\cite{Creutzig:2013hma,Adamovic:2012mn,Creutzig_2017} are different in character and, as we shall see, arise naturally from Lagrangian Higgsless SCFTs.

\subsection{Non-rationality from modular pseudocharacters}
\label{subsec:nonrationality}

The vacuum character of a {strongly rational} VOA is the first component of a vector-valued modular form transforming in a finite-dimensional representation of $SL(2,\mathbb{Z})$. For a strongly finite but non-rational VOA, ordinary irreducible-module characters are not enough to close the modular transformation: one needs~\cite{Miyamoto:2002ar} additional {\it pseudocharacters} that involve $\log q$.\footnote{Pseudocharacters can be interpreted physically as one-point functions of logarithmic intertwiners on the torus~\cite{Creutzig_2017}.} A clean sufficient condition for non-rationality follows: if logarithms of $q$ appear in the $SL(2,\mathbb{Z})$ orbit of the vacuum character, the VOA is non-rational.

Concretely, writing $\tilde q = e^{-2\pi i / \tau}$ for the $S$-transformed nome, the vacuum character expands as
\begin{equation}\label{eq:htasymp}
\chi_0(q) \;=\; \sum_{i\in\text{characters}} S_{0i}\,\chi_i(\tilde q) \;+\; \sum_{\tilde i\in\text{pseudochars}} S_{0\tilde i}\,\pchi_{\tilde i}(\tilde q)\,,
\end{equation}
and if the second sum is non-empty the VOA is non-rational.

When $\mathcal{I}_{\rm Schur}(q) = q^{-c_{\rm 4d}/2}\chi_0(q)$ admits a closed-form expression in terms of functions with known modular properties (typically Jacobi theta functions or similar), one can $S$-transform it directly and read off any $\log\tilde q$ terms. This is the cleanest method when available. Fortunately, the concrete examples that we study in  Section~\ref{sec:examples} fall in this class.

When no closed form is at hand, one can study the $\tau\to 0$ (equivalently $\tilde q\to 0$) limit of the Schur index. Building on~\cite{DiPietro:2014bca,Buican:2015ina,ArabiArdehali:2015ybk}, the analysis of~\cite{ArabiArdehali:2023bpq} (which we summarize in Appendix~\ref{app:htl}) gives the schematic asymptotic form
\begin{equation}\label{eq:htlimit}
q^{c_{\rm 4d}/2}\,\mathcal{I}_{\rm Schur}(q) \;=\; \sum_{n\geq 0}\,\sum_{\kappa_n} \tilde q^{\,2(a_{\rm 4d}-c_{\rm 4d}) + L^{(n)}_*(\kappa_n)}\;P_{\dim\kappa_n}(\log\tilde q)\,,
\end{equation}
where the sum runs over the (higher) Rains functions and their critical points $\kappa_n$, and $P_d(\log\tilde q)$ is a polynomial in $\log\tilde q$ of degree at most $d$. Logarithms in~(\ref{eq:htasymp}) translate into terms with $\dim\kappa_n > 0$ in~(\ref{eq:htlimit}); in Appendix~\ref{app:htl} we give a general argument for the existence of at least one such term in the Lagrangian Higgsless theories of interest. The argument does not control the coefficients of the logs, so it is logically possible for a coincidence to cancel them. We conjecture that such a cancellation never occurs, and that as a consequence all Higgsless Lagrangian SCFTs give rise to strongly finite  non-rational VOAs.

Yet a third strategy to detect logs uses the fact that the vacuum character of a quasi-lisse VOA solves a monic {\it modular linear differential equation} (MLDE)~\cite{Beem:2017ooy,Arakawa:2016hkg}.  One can study the space of solutions of this MLDE and check whether  logarithmic ones occur. Some care is required: an MLDE solution that contains $\log q$ is not by itself proof of non-rationality. In general, the MLDE has both ``true'' logarithmic solutions, which correspond to pseudocharacters and therefore lie in the modular orbit of $\chi_0$, and {\it spurious} solutions, which solve the MLDE but do not appear in the $SL(2,\mathbb{Z})$ orbit of the vacuum character; spurious solutions are typically signaled by non-integer $q$-expansion coefficients. To use the MLDE as a rigorous diagnostic, one should therefore verify that a logarithmic solution is genuinely realized in~(\ref{eq:htasymp}), either by matching its $q$-expansion against the high-temperature expansion of $\chi_0$ or by an independent identification.

\subsection{Diagnostics for Higgslessness in Lagrangian theories}
\label{subsec:diagnostics}
\label{sec:scindex}

We now collect the practical tools that we use in Section~\ref{sec:classrec} to identify candidate Higgsless theories within the Lagrangian classification of~\cite{Bhardwaj:2013qia}. We restrict in what follows to {\it conformal} Lagrangian $\mathcal{N}=2$ gauge theories and assume the geometrization conjecture (\ref{eq:HBgeo}), in other terms we assume  generators of the Higgs branch chiral ring are never nilpotent.

For a Lagrangian theory with half-hypermultiplets $Q^{i=1,\ldots,\dim\rho}$ in a pseudoreal representation $\rho$ of a semisimple gauge group $G$, the Higgs branch is the hyperk\"ahler quotient~\cite{Hitchin:1986ea}
\begin{equation}\label{eq:hkquot}
\mathcal{M}_H \;=\; \mathcal{X}_Q /\!\!/\!\!/ G \;:=\; \{Q \;|\; F^A = 0,\; D^A = 0\}\big/G\;=\;\{Q\;|\; F^A=0\}\big/G_\mathbb{C}\,,
\end{equation}
where $F^A = Q^i Q^j (T_\rho^A)_{ij}$ and $D^A = Q^i (Q^\dagger)_j \Omega^{jk} (T_\rho^A)_{ik}$ are the F- and D-terms ($A = 1,\ldots,\dim G$), and the second equality is the standard fact that the hyperk\"ahler quotient by $G$ equals the holomorphic quotient by the complexified gauge group $G_\mathbb{C}$ on the F-term locus. The Higgs chiral ring $\mathcal{R}_H$ is the ring of $G$-invariant polynomials in the hypermultiplet scalars, modulo F-terms.

\paragraph{Necessary conditions.}

In our analysis in Section~\ref{sec:classrec}, we will find it very useful to impose from the outset
a simple necessary condition for Higgslessness: absence of continuous flavor symmetries. Indeed, by Noether's theorem, a continuous flavor symmetry
of $\mathcal{T}$ implies the existence of a conserved current, which
sits in the $\hat{\mathcal{B}}_1$ moment-map multiplet, whose bottom component is a non-trivial element of $\mathcal{R}_H$. A Higgsless theory therefore has no continuous flavor symmetry. This forces each matter representation to be {\it half}-hypermultiplets (rather than full hypermultiplets), since a full hyper in any representation carries a $\U(1)$ baryonic symmetry.

Another obvious necessary condition  for Higgslessness is that the number $n_h$ of hypermultiplets 
 should not exceed the number $n_v$ of vector multiplets. 
For a Lagrangian theory with $n_h$ hypermultiplets (counted in quaternionic dimension; so a half-hyper in a representation $\rho$ counts as $\tfrac12\dim\rho$ hypers) and $n_v$ vector multiplets, the naive constraint counting gives
\begin{equation}\label{eq:naive_dim}
\dim_\mathbb{H}\mathcal{M}_H\;\stackrel{\text{naive}}{=} \; n_h - n_v \; = 
24\bigl(c_{\rm 4d}-a_{\rm 4d}\bigr)\,,
\end{equation}
where the central charges are $a_{\rm 4d} = (5 n_v + n_h)/24$ and $c_{\rm 4d} = (2 n_v + n_h)/12$. The right-hand side counts $4 n_h$ real coordinates minus $3 n_v$ F- and D-term constraints minus $n_v$ gauge identifications; in particular, the constraints may be redundant, in which case the actual $\dim_\mathbb{H}\mathcal{M}_H$ is larger than $n_h - n_v$. The naive count is therefore a {\it lower bound}, and we record the consequence:
\begin{equation}\label{eq:naive_dim_bound}
n_h - n_v \;>\;0\quad\Longrightarrow\quad \mathcal{M}_H \text{ is non-trivial}.
\end{equation}
A Higgsless Lagrangian theory must therefore have $n_h\leq n_v$, equivalently $c_{\rm 4d}-a_{\rm 4d}\leq 0$, but the converse is {\it not} true.
The difference $c_{\rm 4d}-a_{\rm 4d}$ must be matched between UV and IR SCFTs related by the RG flow triggered by moving on the Higgs branch,
as it is proportional to the cubic anomaly for $U(1)_r$ (see e.g.~\cite{Rastelli:2023sfk}). 
However 
 a theory with $c_{\rm 4d}-a_{\rm 4d}\leq 0$ may still have a non-trivial Higgs branch, provided the IR effective field theory at a generic point of that branch includes massless vector multiplets. We use~(\ref{eq:naive_dim_bound}) at one point in Section~\ref{sec:classrec} to rule out a candidate quiver with $n_h - n_v = 1$ (see quiver in equation \eqref{eq:g2g2usp8so6}).

\paragraph{Higgs Hilbert series and Hall-Littlewood index.}
The Higgs chiral ring  $\mathcal{R}_H$ is finitely generated as a $\mathbb{Z}_{\geq 0}$-graded ring,
where the grading is given by the Cartan generator of the $SU(2)_R$ symmetry.
 Its {\it Hilbert series} is the formal generating function $\mathcal{I}_{\rm Higgs}(t) := \sum_{n\geq 0}\dim(\mathcal{R}_H)_n\,t^{n/2}$, which can always be brought to the rational form
\begin{equation}\label{eq:hilbser}
\mathcal{I}_{\rm Higgs}(t) \;=\; \frac{\mathtt{P}(t)}{\prod_i (1 - t^{w_i})}\,,
\end{equation}
where the denominator encodes a set of generators of weights $w_i$ and the numerator $\mathtt{P}(t)$ encodes the relations among them.
The series~(\ref{eq:hilbser}) is the cleanest possible diagnostic:
\begin{equation}\label{eq:hilbhiggless}
\mathcal{I}_{\rm Higgs}(t) \;=\; 1\quad\Longleftrightarrow\quad \mathcal{T}\text{ is Higgsless}.
\end{equation}
Indeed, since the coefficients of $\mathcal{I}_{\rm Higgs}$ count graded dimensions and are therefore non-negative, $\mathcal{I}_{\rm Higgs}(t)=1$ holds if and only if $\mathcal{R}_H=\mathbb{C}$. In turn, $\mathcal{R}_H=\mathbb{C}$ is equivalent to Higgslessness: if $\mathcal{M}_H$ is a point, then $\mathbb{C}[\mathcal{M}_H]=\mathbb{C}$, and the geometrization conjecture~(\ref{eq:HBgeo}) (assumed throughout this section) excludes nilpotent elements, so that $\mathcal{R}_H=\mathbb{C}[\mathcal{M}_H]=\mathbb{C}$; conversely, $\mathcal{R}_H=\mathbb{C}$ trivially implies that $\mathcal{M}_H$ is a point.
For Lagrangian theories the Hilbert series~(\ref{eq:hilbser}) can in principle be computed algorithmically using \texttt{Macaulay2}~\cite{M2}: one inputs the hypermultiplet scalars and the F-term ideal, computes the Hilbert series of the quotient ring (refined by gauge fugacities), and then performs the contour integral over the gauge fugacities to project to $G$-invariants. In practice, both steps become prohibitive as the rank of the theory grows.

A cheaper but less direct diagnostic is the {\it Hall-Littlewood (HL) limit} of the superconformal index. The HL chiral ring is the ring of bottom components of $\hat{\mathcal{B}}_R$ {\it and} $\bar{\mathcal{D}}_{R(j_1,0)}$ multiplets (the ``HL'' multiplets in the notation of~\cite{Beem:2013sza}); $\mathcal{R}_H$ sits inside it as the $j_1=0$, $r=0$ subring. The unflavored HL index takes the form
\begin{equation}\label{eq:hlgenform1}
\mathcal{I}_{\rm HL}(t) \;=\; \mathtt{P}_{\rm HL}(t)\;\frac{\prod_{i\in {\rm HL}^-}(1-t^{h_i})}{\prod_{i\in {\rm HL}^+}(1-t^{h_i})}\,,
\end{equation}
in close analogy with the Higgs branch Hilbert series~(\ref{eq:hilbser}): ${\rm HL}^+$ and ${\rm HL}^-$ stand for {\it bosonic} and {\it fermionic} HL generators respectively, and the polynomial $\mathtt{P}_{\rm HL}(t)$ encodes the relations among them (and their syzygies). Crucially, every HL generator outside $\mathcal{R}_H$ contains at least one gaugino (a $\lambda^1_+$ or $\tilde\lambda^1_{\dot+}$, see Table~\ref{tab:schurmultipletslagletters}) and is therefore nilpotent in $\mathcal{R}_\mathcal{V}$.
The HL diagnostic for Higgslessness reads
\begin{equation}\label{eq:hlhiggless}
\mathcal{I}_{\rm HL}(t)\text{ is a polynomial in }t^{1/2}\quad\Longleftrightarrow\quad \mathcal{T}\text{ is Higgsless},
\end{equation}
where each direction holds under a qualification that we now spell out.  First, the implication ``$\Leftarrow$'' uses the geometrization conjecture~(\ref{eq:HBgeo}): if $\mathcal{R}_H$ were allowed to contain nilpotents, a polynomial $\mathcal{I}_{\rm HL}$ would be consistent with a non-trivial nilpotent contribution to $\mathcal{R}_H$, which is excluded by~(\ref{eq:HBgeo}). Second, the implication ``$\Rightarrow$'' requires that no accidental cancellation between numerator and denominator of~(\ref{eq:hlgenform1}) makes $\mathcal{I}_{\rm HL}$ a polynomial without the theory being Higgsless. For theories without continuous flavor symmetry, which is the only case relevant for us, such cancellations seem  difficult to engineer and we know of no counterexample.

The HL index is  much cheaper to compute than $\mathcal{I}_{\rm Higgs}$, since it can be written as a contour integral over the index of the free theory rather than requiring a Gr\"obner-basis computation. The price is the looser correspondence with the Higgs branch encoded in the two qualifications above. In practice, we use $\mathcal{I}_{\rm HL}$ throughout the case analysis of Section~\ref{sec:classrec} and the appendices, and corroborate with $\mathcal{I}_{\rm Higgs}$ in the few cases where both are tractable.

\section{Classifying Higgsless Lagrangian SCFTs}
\label{sec:classrec}

Four-dimensional $\mathcal{N}=2$ superconformal Lagrangian gauge theories were classified by Bhardwaj and Tachikawa~\cite{Bhardwaj:2013qia}.  In this section we aim to identify the subset of these theories which have a trivial Higgs branch.

\subsection{Strategy and main result}
\label{subsec:strategy}

Two conceptual ingredients drive the analysis.
The first is the {\it necessary} condition for Higgslessness discussed in Section~\ref{subsec:diagnostics}: a Higgsless theory cannot carry any continuous flavor symmetry, which for a Lagrangian gauge theory means in particular that the matter content can only consist of half-hypermultiplets. 
The second is a {\it sufficient} condition for a theory to have a Higgs branch. To prove that a given theory has a Higgs branch, it is enough to identify a single gauge-invariant operator built from the half-hypermultiplets that is not set to zero by the F-term constraints. Such an operator is then a non-trivial generator of the Higgs chiral ring (and under the Higgs branch geometrization conjecture, it is non-nilpotent). 

Our strategy is to start from the Bhardwaj-Tachikawa classification \cite{Bhardwaj:2013qia}, restrict to candidates with no continuous flavor symmetry, and then, for each candidate, attempt to identify a non-vanishing Higgs branch operator. The main simplification used in this search is to look for such operators inside suitable \textit{subquivers}. By a subquiver we mean the quiver obtained by restricting to a subset of gauge nodes and retaining only those half-hypermultiplets whose gauge charges lie entirely within this subset. All half-hypermultiplets connecting this subset to the rest of the quiver are deleted. See section \ref{subsec:subquiver} for more details on the definition of subquivers. With this definition in mind, the key tool we use is the following lemma. 

\medskip
\noindent
{\it \bf Subquiver lemma:} {\it If a subquiver of a Lagrangian SCFT admits a gauge-invariant operator built from half-hypermultiplets that is not set to zero by the F-terms of the sub-quiver, then the full theory has a non-trivial Higgs branch.}

\medskip
\noindent
A proof of this lemma is given in Section \ref{subsec:subquiver}. The lemma lets us rule out broad families of theories at once, by identifying a small recurring sub-pattern on which a Higgs branch operator can be constructed. A caveat is that, at low ranks, additional relations among invariant tensors may appear, relating our proposed non-vanishing Higgs branch operator to an expression built from simpler operators that vanish. However, since we explicitly check the existence of the proposed Higgs operators in some examples starting from the quivers with lowest possible ranks, this allows us to identify the exceptional cases and gain some confidence in the remaining higher-rank examples. When we cannot propose a Higgs branch operator for a theory, we consider it a candidate Higgsless Lagrangian SCFT. The outcome of the case-by-case analysis in Section~\ref{subsec:caseanalysis} is the following

\medskip
\noindent
{\it \bf Result:} {\it The complete list of candidate interacting Higgsless Lagrangian SCFTs is} 
\begin{enumerate}
\item[(i)] {\it $\USp(4)$ with a half-hypermultiplet in the $\mathbf{16}$;}
\item[(ii)] {\it $\SU(3)\times\SU(2)$ with half-hypermultiplets in the $\mathbf{8}\times\mathbf{2}$;}
\item[(iii)] {\it For each $m\in 4\mathbb{Z}_{>0}$, the trivalent quiver with central $\USp(m)$ node}:

\hspace*{-0.9cm}
\begin{tikzpicture}[baseline={(0,-0.5ex)}]
		\node        (circle1)      at (0,0)   { $\USp(m)$};
		\node        (circle3)    at (1.8,0)    {$\SO(m)$};
		\node        (circle4)    at (-2,0)    {$\SO(\frac{m+4}{2})$};
		\node        (circle5)    at (0,1)    {$\SO(\frac{m+4}{2})$};
		\node        (circle6)    at (4,0)    {$\USp(m-4)$};
		\node        (circle7)    at (6.5,0)    {$\SO(m-4)$};
		\node        (circle8)    at (8.4,0)    {$\cdots$};
		\node        (circle9)    at (9.8,0)    {$\USp(4)$};
		\node        (circle10)    at (11.6,0)    {$\SO(4)$};
		\draw[-] (circle1.east)  -- (circle3.west) node[pos=0.5, above, inner sep=1pt] {$q_3$};
		\draw[-] (circle1.north) -- (circle5.south)node[pos=0.5, right, inner sep=1pt] {$q_2$};
		\draw[-] (circle1.west) -- (circle4.east)node[pos=0.5, above, inner sep=1pt] {$q_1$};
		\draw[-] (circle3.east) -- (circle6.west)node[pos=0.5, above, inner sep=1pt] {$q_4$};
		\draw[-] (circle6.east) -- (circle7.west)node[pos=0.5, above, inner sep=1pt] {$q_5$};
		\draw[-] (circle7.east) -- (circle8.west);
		\draw[-] (circle8.east) -- (circle9.west);
		\draw[-] (circle9.east) -- (circle10.west)node[pos=0.5, above, inner sep=1pt] {$q_{k+2}$} ;
\end{tikzpicture}

\item[(iv)] {\it The quiver $\mathbf{\frac{1}{2}asym3}\,$--$\,\USp(8)\,$--$\,\SO(6)$.}
\end{enumerate}
{\it Theories (i) and (ii) will be proved to be Higgsless in Section~\ref{sec:examples}. For the infinite family (iii), we present a strong but non-rigorous argument in Section \ref{subsec:treegraphs} and show that the HL index for the first two members truncates. We expect the whole infinite sequence to be Higgsless. For theory (iv), we present a partial argument in Section \ref{subsec:single_node} and show that its HL index also truncates.
}

\medskip
\noindent
Table~\ref{tab:higgsless_summary} summarizes the list, together with some basic data and the status of the evidence in each case.

\begin{table}[t]
\centering
\renewcommand{\arraystretch}{2}
\resizebox{\textwidth}{!}{
\begin{tabular}{|c|c|c|c|c|}
\hline
Theory & $n_h$ & $n_v$ & $c_{\rm 2d}$ & Evidence for Higgslessness \\
\hline\hline
(i) $\USp(4)-\tfrac12\,\mathbf{16}$ & $8$ & $10$ & $-28$ & \makecell{$\mathcal{I}_{\text{Higgs}}=1$, $\mathcal{I}_{\text{HL}}$ truncation, \\ $C_2$-cofinite VOA (Section~\ref{sec:usp416b2})} \\
\hline
(ii) $\SU(3)\frac{\ \tfrac12\,\left(\mathbf{8}\times\mathbf{2}\right)\ }{}\SU(2)$  & $8$ & $11$ & $-30$ & \makecell{$\mathcal{I}_{\text{Higgs}}=1$, $\mathcal{I}_{\text{HL}}$ truncation, \\ $C_2$-cofinite VOA (Section~\ref{sec:su3su282})} \\
\hline
(iii) trivalent $\USp(m)$ family, $m\in4\mathbb{Z}_{>0}$ & \multicolumn{2}{c|}{$n_h-n_v=-\tfrac{m+4}{2}$} & $-(2n_v+n_h)$ & \makecell{systematic operator enumeration,\\ $\mathcal{I}_{\text{HL}}$ truncation for $m=4,8$} \\
\hline
(iv) $\mathbf{\tfrac{1}{2}asym3}$--$\USp(8)$--$\SO(6)$ & $48$ & $51$ & $-150$ & \makecell{partial operator enumeration,\\ $\mathcal{I}_{\text{HL}}$ truncation} \\
\hline
\end{tabular}}
\caption{The candidate interacting Higgsless Lagrangian SCFTs, with the number of hypermultiplets and vector multiplets, the two-dimensional central charge $c_{\rm 2d}=-12c_{\rm 4d}=-(2n_v+n_h)$, and the status of the evidence for the triviality of the Higgs branch.}
\label{tab:higgsless_summary}
\end{table}

\medskip
\noindent
The remainder of this section justifies this result.

\subsection{Review of the Bhardwaj-Tachikawa classification}
\label{subsec:BTreview}

In this subsection, we summarize the  classification~\cite{Bhardwaj:2013qia}, which also organizes our search. A glossary of terms is collected in Appendix \ref{app:gloss} for the reader's convenience.

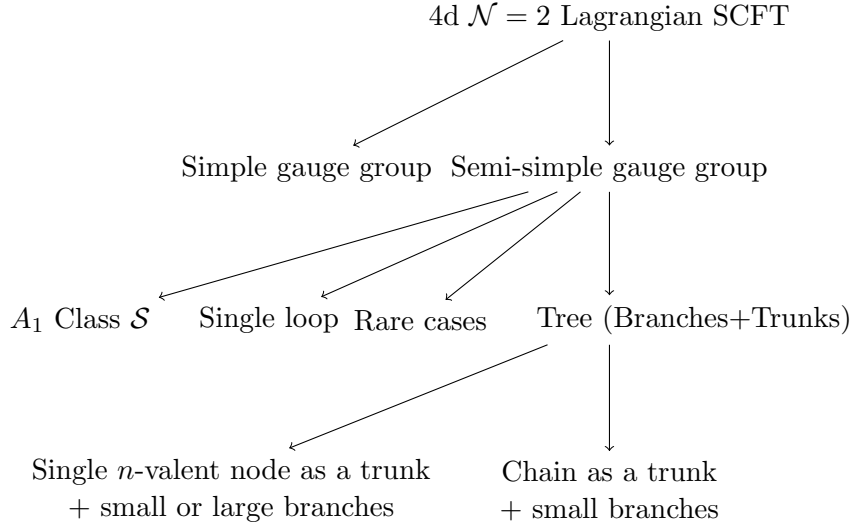
\begin{figure}[t]
			\centering
			\begin{tikzpicture}[]

				\node        (node0)    at (2,1)    { 4d $\mathcal{N}=2$ Lagrangian SCFT};
				\node       (nodeA)    at (-2,-1)    {Simple gauge group};
				\node       (nodeB)    at (2,-1)    {Semi-simple gauge group};		
				
				\node       (nodeC)    at (-5,-3)    {$A_1$ Class $\mathcal{S}$};		
				\node       (nodeD)    at (-2.5,-3)    {Single loop};				
				\node       (nodeE)    at (-0.5,-3)    {Rare cases};
				\node       (nodeF)    at (2,-3)    {\;\;\;\;\;\;\;\;\;\;\;\;\;\;\;\;\;\;\;\;\;Tree (Branches+Trunks)};
				
				\node       (nodeG)    at (-3,-5)    {Single $n$-valent node as a trunk };
				\node       (nodeG2)    at (-3,-5.5)    {+ small or large branches};
				\node       (nodeH)    at (2,-5)    {Chain as a trunk };
				\node       (nodeH2)    at (2,-5.5)    {+ small branches};
				
				\draw[->] (node0) -- (nodeA);
				\draw[->] (node0) -- (nodeB);
				
				\draw[->] (nodeB) -- (nodeC);
				\draw[->] (nodeB) -- (nodeD);
				\draw[->] (nodeB) -- (nodeE);
				\draw[->] (nodeB) -- (nodeF);		
				
				\draw[->] (nodeF) -- (nodeG);
				\draw[->] (nodeF) -- (nodeH);		
				
			\end{tikzpicture}
			\caption{Figure summarizing the strategy for classification of 4d $\mathcal{N}=2$ Lagrangian SCFTs followed in \cite{Bhardwaj:2013qia}. See text for more details.}
			\label{fig:classification}
		\end{figure}

A Lagrangian $\mathcal{N}=2$ gauge theory is specified by a semisimple gauge group $G=\prod_{i=1}^k G_i$ together with half-hypermultiplets in a pseudoreal representation of $G$. The conformal theories with simple gauge group are tabulated directly in~\cite{Bhardwaj:2013qia}. The theories with genuinely semi-simple gauge group are organized by viewing them as built from {\it $n$-gons}, that is, half-hypermultiplets transforming in irreducible representations of $n$ simple factors of $G$. Only $n\leq 3$ is possible. The only non-degenerate $3$-gon\footnote{The $\SU(2)^2-\USp(4)$ $3$-gon is considered as an $\SO(4)-\USp(4)$ $2$-gon.} is the $\SU(2)^3$ tri-fundamental, and theories that contain at least one such tri-fundamental are the $A_1$ class $\mathcal{S}$ theories of~\cite{Gaiotto:2009we, Gaiotto:2009hg}. The remaining semi-simple Lagrangian theories are built out of $1$-gons and $2$-gons, and fall into three families:
\begin{enumerate}
\item[--] {\it rare cases}: three sporadic theories, namely
$\SU(3)\times\SU(2)$ with a half-hyper in $\mathbf{8}\times\mathbf{2}$, $\SO(9)\times\USp(6)$ with a half-hyper in $\mathbf{S}\times\mathbf{6}$ and a full hyper in $\mathbf{9}\times\mathbf{1}$, and $G_2\times\SU(4)$ with a full hyper in $\mathbf{7}\times\mathbf{4}$;
\item[--] {\it single loops};
\item[--] {\it tree graphs}.
\end{enumerate}
Theories built out of $1$-gons and $2$-gons admit a quiver diagram presentation, with nodes corresponding to gauge or flavor groups and edges to bi-fundamental hypermultiplets. A {\it tree graph} is built by attaching {\it branches} to the ends of a {\it trunk} such that each gauge node is balanced (vanishing one-loop $\beta$-function). The trunk is part of the \textit{main subgraph}, by which we mean the part of the
quiver whose hypermultiplets transform in bi-fundamental representations of
adjacent nodes. The trunk is characterized as the part of this subgraph along
which the {\it current} vanishes\footnote{The precise definition of current is
given in \cite{Bhardwaj:2013qia} and recalled in Appendix~\ref{app:gloss}.}. There are two types of trunks, namely a \text{chain} of nodes or a \text{single node}, and a finite list of allowed branches, distinguished as {\it small} or {\it large} according to the sign of the \textit{branch current} \footnote{See Appendix \ref{app:gloss} for the definition of branch current.}. Only small branches can be attached to the ends of a chain trunk. The classification is summarized in Figure \ref{fig:classification}.

\subsection{Subquiver lemma}
\label{subsec:subquiver}
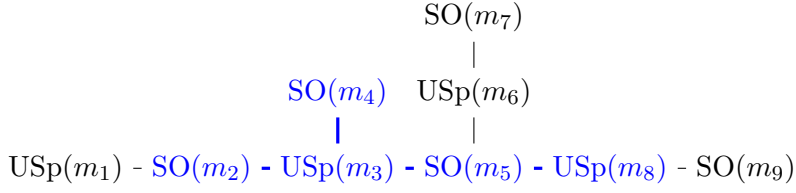
\begin{figure}[t]
			\centering
			\begin{tikzpicture}[]

				\node        (circle1)    at (-1.8,0)    { $\USp(m_1)$};
				\node        (circle2)    at (0,0)    { $\textcolor{blue}{\SO(m_2)}$};
				\node        (circle3)    at (1.8,1)    { $\textcolor{blue}{\SO(m_4)}$};
				\node        (circle4)    at (1.8,0)    {$\textcolor{blue}{\USp(m_3)}$};
				\node        (circle5)   at (3.6,0)    {$\textcolor{blue}{\SO(m_5)}$};
				\node        (circle6)   at (3.6,1)    {$\USp(m_6)$};
				\node        (circle7)   at (3.6,2)    {$\SO(m_7)$};
				\node        (circle8)    at (5.4,0)    { $\textcolor{blue}{\USp(m_{8})}$};
				\node        (circle9)   at (7.2,0)    {$\SO(m_9)$};
				
				\draw[-] (circle1.east) -- (circle2.west);
				\draw[blue,-,very thick] (circle2.east) -- (circle4.west);
				\draw[blue,-,very thick] (circle4.east) -- (circle5.west);
				\draw[blue,-,very thick] (circle5.east) -- (circle8.west);
				\draw[-] (circle8.east) -- (circle9.west);
				\draw[blue,-,very thick] (circle4.north) -- (circle3.south);
				\draw[-] (circle5.north) -- (circle6.south);
				\draw[-] (circle6.north) -- (circle7.south);
			\end{tikzpicture}
			\caption{An example of a sub-quiver (colored blue) inside a full quiver.}
			\label{fig:subquiver}
		\end{figure}

In order to ease our search, we will use the subquiver lemma, mentioned in Section \ref{subsec:strategy}. Here, we provide a proof of the subquiver lemma. Let $\mathcal{T}$ be a Lagrangian SCFT that admits a quiver description. A subquiver $\mathcal{T}_{\rm sub}\subset \mathcal{T}$, is a quiver obtained by deleting all half-hypers connecting $\mathcal{T}_{\rm sub}$ to the rest of $\mathcal{T}$, while keeping the gauge nodes of $\mathcal{T}_{\rm sub}$. Importantly, the removed gauge nodes are not converted into flavor symmetries of $\mathcal{T}_{\rm sub}$, since the half-hypers transformed under them are removed as well. An illustrative example of a subquiver $\mathcal{T}_{\rm sub}\subset \mathcal{T}$ is given in Figure \ref{fig:subquiver}. We denote the half-hypers in the subquiver by $q_{\rm sub}$ and all remaining half-hypers by $q_{\rm rest}$.  Let $R_{\rm sub}=\mathbb{C}[q_{\rm sub}]$ and $R=\mathbb{C}[q_{\rm sub},q_{\rm rest}]$, and let $I_{\rm sub}\subset R_{\rm sub}$ and $I\subset R$ be the corresponding ideals generated by the corresponding F-terms. 

The basic statement of the lemma is the following: if $\mathcal{O}\in R_{\rm sub}$ is gauge invariant under the gauge group of $\mathcal{T}_{\rm sub}$ and is non-zero in the quotient $R_{\rm sub}/I_{\rm sub}$, then the same operator $\mathcal{O}$ is a nonzero gauge-invariant operator in the full quotient $R/I$. To show this, let us define the map $\rho$, which sets $q_{\text{rest}}\to0$. Under this map
\begin{equation}
\rho: \mathbb{C}[q_{\text{sub}},q_{\text{rest}}]\to\mathbb{C}[q_{\text{sub}}]~,~ \rho(I)= I_{\text{sub}},
\end{equation}
$\rho$ maps the ideal generated by F-terms of the full quiver to that of the subquiver. This follows simply from the fact that F-terms are quadratic in the hypermultiplets: setting $q_{\text{rest}}$ to zero maps the F-terms at the nodes of $\mathcal{T}_{\rm sub}$ to the F-terms of the subquiver, and annihilates the F-terms at the remaining nodes. An operator $O\in I$ can be written as
\begin{equation}
    O=\sum_i g_i h_i~,
\end{equation}
where $g_i\in R$ and $h_i$ are the generators of the ideal, \textit{i.e.}\ the F-terms. Under the map $\rho$,
\begin{equation}
    \rho(O)=\sum_i\rho(g_i)\rho(h_i)~,
\end{equation}
and $\rho(h_i)\in I_{\rm sub}$. Therefore, $\rho(O)\in I_{\text{sub}}$. In simple terms, if an operator $O$ is set to zero by the F-terms of the full quiver, its image $\rho(O)$ is set to zero by the F-terms of the subquiver. By contraposition, if there exists an operator $\mathcal{O}$ built from $q_{\rm sub}$ alone (so that $\rho(\mathcal{O})=\mathcal{O}$) that is not set to zero by the F-terms of the subquiver, it cannot be set to zero by the F-terms of the full quiver.

It is important to note that $\mathcal{O}$ need not be a Higgs branch operator of the subquiver itself. The subquiver is generically not conformal (it is conformal only if $\mathcal{T}_{\text{sub}}=\mathcal{T}$), so $\mathcal{O}$ may well be nilpotent in $R_{\rm sub}/I_{\rm sub}$. However, since the full theory $\mathcal{T}$ {\it is} conformal, the Higgs branch geometrization conjecture~(\ref{eq:HBgeo}) implies that $\mathcal{O}$, viewed as an operator of $\mathcal{T}$, is non-nilpotent, and hence a bona fide Higgs branch operator: the full theory has a non-trivial Higgs branch.

In the rest of this section we deal mostly with bi-fundamental half-hypers of $\SO\times \USp$. For ease of reference, the F-term equations at a $\USp$ and at an $\SO$ gauge node read
\begin{equation}\label{eq:Fterms}
\sum_A (q)^{i_A}_a\,(q)^{j_A}_b\,\delta_{i_A j_A} = 0\quad(\USp\text{ node}),\qquad
\sum_A (q)^{i}_{a_A}\,(q)^{j}_{b_A}\,\Omega^{a_A b_A} = 0\quad(\SO\text{ node}),
\end{equation}
where $A$ labels the $\SO$ ($\USp$) gauge groups attached to the $\USp$ ($\SO$) node in question. For more general representations the bilinears $(qq)$ must be projected onto the singlet of the flavor group and the adjoint of the gauge group.

\subsection{Diagrammatic notation}
\label{subsec:diagnotation}

The Higgs branch operators we shall construct involve many index contractions. To control the bookkeeping, we adopt a simple diagrammatic notation, in which a half-hypermultiplet 
$\tikz[baseline=-0.5ex]{
	\node[draw,minimum size=1mm] (a) at (0,0) {};
	\node[draw,minimum size=1mm,fill] (b) at (0.7,0) {};
	\draw (a)--(b);
}$
is drawn as an edge between two boxes, with the unfilled (filled) box representing the $\SO$ ($\USp$) factor of its gauge representation. When an index on a box is fully contracted with another index, the box is replaced by a circle of the same color convention. For example, the gauge singlet $q^a_i p^b_j \delta_{ab}\Omega^{ij}$ is represented as
\begin{equation}
q^a_i p^b_j\, \delta_{ab}\Omega^{ij}\;=\;
\begin{tikzpicture}[baseline={(0,-0.5ex)}]
	\node[draw,circle,minimum size=1mm] (c1) at (0,0) {};
	\node[draw,circle,minimum size=1mm,fill] (c2) at (1,0) {};
	\draw[-] (c1) to[bend left=40] node[midway, above, inner sep=1pt]{$q$} (c2);
	\draw      (c1) to[bend right=40] node[midway, below, inner sep=1pt]{$p$} (c2);
\end{tikzpicture}\,.
\end{equation}
A gauge-invariant operator corresponds to a closed diagram. The F-term constraints~(\ref{eq:Fterms}) take the diagrammatic form
\begin{equation}
\centering
\begin{tikzpicture}[baseline={(0,-0.5ex)}]
	\node () at (-7.4,0.6)  {\Large$\displaystyle \sum_{I}$};
	\node[scale=2] () at (-6.6,0.8)  {\Huge$($};
	\node[draw,circle,minimum size=1mm,fill] (b1) at (-6,0) {};
	\node[draw,minimum size=1mm] (b2) at (-6,1.5) {};
	\node[draw,minimum size=1mm] (b3) at (-4.5,0) {};
	\draw[-] (b1) -- (b2) node[pos=0.5, left, inner sep=1pt] {$p_I$};
	\draw[-] (b1) -- (b3) node[pos=0.5, below, inner sep=1pt] {$q_I$};
	\node[scale=2] () at (-3.4,0.8)  {\Huge$)_{\text{\tiny Adj}}$};
	\node () at (-3,0.8)  {\Large$=0\,,$};

	\node () at (-1.4,0.6)  {\Large$\displaystyle \sum_{I}$};
	\node[scale=2] () at (-0.6,0.8)  {\Huge$($};
	\node[draw,circle,minimum size=1mm] (c1) at (0,0) {};
	\node[draw,minimum size=1mm,fill] (c2) at (0,1.5) {};
	\node[draw,minimum size=1mm,fill] (c3) at (1.5,0) {};
	\draw[-] (c1) -- (c2) node[pos=0.5, left, inner sep=1pt] {$p_I$};
	\draw[-] (c1) -- (c3) node[pos=0.5, below, inner sep=1pt] {$q_I$};
	\node[scale=2] () at (2.4,0.8)  {\Huge$)_{\text{\tiny Adj}}$};
	\node () at (3,0.8)  {\Large$=0\,,$};
\end{tikzpicture}
\end{equation}
where the subscript ``$\mathrm{Adj}$'' instructs us to project the open indices on the squares onto the adjoint representation. The sum runs over all hypermultiplets attached to the node, and the central circle is filled or unfilled according to whether the central gauge group is $\USp$ or $\SO$. (When the open indices are simply contracted using the invariant tensor in the fundamental representation, no adjoint projection is required.) Consider the application of the following F-term diagrammatically
\begin{equation}
\centering
\begin{tikzpicture}[baseline={(0,-0.5ex)}]
	\node[draw,circle,minimum size=1mm] (c1) at (0,0) {};
	\node[draw,minimum size=1mm,fill] (c2) at (0,1.5) {};
	\node[draw,minimum size=1mm,fill] (c3) at (1.5,0) {};
	\draw[-] (c1) -- (c2) node[pos=0.5, left, inner sep=1pt] {$p$};
	\draw[-] (c1) -- (c3) node[pos=0.5, below, inner sep=1pt] {$q$};
	\node () at (2.3,0.8)  {\Large$=0\,,$};
\end{tikzpicture}
\end{equation}
Contracting the filled squares above, the diagram
$\begin{tikzpicture}[baseline={(0,-0.5ex)}]
	\node[draw,circle,minimum size=1mm] (c1) at (0,0) {};
	\node[draw,circle,minimum size=1mm,fill] (c2) at (0.8,0) {};
	\draw[-] (c1) to[bend left=40] node[midway, above, inner sep=1pt]{$q$}(c2);
	\draw      (c1) to[bend right=40] node[midway, below, inner sep=1pt]{$p$}(c2);
\end{tikzpicture}$
is set to zero.

\subsection{Case-by-case analysis}
\label{subsec:caseanalysis}

We now go through the families of the Bhardwaj-Tachikawa classification in turn, in each case applying the strategy discussed in Section~\ref{subsec:strategy}: restrict to candidates without continuous flavor symmetry, then test the remaining candidates for the existence of a gauge-invariant Higgs branch operator that survives the F-terms.

\subsubsection{Simple gauge group, $A_1$ class $\mathcal{S}$, rare cases}

The Higgsless theories in three of the Bhardwaj-Tachikawa families can be identified using only the flavor symmetry constraint.

\paragraph{Simple gauge group.}
The only conformal Lagrangian theory with a simple gauge group and no continuous flavor symmetry is $\USp(4)$ with a half-hypermultiplet in the $\mathbf{16}$. It is Higgsless, as will be shown in Section~\ref{sec:usp416b2}.

\paragraph{$A_1$ class $\mathcal{S}$.}
All $A_1$ class $\mathcal{S}$ theories have a Higgs branch. Punctured theories carry flavor symmetries from the punctures, and the Higgs branch corresponding to Riemann surfaces without punctures was determined in \cite{Hanany:2010qu}.

\paragraph{Rare cases.}
Of the three sporadic rare cases, only $\SU(3)\times\SU(2)$ with a half-hypermultiplet in $\mathbf{8}\times\mathbf{2}$ is free of continuous flavor symmetry. It is Higgsless, as will be shown in Section \ref{sec:su3su282}.

\subsubsection{Single loops}
\label{subsec:singleloop}

The only loop quiver with no continuous flavor symmetry is the $\SO(m)$--$\USp(m-2)$ loop with bi-fundamental half-hypermultiplets, shown in Figure~\ref{fig:loop}.\footnote{For the loop with an $\SO(8)$--$\USp(6)$ edge, the bi-fundamental can be in any of $\tfrac{1}{2}\mathbf{8}_v\!\times\!\mathbf{vect}$, $\tfrac{1}{2}\mathbf{8}_s\!\times\!\mathbf{vect}$, or $\tfrac{1}{2}\mathbf{8}_c\!\times\!\mathbf{vect}$.}
\begin{figure}[t]
\centering
\begin{tikzpicture}[]
	\node        (circle1)    at (-0.5,0)    { $\SO(m)_1$};
	\node       (circle2)    at (2,0)    {$\USp(m-2)_1$};
	\node (A) at (4,0) {$\cdots$};
	\node        (circle3)   at (6,0)    {$\SO(m)_n$};
	\node        (circle4)    at (6,1)    { $\USp(m-2)_n$};
	\node (B) at (4,1) {$\cdots$};
	\node       (circle5)    at (2,1)    {$\SO(m)_k$};
	\node        (circle6)   at (-0.5,1)    {$\USp(m-2)_k$};
	\draw[-] (circle1.east) -- (circle2.west);
	\draw[-] (circle2.east) -- (A.west);
	\draw[-] (A.east) -- (circle3.west);
	\draw[-] (circle3.north) -- (circle4.south);
	\draw[-] (circle6.east) -- (circle5.west);
	\draw[-] (circle5.east) -- (B.west);
	\draw[-] (circle4.west) -- (B.east);
	\draw[-] (circle1.north) -- (circle6.south);
\end{tikzpicture}
\caption{An $\SO(m)$--$\USp(m-2)$ loop of length $2k$ with bi-fundamental half-hypermultiplets.}
\label{fig:loop}
\end{figure}
For any such loop, the operator
\begin{align}
\mathcal{O}_{\rm loop}\;=\; (q_1)^{i_1}_{a_1}\,\delta_{i_1 j_1}\,(q_2)^{i_2}_{b_1}\,\Omega^{a_1 b_1}\,(q_3)^{j_2}_{a_2}\,\delta_{i_2 j_2}\,(q_4)^{i_3}_{b_2}\,\Omega^{a_2 b_2}\,\cdots\,(q_{2k-1})^{j_k}_{a_k}\,\delta_{i_k j_k}\,(q_{2k})^{j_1}_{b_k}\,\Omega^{a_k b_k}
\end{align}
is gauge-invariant and has dimension $E=2k$. Here $(q_{2n})^{i_{n+1}}_{a_n}$ denotes the half-hyper between $\USp(m-2)_n$ and $\SO(m)_{n+1}$, and $(q_{2n-1})^{i_n}_{a_n}$ the half-hyper between $\SO(m)_n$ and $\USp(m-2)_n$, with $i_n$ a $\SO(m)_n$ fundamental and $a_n$ a $\USp(m-2)_n$ fundamental, $n=1,\ldots,k$ (cyclically, $i_{k+1}\equiv i_1$). In our diagrammatic notation,
\begin{equation}
\mathcal{O}_{\rm loop}\;=\;
	\begin{tikzpicture}[baseline={(2,0.5)}]
		\node[draw,circle,minimum size=1mm,fill]        (circlenm1)      at (3,0)  {} ;
		\node[draw,circle,minimum size=1mm]        (circlen)      at (1.5,0)  {} ;
		\node[draw,circle,minimum size=1mm,fill]        (circle1)      at (0,0)  {} ;
		\node[draw,circle,minimum size=1mm]        (circle2)      at (0,1.5)  {} ;
		\node[draw,circle,minimum size=1mm,fill]        (circle3)    at (1.5,1.5)    {};
		\node[draw,circle,minimum size=1mm]        (circle4)    at (3,1.5)    {};
		\node[draw,circle,minimum size=1mm,fill]        (circle5)    at (4.5,1.5)    {};
		\node[draw,circle,minimum size=1mm]        (circle6)    at (6,1.5)    {};
		\node[draw,circle,minimum size=1mm,fill]        (circle7)    at (6,0)    {};
		\node[draw,circle,minimum size=1mm]        (circle8)    at (4.5,0)    {};
		\draw[-] (circle1) -- (circle2) node[pos=0.5, left, inner sep=1pt] {$q_1$};
		\draw[-] (circle2) -- (circle3) node[pos=0.5, above, inner sep=1pt] {$q_2$};
		\draw[-] (circle3) -- (circle4) node[pos=0.5, above, inner sep=1pt] {$q_3$};
		\draw[dashed] (circle4) -- (circle5) node[pos=0.5, above, inner sep=1pt] {};
		\draw[-] (circle5) -- (circle6) node[pos=0.5, above, inner sep=1pt] {$q_n$};
		\draw[-] (circle6) -- (circle7) node[pos=0.5, right, inner sep=1pt] {$q_{n+1}$};
		\draw[-] (circle7) -- (circle8) node[pos=0.5, below, inner sep=1pt] {$q_{n+2}$};
		\draw[dashed] (circle8) -- (circlenm1) node[pos=0.5, below, inner sep=1pt] {};
		\draw[-] (circlenm1) -- (circlen) node[pos=0.5, below, inner sep=1pt] {$q_{2k-1}$};
		\draw[-] (circlen) -- (circle1) node[pos=0.5, below, inner sep=1pt] {$q_{2k}$};
	\end{tikzpicture}
\end{equation}
\noindent
It is not annihilated by the F-terms in equation \eqref{eq:Fterms} because the half-hypers on each edge are contracted with half-hypers on the adjacent edges, unlike in the F-terms, where two half-hypers on the same edge are contracted. Therefore, every single-loop theory has a non-trivial Higgs branch. The Hall-Littlewood indices of several low-rank examples, computed in Appendix \ref{app:hlindquiv}, do not truncate, consistent with this conclusion.

\subsubsection{Tree graphs}
\label{subsec:treegraphs}

A tree graph is built by attaching {\it branches} to the ends of a {\it trunk} such that every gauge node is balanced. The trunks built from half-hypermultiplets are
\begin{enumerate}
\item[--] an $\SO(m)$--$\USp(m-2)$ chain of length $2k$;
\item[--] a single $\SO(m)$, $\USp(m)$, or $G_2$ node.
\end{enumerate}
The full list of admissible branches\footnote{A number of branches in the list of \cite{Bhardwaj:2013qia} are related by $\SO(8)$ triality or by the $\SO(5)\cong\USp(4)$ isomorphism, and need not be considered separately. We discuss these dualities in Appendix \ref{app:branches}.} are given in \cite{Bhardwaj:2013qia}, and the subset relevant to us is collected in Appendix \ref{app:branches}. We follow the convention of treating a $1$-gon half-hyper attached to a gauge node\footnote{The $1$-gon half-hyper may be in the fundamental of a $\USp(m)$ node, or in one of the special representations $\mathbf{asym3}$ of $\USp(6)$ or $\USp(8)$, $\mathbf{S}$ of $\SO(11)$ or $\SO(13)$, and $\mathbf{S}$ or $\mathbf{C}$ of $\SO(12)$} as a separate {\it node} of the quiver, although \cite{Bhardwaj:2013qia} does not. A node attached to $k$ branches is called {\it $k$-valent}. If a $1$-gon half-hyper appears with multiplicity greater than one at the same node, the resulting theory has at least $\SO(2)$ flavor symmetry and is discarded.

We now treat the two types of trunk in turn.

\paragraph{Chain trunk.}
Any tree graph with an $\SO(m)$--$\USp(m-2)$ chain trunk contains one of the three subgraphs in Figure~\ref{fig:chain}.\footnote{If the outer $\SO(n_1)$ and/or $\SO(n_k)$ node is replaced by a single half-hyper ($n_1=1$ and/or $n_k=1$), or by one of the admissible special branches, the same subgraph is still present.} On any such subgraph, the operator
\begin{equation}\label{eq:HBop_chain}
\mathcal{O}_{\rm chain} \;=\; (q_n)^a_i\,(q_n)^a_j\,(q_n)^b_i\,(q_n)^b_j\,,\qquad 2\leq n\leq k-2\,,
\end{equation}
is gauge-invariant and not annihilated by the F-terms. For clear notation, we will not specify the invariant 2-tensor used to contract indices for $\SO$ or $\USp$ gauge nodes. The contraction of two repeated indices is always with their respective invariant 2-tensor for the respective representation. Diagrammatically, 
\begin{equation}
\mathcal{O}_{\rm chain}\;=\;
	\begin{tikzpicture}[baseline={(2,0.5)}]
		\node[draw,circle,minimum size=1mm]        (circlen1)      at (0,0)  {} ;
		\node[draw,circle,minimum size=1mm,fill]        (circle2)      at (1.5,0)  {} ;
		\node[draw,circle,minimum size=1mm]        (circle3)      at (1.5,1.5)  {} ;
		\node[draw,circle,minimum size=1mm,fill]        (circle4)      at (0,1.5)  {} ;
		\draw[-] (circle1) -- (circle2) node[pos=0.5, below, inner sep=1pt] {$q_n$};
		\draw[-] (circle2) -- (circle3) node[pos=0.5, right, inner sep=1pt] {$q_n$};
		\draw[-] (circle3) -- (circle4) node[pos=0.5, above, inner sep=1pt] {$q_n$};
		\draw[-] (circle4) -- (circle1) node[pos=0.5, left, inner sep=1pt] {$q_n$};
	\end{tikzpicture}~.
\end{equation}
\noindent
Since the current in the chain trunk is zero and all remaining linear branches have non-zero current, conformal invariance requires adding an additional single $1$-gon half-hypermultiplet at one or more $\USp(m)$ nodes when attaching any of them to either end of the chain. There are also branches with further branching that have zero current and can be directly attached to the chain trunk. Therefore, any graph constructed from this type of trunk always contains one of the graphs in Figure \ref{fig:chain} as a subgraph, and by the sub-quiver lemma, every chain-trunk tree graph has a Higgs branch. We computed the HL index for some graphs of this type with the lowest possible rank in Appendix \ref{app:O_chain}. In the lowest rank examples, the graphs are dual to the lowest rank loop graphs and their Higgs branch generators are known to include a dimension four operator. We propose that this operator is present for more general chain type subgraphs. 

\begin{figure}[t]
	\centering
	\begin{tikzpicture}[]
		\node        (circle1)    at (0,0)    { $\SO(m_1)$};
		\node        (circleu1)    at (2,1.3)    { $\SO(n_1)$};
		\node       (circle2)    at (2,0)    {$\USp(m_2)$};
		\node        (circle3)   at (4,0)    {$\SO(m_3)$};
		\node (A) at (6,0) {$\cdots$};
		\node        (circle4)    at (8,0)    { $\SO(m_{k-2})$};
		\node       (circle5)    at (11,0)    {$\USp(m_{k-1})$};
		\node        (circle6)   at (13.5,0)    {$\SO(m_k)$};
		\node        (circleu2)    at (11,1.3)    { $\SO(n_k)$};
		\draw[-] (circle1.east) -- (circle2.west) node[pos=0.5, above] {$q_1$};
		\draw[-] (circle2.east) -- (circle3.west) node[pos=0.5, above] {$q_2$};
		\draw[-] (circle3.east) -- (A.west) node[pos=0.5, above] {$q_3$};
		\draw[-] (A.east) -- (circle4.west) node[pos=0.5, above] {$q_{k-3}$};
		\draw[-] (circle4.east) -- (circle5.west) node[pos=0.5, above] {$q_{k-2}$};
		\draw[-] (circle5.east) -- (circle6.west) node[pos=0.5, above] {$q_{k-1}$};
		\draw[-] (circle2.north) -- (circleu1.south) node[pos=0.5, right] {$\tilde{q}_1$};
		\draw[-] (circle5.north) -- (circleu2.south) node[pos=0.5, right] {$\tilde{q}_{k-1}$};
	\end{tikzpicture}
	\vspace{0.5em}
	
	\begin{tikzpicture}[]
		\node        (circle1)    at (0,0)    { $\SO(m_1)$};
		\node        (circleu1)    at (2,1.3)    { $\SO(n_1)$};
		\node       (circle2)    at (2,0)    {$\USp(m_2)$};
		\node        (circle3)   at (4,0)    {$\SO(m_3)$};
		\node (A) at (6,0) {$\cdots$};
		\node        (circle4)    at (8,0)    { $\USp(m_{k-2})$};
		\node       (circle5)    at (11,0)    {$\SO(m_{k-1})$};
		\node        (circle6)   at (13.5,0)    {$\USp(m_k)$};
		\node        (circleu2)    at (11,1.3)    { $\USp(n_k)$};
		\draw[-] (circle1.east) -- (circle2.west) node[pos=0.5, above] {$q_1$};
		\draw[-] (circle2.east) -- (circle3.west) node[pos=0.5, above] {$q_2$};
		\draw[-] (circle3.east) -- (A.west) node[pos=0.5, above] {$q_3$};
		\draw[-] (A.east) -- (circle4.west) node[pos=0.5, above] {$q_{k-3}$};
		\draw[-] (circle4.east) -- (circle5.west) node[pos=0.5, above] {$q_{k-2}$};
		\draw[-] (circle5.east) -- (circle6.west) node[pos=0.5, above] {$q_{k-1}$};
		\draw[-] (circle2.north) -- (circleu1.south) node[pos=0.5, right] {$\tilde{q}_1$};
		\draw[-] (circle5.north) -- (circleu2.south) node[pos=0.5, right] {$\tilde{q}_{k-1}$};
	\end{tikzpicture}
	\vspace{0.5em}
	
	\begin{tikzpicture}[]
		\node        (circle1)    at (0,0)    { $\USp(m_1)$};
		\node        (circleu1)    at (2,1.3)    { $\USp(n_1)$};
		\node       (circle2)    at (2,0)    {$\SO(m_2)$};
		\node        (circle3)   at (4,0)    {$\USp(m_3)$};
		\node (A) at (6,0) {$\cdots$};
		\node        (circle4)    at (8,0)    { $\USp(m_{k-2})$};
		\node       (circle5)    at (11,0)    {$\SO(m_{k-1})$};
		\node        (circle6)   at (13.5,0)    {$\USp(m_k)$};
		\node        (circleu2)    at (11,1.3)    { $\USp(n_k)$};
		\draw[-] (circle1.east) -- (circle2.west) node[pos=0.5, above] {$q_1$};
		\draw[-] (circle2.east) -- (circle3.west) node[pos=0.5, above] {$q_2$};
		\draw[-] (circle3.east) -- (A.west) node[pos=0.5, above] {$q_3$};
		\draw[-] (A.east) -- (circle4.west) node[pos=0.5, above] {$q_{k-3}$};
		\draw[-] (circle4.east) -- (circle5.west) node[pos=0.5, above] {$q_{k-2}$};
		\draw[-] (circle5.east) -- (circle6.west) node[pos=0.5, above] {$q_{k-1}$};
		\draw[-] (circle2.north) -- (circleu1.south) node[pos=0.5, right] {$\tilde{q}_1$};
		\draw[-] (circle5.north) -- (circleu2.south) node[pos=0.5, right] {$\tilde{q}_{k-1}$};
	\end{tikzpicture}
	\caption{Subgraphs whose corresponding theories admit the Higgs branch operator~(\ref{eq:HBop_chain}). The outer $\SO(n_1)$ and/or $\SO(n_k)$ nodes may be replaced by single half-hypers (i.e.\ $n_1=1$ and/or $n_k=1$), or by any of the admissible special branches.}
	\label{fig:chain}
\end{figure}

\paragraph{Single-node trunk.}\label{subsec:single_node}
We split into three sub-cases according to the central node and the type of branches attached. A balanced $2$-valent single-node trunk with small branches is, by definition, equivalent to a chain trunk, and has been treated. The remaining sub-cases are:
(a) a single-node trunk with a large branch; (b) a $G_2$ trunk; (c) an $\SO(m)$ or $\USp(m)$ trunk with valence $\geq 3$.

\medskip
\noindent
{\it (a) Single-node trunk with a large branch.} 
The only large branch without flavor symmetry is $-\USp(8)-\mathbf{\tfrac{1}{2}asym3}$, and the only balanced graph it builds is the quiver
\begin{equation}\label{eq:undecidedquiver}
\begin{tikzpicture}[baseline={(0,-0.5ex)}]
		\node         (circle1)      at (0,0)   { $\mathrm{USp}(8)$};
		\node         (circle2)    at (1.8,0)    {$\mathrm{SO}(6)$};
		\node         (square1)    at (-2,0)    {$\mathbf{\tfrac{1}{2}asym3}$};
		\draw[-] (square1.east) -- (circle1.west);
		\draw[-] (circle1.east) -- (circle2.west);
\end{tikzpicture}
\end{equation}

A potential group-theoretic argument for the theory $\mathbf{\tfrac{1}{2}asym3}-\USp(8)-\SO(6)$ being Higgsless is as follows. Let $Q_{[abc]}$, with $Q_{[abc]}\Omega^{ab}=0$, denote the half-hypermultiplet in the rank-three antisymmetric representation $\mathbf{48}$ of $\USp(8)$, and let $q_a^i$ denote the bifundamental half-hypermultiplet. The F-term constraints are
\begin{align}
    q_a^{[i}q_b^{j]}\Omega^{ab} &=0 ,
    \label{eq:SO6-Fterm}\\
    q_a^i q_b^j\delta_{ij}
    + Q_{[acd]}Q_{[bef]}\Omega^{ce}\Omega^{df} &=0 .
    \label{eq:USp8-Fterm}
\end{align}

Here we discuss the structure of possible gauge-invariant Higgs-branch operators in this theory. We argue that this theory is Higgsless. Our analysis is not complete, but we systematically exclude several simple classes of operators, together with an explicit check for operators up to low dimensions.

Let us first consider gauge invariants built only from the bifundamental fields $q^i_a$. Since all $\USp(8)$ indices have to be contracted using the symplectic form $\Omega$, any such operator contains bilinear of the form
\begin{equation}
	\Omega^{ab} q^i_a q^j_b~,
\end{equation}
which is precisely the $\SO(6)$ F-term relation \eqref{eq:SO6-Fterm}. Hence, singlets comprising of only $q$ are absent in the chiral ring.

Next consider gauge invariants involving both $q^i_a$ and $Q_{[abc]}$. $\SO(6)$ indices are contracted using $\delta_{ij}$ and $\epsilon_{i_1\ldots i_6}$. Let us first examine the contractions with epsilons. We first note that the product of two epsilons can be written as a sum of products of the symmetric bilinear tensors, so we restrict to operators that contain at most one epsilon tensor for each gauge node. A single epsilon shows up in the gauge invariant as
\begin{equation}
	\epsilon_{i_1\ldots i_6}
	q^{i_1}_{a_1}\cdots q^{i_6}_{a_6}\dots~.
\end{equation}
Antisymmetry in the $\SO(6)$ indices forces the $\USp(8)$ indices $a_1,\ldots,a_6$ to be fully antisymmetrized. For $\USp(8)$, any rank-$6$ tensor constructed out of fundamentals can be written with an explicit symplectic form,
\begin{equation}
	A_{[a_1\ldots a_6]}
	=
	\Omega_{[a_1a_2} B_{a_3a_4a_5a_6]} ,
\end{equation}
for some four-index tensor $B$. Thus any such contraction reduces to terms containing an $\Omega$-contraction between two $q$'s. These terms are then set to zero by the $\SO(6)$ F-term relation \eqref{eq:SO6-Fterm}. Therefore the  $\SO(6)$ contractions with epsilon does not produce a nonzero Higgs branch operator.

We are then left with contractions involving only $\delta_{ij}$ on the $\SO(6)$ indices. In this case, pairs of $q$'s appear through
\begin{equation}
	q^i_a q^j_b\delta_{ij} ~.
\end{equation}
By the $\USp(8)$ F-term relation \eqref{eq:USp8-Fterm}, such bilinears can be exchanged for combinations of the $Q_{[abc]}$'s. Therefore the remaining problem reduces to understanding gauge invariants built purely from $Q_{[abc]}$.

Some large classes of such invariants vanish immediately. Since $Q_{[abc]}$ is $\Omega$-traceless	($\Omega^{ab} Q_{[abc]}=0$), all $\Omega$-contractions must connect indices belonging to different $Q$'s. If two $Q$'s are connected by three $\Omega$'s, the corresponding contraction is antisymmetric under exchange of the two $Q$'s and hence such an operator vanishes. If in the invariant, for every $Q$ there exists another $Q$, such that their indices are contracted by $\Omega$'s, the contraction contains the same tensor structure that appears in the $\USp(8)$ F-term relation \eqref{eq:USp8-Fterm}. It can therefore be rewritten in terms of $q$-bilinears and subsequently vanishes using the $\SO(6)$ F-term relation.

The remaining possibility is that the $Q$'s are contracted with either one or two $\Omega$, but in a way that it does not belong to any of the case discussed above. We do not have a general argument excluding every such singlet. However, we have explicitly checked all candidate operators up to conformal dimension $8$, and we find that they either vanish by the above symmetry and F-term arguments, or reduce to operators already shown to vanish. More explicitly, up to conformal weight $8$, the singlets constructed purely out of $Q_{[abc]}$, which are non-vanishing before application of F-terms, appear only at conformal weight $4$ and $8$.  Their explicit expressions for the primary invariants (\textit{i.e.} they cannot be written as products of lower weight invariants) are as follows
\begin{align}
	E=4:\;\;
	& \mathsf{M}_{a_1 b_1}\mathsf{M}_{a_2 b_2}\,
	\Omega^{b_1a_2}\Omega^{b_2a_1} ,
	\\[0.5em]
	E=8:\;\;
	& \mathsf{M}_{a_1 b_1}\mathsf{M}_{a_2 b_2}
	Q_{[c_1 d_1 e_1]}Q_{[c_2 d_2 e_2]}\mathsf{M}_{a_4 b_4}\,
	\Omega^{b_1a_2}\Omega^{b_2c_1}\Omega^{d_1d_2}
	\Omega^{e_1a_4}\Omega^{b_4e_2}\Omega^{c_2a_1},
	\\[0.5em]
	E=8:\;\;
	& \mathsf{M}_{a_1 b_1}\mathsf{M}_{a_2 b_2}
	\mathsf{M}_{a_3 b_3}\mathsf{M}_{a_4 b_4}\,
	\Omega^{b_1a_2}\Omega^{b_2a_3}
	\Omega^{b_3a_4}\Omega^{b_4a_1}.
\end{align}
Here $\mathsf{M}_{ab}:=Q_{a c d}Q_{b e f}\Omega^{ce}\Omega^{df}$.
All these are set to zero by F-terms and thus, up to this order, the chiral ring contains no nontrivial Higgs-branch generators. 

Let us also note that certain contractions, which a priori appear to be independent, due to tensor identities, can in fact equal to products of lower order invariants. An example of such a phenomenon is the case of the following conformal weight $8$ invariant, which we denote as $\mathcal{O}_{Q^8}$. 
\begin{equation}
	\begin{split}
 \mathcal O_{Q^8}=&\;Q_{[x r_1 r_2]}Q_{[y s_1 s_2]}Q_{[z t_1 t_2]}Q_{[w u_1 u_2]}Q_{[a_1a_2a_3]}Q_{[b_1b_2b_3]}Q_{[c_1c_2c_3]}Q_{[d_1d_2d_3]}\notag\\
	&\times\Omega^{r_1s_1}\Omega^{r_2s_2}\Omega^{t_1u_1}\Omega^{t_2u_2}\Omega^{x a_1}\Omega^{y b_1}\Omega^{z c_1}\Omega^{w d_1}\notag\\
	&\times\Omega^{a_2 c_2}\Omega^{a_3 d_2}\Omega^{b_2 c_3}\Omega^{b_3 d_3}~.
	\end{split}
\end{equation}
From explicit expansion of this operator, we check that $\mathcal O_{Q^8}$ is proportional to \newline $(\mathsf{M}_{a_1 b_1}\mathsf{M}_{a_2 b_2}\Omega^{b_1a_2}\Omega^{b_2a_1})^2$, which is the square of $E=4$ primary invariant mentioned above. Since the latter vanishes in the quotient by the F-term relations, $\mathcal O_{Q^8}$ also vanishes.

As independent supporting evidence, we computed the Hall--Littlewood index, obtaining
\begin{equation}
	\mathcal I_{\mathrm{HL}}(t)=1+2t^5 .
\end{equation}
The absence of terms up to order $t^4$ is consistent with the direct operator analysis above. The two states contributing $+2t^5$ are bosonic HL chiral ring generators which, in a Higgsless theory, must descend from $\overline{\mathcal{D}}$-type multiplets (they necessarily contain gauginos, cf.\ Table~\ref{tab:schurmultipletslagletters}) and are therefore nilpotent in $\mathcal{R}_\mathcal{V}$; they do not signal Higgs branch generators. The above arguments along with the truncated HL index provide strong evidence that this theory is Higgsless.

\medskip
\noindent
{\it (b) $G_2$ trunk.} A trunk consisting of a single $G_2$ node admits only finitely many balanced tree graphs.\footnote{The two branches $\cdots-\USp(8)-G_2$ and $\cdots-\USp(6)-G_2-\USp(2)-\mathbf{1}$ are classified as branches with special representations as in \cite{Bhardwaj:2013qia}.} They are enumerated in Appendix \ref{app:G_2}, where we also compute their HL indices. None of them truncates, so each such theory has a Higgs branch.

\medskip
\noindent
{\it (c) $\SO(m)$ or $\USp(m)$ trunk with valence $\geq 3$.} 

\subparagraph{(c.1) single $\SO(m)$ or $\USp(m)$ trunk with valence $\geq 4$.}
Any quiver with a $\SO(m)$ or $\USp(m)$ node with valency $\geq 4$ contains one of the subgraphs of Figure \ref{fig:4valent}. For these subgraphs, the operator
\begin{equation}
\mathcal{O}_{n\geq 4\text{-valent}} \;=\; (q_1)^\alpha_i\,(q_1)^\beta_i\,(q_2)^\alpha_a\,(q_2)^\beta_a\,,
\end{equation}
with $\alpha,\beta$ the central-node indices and $i,a$ those of the neighboring nodes, is gauge-invariant and not annihilated by the F-terms.  Diagrammatically, when the central node is $\SO$,
\begin{equation}
\mathcal{O}_{n\geq 4\text{-valent}}\;=\;
	\begin{tikzpicture}[baseline={(0,0.5)}]
		\node[draw,circle,minimum size=1mm,fill]        (circle1)      at (0,0)  {} ;
		\node[draw,circle,minimum size=1mm]        (circle2)      at (0,1.5)  {} ;
		\node[draw,circle,minimum size=1mm,fill]        (circle3)    at (1.5,1.5)    {};
		\node[draw,circle,minimum size=1mm]        (circle4)    at (1.5,0)    {};
		\draw[-] (circle1) -- (circle2) node[pos=0.5, left, inner sep=1pt] {$q_1$};
		\draw[-] (circle2) -- (circle3) node[pos=0.5, above, inner sep=1pt] {$q_2$};
		\draw[-] (circle3) -- (circle4) node[pos=0.5, right, inner sep=1pt] {$q_2$};
		\draw[-] (circle4) -- (circle1) node[pos=0.5, below, inner sep=1pt] {$q_1$};
	\end{tikzpicture}
\end{equation}
and similarly for a $\USp$ central node with filled and unfilled circles exchanged. All such theories therefore have a Higgs branch. The HL indices computed in Appendix \ref{app:O_4valent} show a denominator factor signaling a  dimension four generator, consistent with $\mathcal{O}_{n\geq 4\text{-valent}}$.

\begin{figure}
\centering
\begin{tikzpicture}[]
		\node         (circle1)      at (0,0)   { $\SO(m)$};
		\node         (circle2)    at (-1,1)    {$\USp(n_2)$};
		\node         (circle3)    at (2,0)    {$\USp(n_4)$};
		\node         (circle4)    at (-2,0)    {$\USp(n_1)$};
		\node         (circle5)    at (1,1)    {$\USp(n_3)$};
		\draw[-] (circle1.north) -- (circle2.south) node[pos=0.5, above, inner sep=1pt] {$q_2$};
		\draw[-] (circle1.east) -- (circle3.west) node[pos=0.5, above] {$q_4$};
		\draw[-] (circle1.north) -- (circle5.south) node[pos=0.5, above, inner sep=1pt] {$q_3$};
		\draw[-] (circle1.west) -- (circle4.east) node[pos=0.5, above] {$q_1$};
\end{tikzpicture};
\hspace{1em}
\begin{tikzpicture}[]
		\node         (circle1)      at (0,0)   { $\USp(m)$};
		\node         (circle2)    at (-1,1)    {$\SO(n_2)$};
		\node         (circle3)    at (2,0)    {$\SO(n_4)$};
		\node         (circle4)    at (-2,0)    {$\SO(n_1)$};
		\node         (circle5)    at (1,1)    {$1$};
		\draw[-] (circle1.north) -- (circle2.south) node[pos=0.5, above, inner sep=1pt] {$q_2$};
		\draw[-] (circle1.east) -- (circle3.west) node[pos=0.5, above] {$q_4$};
		\draw[-] (circle1.north) -- (circle5.south) node[pos=0.4, above, inner sep=1pt] {$q_3$};
		\draw[-] (circle1.west) -- (circle4.east) node[pos=0.5, above] {$q_1$};
\end{tikzpicture}.
\caption{Subgraphs of an $n\geq 4$-valent $\SO$ or $\USp$ node.}
\label{fig:4valent}
\end{figure}

\subparagraph{(c.2) trivalent single $\SO(m)$ or $\USp(m)$ trunk.}
This is the most delicate case which yields an infinite family of candidate Higgsless theories. We further split our discussion according to whether the trivalent quiver contains the subgraph of Figure~\ref{fig:3valent}.

If a trivalent quiver contains the subgraph of Figure \ref{fig:3valent}, we propose the Higgs branch operator
\begin{equation}\label{eq:HBop_3valent}
\mathcal{O}_{3\text{-valent}} \;=\; (q_1)^\sigma_i\,(q_1)^\alpha_i\,(q_2)^\alpha_a\,(q_2)^\beta_a\,(q_1)^\beta_j\,(q_1)^\gamma_j\,(q_2)^\gamma_b\,(q_2)^\sigma_b\,,
\end{equation}
which is gauge-invariant and not annihilated by the F-terms\footnote{We identify one exception at low ranks that is the trivalent graph $\SO(5)-(\USp(2)-\SO(3))^3$ which is dual to a smaller graph $\USp(4)-\SO(4)^3$. The Higgs operator $\mathcal{O}_{3\text{-valent}}$ vanishes for this quiver because the trace relations relate it to $({q_1}^\beta_i{q_1}^\alpha_i{q_2}^\alpha_a {q_2}^\beta_a)^2 $ which vanishes by the F-term equations.}. The corresponding diagram, for an $\SO$ central node, is
\begin{equation}
\mathcal{O}_{3\text{-valent}}\;=\;
        \begin{tikzpicture}[baseline={(0,0.5)}]
        \node[draw,circle,minimum size=1mm,fill]        (circle1)      at (0,0)  {} ;
        \node[draw,circle,minimum size=1mm]        (circle2)      at (0,1.5)  {} ;
        \node[draw,circle,minimum size=1mm,fill]        (circle3)    at (1.5,1.5)    {};
        \node[draw,circle,minimum size=1mm]        (circle4)    at (3,1.5)    {};
        \node[draw,circle,minimum size=1mm,fill]        (circle5)    at (4.5,1.5)    {};
        \node[draw,circle,minimum size=1mm]        (circle6)    at (4.5,0)    {};
        \node[draw,circle,minimum size=1mm,fill]        (circle7)    at (3,0)    {};
        \node[draw,circle,minimum size=1mm]        (circle8)    at (1.5,0)    {};
        \draw[-] (circle1) -- (circle2) node[pos=0.5, left, inner sep=1pt] {$q_1$};
        \draw[-] (circle2) -- (circle3) node[pos=0.5, above, inner sep=1pt] {$q_2$};
        \draw[-] (circle3) -- (circle4) node[pos=0.5, above, inner sep=1pt] {$q_2$};
        \draw[-] (circle4) -- (circle5) node[pos=0.5, above, inner sep=1pt] {$q_1$};
        \draw[-] (circle5) -- (circle6) node[pos=0.5, right, inner sep=1pt] {$q_1$};
        \draw[-] (circle6) -- (circle7) node[pos=0.5, below, inner sep=1pt] {$q_2$};
        \draw[-] (circle7) -- (circle8) node[pos=0.5, below, inner sep=1pt] {$q_2$};
        \draw[-] (circle8) -- (circle1) node[pos=0.5, below, inner sep=1pt] {$q_1$};
        \end{tikzpicture}
\end{equation}
\noindent
(with the type of circles exchanged for a $\USp$ central node). HL-index computations for two examples are recorded in Appendix \ref{app:hlindquiv}. By the sub-quiver lemma, every trivalent quiver containing the subgraph of Figure~\ref{fig:3valent} has a Higgs branch.

\begin{figure}
\centering
\begin{tikzpicture}[baseline={(0,-0.5ex)}]
		\node        (circle1)      at (0,0)   { $\SO(m)$};
		\node        (circle3)    at (2.1,0)    {$\USp(n_3)$};
		\node        (circle4)    at (-2.1,0)    {$\USp(n_1)$};
		\node        (circle5)    at (0,1)    {$\USp(n_2)$};
		\node        (circle6)    at (4,0)    {$\SO(l_1)$};
		\node        (circle8)    at (-4,0)    {$\SO(l_2)$};
		\draw[-] (circle1.east) -- (circle3.west) node[pos=0.5, above, inner sep=1pt] {$q_3$};
		\draw[-] (circle1.north) -- (circle5.south) node[pos=0.5, right, inner sep=1pt] {$q_2$};
		\draw[-] (circle1.west) -- (circle4.east) node[pos=0.5, above, inner sep=1pt] {$q_1$};
		\draw[-] (circle3.east) -- (circle6.west);
		\draw[-] (circle4.west) -- (circle8.east);
\end{tikzpicture} 
\vspace{0.6em}

\begin{tikzpicture}[baseline={(0,-0.5ex)}]
		\node       (circle1)      at (0,0)   { $\USp(n)$};
		\node      (circle3)    at (2.1,0)    {$\SO(m_3)$};
		\node      (circle4)    at (-2.1,0)    {$\SO(m_1)$};
		\node       (circle5)    at (0,1)    {$\SO(m_2)$};
		\node      (circle6)    at (4,0)    {$\USp(l_1)$};
		\node      (circle8)    at (-4,0)    {$\USp(l_2)$};
		\draw[-] (circle1.east) -- (circle3.west) node[pos=0.5, above, inner sep=1pt] {$q_3$};
		\draw[-] (circle1.north) -- (circle5.south) node[pos=0.5, right, inner sep=1pt] {$q_2$};
		\draw[-] (circle1.west) -- (circle4.east) node[pos=0.5, above, inner sep=1pt] {$q_1$};
		\draw[-] (circle3.east) -- (circle6.west);
		\draw[-] (circle4.west) -- (circle8.east);
\end{tikzpicture}
\caption{Subgraphs that contain a trivalent $\SO$ or $\USp$ node and admit the Higgs branch operator (\ref{eq:HBop_3valent}).}
\label{fig:3valent}
\end{figure}

\begin{figure}
\centering
\begin{tikzpicture}[baseline={(0,-0.5ex)}]
		\node[draw,circle,minimum size=8mm]        (circle1)      at (0,0)  {} ;
		\node[draw,circle,minimum size=8mm]        (circle3)    at (1.5,0)    {};
		\node[draw,circle,minimum size=8mm]        (circle4)    at (-1.5,0)    {};
		\node[draw,circle,minimum size=8mm]        (circle5)    at (0,1.2)    {};
		\node[draw,circle,minimum size=8mm]       (circle6)    at (3,0)    {};
		\node        (circle8)    at (4,0)    {$\cdots$};
		\node[draw,circle,minimum size=8mm]        (circle9)    at (5,0)    {};
		\node[draw,circle,minimum size=8mm]       (circle10)    at (6.5,0)    {};
		\draw[-] (circle1.east) -- (circle3.west) node[pos=0.5, above, inner sep=1pt] {$3$};
		\draw[-] (circle1.north) -- (circle5.south)node[pos=0.5, right, inner sep=1pt] {$2$};
		\draw[-] (circle1.west) -- (circle4.east)node[pos=0.5, above, inner sep=1pt] {$1$};
		\draw[-] (circle3.east) -- (circle6.west);
		\draw[-] (circle6.east) -- (circle8.west);
		\draw[-] (circle8.east) -- (circle9.west);
		\draw[-] (circle9.east) -- (circle10.west);
\end{tikzpicture}
\caption{Schematic form of the remaining trivalent quivers. The central node is $\USp(m)$ or $\SO(m)$. Branches must be chosen from the list of relevant branches in Appendix \ref{app:branches} without further branching. Branches 1 and 2 have length-two, and branch 3 has at least length-two.}
\label{fig:possible3valent}
\end{figure}
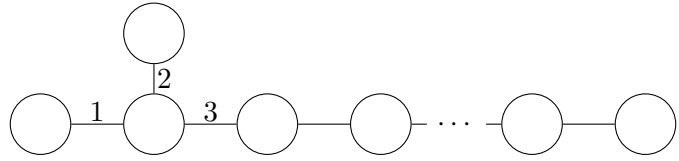

\medskip
The remaining trivalent quivers are precisely those of the schematic form shown in Figure \ref{fig:possible3valent}. If branches 1 and 2 contain more than one additional node, the graph would contain the subgraph shown in Figure \ref{fig:3valent}. If there is any further branching in branch 3, the graph would contain the subgraph shown in Figure \ref{fig:chain}. 

We analyze these in three sub-cases, according to the type of central node and the nature of branches 1 and 2.

\medskip
\noindent
{\bf Case A: $\SO$ central, generic branches.} 
We first consider the generic case, by which we mean that branches~1 and~2 are not built from special representations. Once the central node is fixed, conformal invariance fixes the rank of every other node in the quiver.  The ranks along branch~3 are forced to grow (Figure~\ref{fig:small_so_3valent}), so the branch cannot terminate without further branching. Hence such a quiver always contains the chain subgraph of Figure~\ref{fig:chain} and has a Higgs branch.

\begin{figure}
\centering
\begin{tikzpicture}[baseline={(0,-0.5ex)}]
		\node        (circle1)      at (0,0)   { $\SO(m)$};
		\node        (circle3)    at (1.8,0)    {$\USp(m)$};
		\node        (circle4)    at (-2,0)    {$\USp(\frac{m-4}{2})$};
		\node        (circle5)    at (0,1)    {$\USp(\frac{m-4}{2})$};
		\node        (circle6)    at (4,0)    {$\SO(m+4)$};
		\node        (circle7)    at (6,0)    {$\cdots$};
		\draw[-] (circle1.east) -- (circle3.west) ;
		\draw[-] (circle1.north) -- (circle5.south);
		\draw[-] (circle1.west) -- (circle4.east);
		\draw[-] (circle3.east) -- (circle6.west);
		\draw[-] (circle6.east) -- (circle7.west);
\end{tikzpicture}
\caption{Trivalent $\SO(m)$ node with branches~1 and~2 of length two and branch~3 of length $>2$. Conformal invariance forces the rank along branch~3 to grow, so the branch cannot terminate without further branching.}
\label{fig:small_so_3valent}
\end{figure}

\medskip
\noindent
{\bf Case B: $\USp$ central, generic branches (The candidate Higgsless family).}~
Let the central node be \(\USp(m)\) (see Figure \ref{fig:usp_3valent}). Now the ranks along branch~3 decrease. For $m\equiv 2\pmod 4$ the chain terminates with $\USp(2)$--$\SO(2)$, which is not conformal. For $m\equiv 0\pmod 4$ the conformal-invariance constraints admit the infinite family of balanced quivers shown in Figure~\ref{fig:usp_3valent}, with $m\in 4\mathbb{Z}_{>0}$. This is the candidate Higgsless family~(iii) of our result statement.

\begin{figure}
\centering
\begin{tikzpicture}[baseline={(0,-0.5ex)}]
		\node        (circle1)      at (0,0)   { $\USp(m)$};
		\node        (circle3)    at (1.8,0)    {$\SO(m)$};
		\node        (circle4)    at (-2,0)    {$\SO(\frac{m+4}{2})$};
		\node        (circle5)    at (0,1)    {$\SO(\frac{m+4}{2})$};
		\node        (circle6)    at (4,0)    {$\USp(m-4)$};
		\node        (circle7)    at (6.5,0)    {$\SO(m-4)$};
		\node        (circle8)    at (8.4,0)    {$\cdots$};
		\node        (circle9)    at (9.8,0)    {$\USp(4)$};
		\node        (circle10)    at (11.6,0)    {$\SO(4)$};
		\draw[-] (circle1.east)  -- (circle3.west) node[pos=0.5, above, inner sep=1pt] {$q_3$};
		\draw[-] (circle1.north) -- (circle5.south)node[pos=0.5, right, inner sep=1pt] {$q_2$};
		\draw[-] (circle1.west) -- (circle4.east)node[pos=0.5, above, inner sep=1pt] {$q_1$};
		\draw[-] (circle3.east) -- (circle6.west)node[pos=0.5, above, inner sep=1pt] {$q_4$};
		\draw[-] (circle6.east) -- (circle7.west)node[pos=0.5, above, inner sep=1pt] {$q_5$};
		\draw[-] (circle7.east) -- (circle8.west);
		\draw[-] (circle8.east) -- (circle9.west);
		\draw[-] (circle9.east) -- (circle10.west)node[pos=0.5, above, inner sep=1pt] {$q_{k+2}$} ;
\end{tikzpicture}
\caption{The infinite family of candidate Higgsless theories with a trivalent $\USp(m)$ trunk, $m\in 4\mathbb{Z}_{>0}$.}
\label{fig:usp_3valent}
\end{figure}

We now argue that the theories in this family admit no Higgs branch operator. We label the half-hypers as in Figure \ref{fig:usp_3valent}: $q_1$, $q_2$ are attached to the two outer $\SO(\frac{m+4}{2})$ nodes of branches 1 and 2, $q_3$ joins the central $\USp(m)$ to the first node of branch 3, and $q_4, q_5,\ldots, q_{k+2}$ continue along branch 3. The argument is conveniently illustrated on the $m=8$ representative in Figure \ref{fig:usp_3valentexample} and extends to the entire family. The key group-theoretic input in the $\epsilon$-tensor step below is $(m+4)/2 > m/2$, i.e. fully antisymmetric tensors with more indices than the rank of the relevant $\USp$ group always reduce to $\Omega$-traces.
\begin{figure}
\centering
\begin{tikzpicture}[baseline={(0,-0.5ex)}]
		\node        (circle1)      at (0,0)   { $\USp(8)$};
		\node        (circle3)    at (2,0)    {$\SO(8)$};
		\node        (circle4)    at (-2,0)    {$\SO(6)$};
		\node        (circle5)    at (0,1.2)    {$\SO(6)$};
		\node        (circle6)    at (4,0)    {$\USp(4)$};
		\node        (circle7)    at (6,0)    {$\SO(4)$};
		\draw[-] (circle1.east)  -- (circle3.west) node[pos=0.5, above, inner sep=1pt] {$q_3$};
		\draw[-] (circle1.north) -- (circle5.south)node[pos=0.5, right, inner sep=1pt] {$q_2$};
		\draw[-] (circle1.west) -- (circle4.east)node[pos=0.5, above, inner sep=1pt] {$q_1$};
		\draw[-] (circle3.east) -- (circle6.west)node[pos=0.5, above, inner sep=1pt] {$q_4$};
		\draw[-] (circle6.east) -- (circle7.west)node[pos=0.5, above, inner sep=1pt] {$q_5$};
\end{tikzpicture}
\caption{The $m=8$ representative of the family of Figure~\ref{fig:usp_3valent}.}
\label{fig:usp_3valentexample}
\end{figure}
The F-terms at the various nodes are:
\begin{align}
	\label{eq:Fterms_SO8}
	\SO(8) ~:&~(q_3)^\alpha_a\,(q_3)^\alpha_b + (q_4)^\alpha_a\,(q_4)^\alpha_b = 0\,,\\
	\USp(4)~:&~  (q_4)^\alpha_a\,(q_4)^\beta_a + (q_5)^\alpha_a\,(q_5)^\beta_a = 0\,,\\
    \label{eq:Fterms_SO4}
	\SO(4)~:&~ (q_5)^\alpha_a\,(q_5)^\alpha_b = 0~,\\
	\label{eq:Fterms_SO6}
	\SO(6)~:&~ (q_1)^\alpha_a\,(q_1)^\alpha_b = 0\,,\\
	\SO(6)~:&~ (q_2)^\alpha_a\,(q_2)^\alpha_b = 0\,,\\
	\label{eq;usp8Fterm}
	\USp(8)~:&~ (q_1)^\alpha_a\,(q_1)^\beta_a + (q_2)^\alpha_a\,(q_2)^\beta_a + (q_3)^\alpha_a\,(q_3)^\beta_a = 0\,.
\end{align}
We organize the enumeration as a reduction procedure, treating first the invariants built solely from the bilinear invariant tensors ($\delta$ for $\SO$ nodes, $\Omega$ for $\USp$ nodes). Once this class of contractions has been exhausted, we turn separately to contractions involving $\epsilon$ tensors.

The contraction of two $q_5$
fields through the terminal node vanishes by the F-term \eqref{eq:Fterms_SO4}, so the only
nonzero possibility is to contract their $\SO(4)$ gauge indices with each other. The remaining free
$\USp(4)$ indices must then be saturated by adjoining $q_4$ fields. Each end of the $q_5$ pair can be joined to a $q_4$ pair only once. A second contraction at the same end can be traded, using the $\USp(4)$ F-term, for a $q_5$ bilinear that vanishes by \eqref{eq:Fterms_SO4}. Closing the loop at this stage produces a contraction that vanishes by the F-terms for the same reason. One must therefore continue by joining $q_3$'s at the free ends of the $q_4$'s. Applying the F-term
relations successively replaces the resulting chain by a purely $q_3$
chain:

\begin{equation}
	\label{eq:q36}
	\begin{tikzpicture}[baseline={(0,0)}]
		\node[draw,circle,minimum size=1mm] (circle1) at (0,0) {};
		\node[draw,circle,minimum size=1mm,fill]      (circle2) at (-1,1) {};
		\node[draw,circle,minimum size=1mm,fill]      (circle3) at (-1,-1) {};
		\node[draw,circle,minimum size=1mm]      (circle4) at (-2,1) {};
		\node[draw,circle,minimum size=1mm]      (circle5) at (-2,-1) {};
		\node[draw,circle,minimum size=1mm,fill]      (circle6) at (-3,1) {};
		\node[draw,circle,minimum size=1mm,fill]      (circle7) at (-3,-1) {};
		
		\node[]      (circlen) at (-4,1) {};
		\node[]      (circlenp1) at (-4,-1) {};
		
		\draw[-] (circle1) to[bend right=40]
		node[pos=0.5, yshift=6pt,above, inner sep=1pt] {$q_5$} (circle2);
		\draw[-] (circle1) to[bend left=40]
		node[pos=0.5, below,yshift=-6pt, inner sep=1pt] {$q_5$} (circle3);
		
		\draw[-] (circle2) to
		node[pos=0.5, yshift=6pt,above, inner sep=1pt] {$q_4$} (circle4);
		\draw[-] (circle3) to
		node[pos=0.5, yshift=6pt,above, inner sep=1pt] {$q_4$} (circle5);	
		
		\draw[-] (circle4) to
		node[pos=0.5, yshift=6pt,above, inner sep=1pt] {$q_3$} (circle6);
		\draw[-] (circle5) to
		node[pos=0.5, yshift=6pt,above, inner sep=1pt] {$q_3$} (circle7);	
		
		\draw[-,dashed] (circle6) to
		node[pos=0.5, below,yshift=-6pt, inner sep=1pt] {} (circlen);
		\draw[-,dashed] (circle7) to
		node[pos=0.5, below,yshift=-6pt, inner sep=1pt] {} (circlenp1);
	\end{tikzpicture}~~
	=
	\begin{tikzpicture}[baseline={(0,0)}]
		\node[draw,circle,minimum size=1mm] (circle1) at (0,0) {};
		\node[draw,circle,minimum size=1mm,fill]      (circle2) at (-1,1) {};
		\node[draw,circle,minimum size=1mm,fill]      (circle3) at (-1,-1) {};
		\node[draw,circle,minimum size=1mm]      (circle4) at (-2,1) {};
		\node[draw,circle,minimum size=1mm]      (circle5) at (-2,-1) {};
		\node[draw,circle,minimum size=1mm,fill]      (circle6) at (-3,1) {};
		\node[draw,circle,minimum size=1mm,fill]      (circle7) at (-3,-1) {};
		
		\node[]      (circlen) at (-4,1) {};
		\node[]      (circlenp1) at (-4,-1) {};
		
		\draw[-] (circle1) to[bend right=40]
		node[pos=0.5, yshift=6pt,above, inner sep=1pt] {$q_3$} (circle2);
		\draw[-] (circle1) to[bend left=40]
		node[pos=0.5, below,yshift=-6pt, inner sep=1pt] {$q_3$} (circle3);
		
		\draw[-] (circle2) to
		node[pos=0.5, yshift=6pt,above, inner sep=1pt] {$q_3$} (circle4);
		\draw[-] (circle3) to
		node[pos=0.5, yshift=6pt,above, inner sep=1pt] {$q_3$} (circle5);	
		
		\draw[-] (circle4) to
		node[pos=0.5, yshift=6pt,above, inner sep=1pt] {$q_3$} (circle6);
		\draw[-] (circle5) to
		node[pos=0.5, yshift=6pt,above, inner sep=1pt] {$q_3$} (circle7);	
		
		\draw[-,dashed] (circle6) to
		node[pos=0.5, below,yshift=-6pt, inner sep=1pt] {} (circlen);
		\draw[-,dashed] (circle7) to
		node[pos=0.5, below,yshift=-6pt, inner sep=1pt] {} (circlenp1);
	\end{tikzpicture}
\end{equation}

Starting instead from a $q_4$ pair, the same reasoning shows that the only allowed subdiagram is the following, which the F-terms again convert into a purely $q_3$ configuration:
\begin{equation}
	\label{eq:q34}
	\begin{tikzpicture}[baseline={(0,0)}]
		\node[draw,circle,minimum size=1mm,fill] (circle1) at (0,0) {};
		\node[draw,circle,minimum size=1mm]      (circle2) at (-1,1) {};
		\node[draw,circle,minimum size=1mm]      (circle3) at (-1,-1) {};
		\node[draw,circle,minimum size=1mm,fill]      (circle4) at (-2,1) {};
		\node[draw,circle,minimum size=1mm,fill]      (circle5) at (-2,-1) {};
		
		\node[]      (circlen) at (-3,1) {};
		\node[]      (circlenp1) at (-3,-1) {};

		\draw[-] (circle1) to[bend right=40]
		node[pos=0.5, yshift=6pt,above, inner sep=1pt] {$q_4$} (circle2);
		\draw[-] (circle1) to[bend left=40]
		node[pos=0.5, below,yshift=-6pt, inner sep=1pt] {$q_4$} (circle3);
		
		\draw[-] (circle2) to
		node[pos=0.5, yshift=6pt,above, inner sep=1pt] {$q_3$} (circle4);
		\draw[-] (circle3) to
		node[pos=0.5, yshift=6pt,above, inner sep=1pt] {$q_3$} (circle5);	
		
		\draw[-,dashed] (circle4) to
		node[pos=0.5, below,yshift=-6pt, inner sep=1pt] {} (circlen);
		\draw[-,dashed] (circle5) to
		node[pos=0.5, below,yshift=-6pt, inner sep=1pt] {} (circlenp1);
	\end{tikzpicture}~~
	=
	\begin{tikzpicture}[baseline={(0,0)}]
		\node[draw,circle,minimum size=1mm,fill] (circle1) at (0,0) {};
		\node[draw,circle,minimum size=1mm]      (circle2) at (-1,1) {};
		\node[draw,circle,minimum size=1mm]      (circle3) at (-1,-1) {};
		\node[draw,circle,minimum size=1mm,fill]      (circle4) at (-2,1) {};
		\node[draw,circle,minimum size=1mm,fill]      (circle5) at (-2,-1) {};
		
		\node[]      (circlen) at (-3,1) {};
		\node[]      (circlenp1) at (-3,-1) {};

		\draw[-] (circle1) to[bend right=40]
		node[pos=0.5, yshift=6pt,above, inner sep=1pt] {$q_3$} (circle2);
		\draw[-] (circle1) to[bend left=40]
		node[pos=0.5, below,yshift=-6pt, inner sep=1pt] {$q_3$} (circle3);
		
		\draw[-] (circle2) to
		node[pos=0.5, yshift=6pt,above, inner sep=1pt] {$q_3$} (circle4);
		\draw[-] (circle3) to
		node[pos=0.5, yshift=6pt,above, inner sep=1pt] {$q_3$} (circle5);	
		
		\draw[-,dashed] (circle4) to
		node[pos=0.5, below,yshift=-6pt, inner sep=1pt] {} (circlen);
		\draw[-,dashed] (circle5) to
		node[pos=0.5, below,yshift=-6pt, inner sep=1pt] {} (circlenp1);
	\end{tikzpicture}
\end{equation}

Let us also note that one cannot consider a chain with more than six $q_3$'s in series, as that part of the singlet is set to zero by successive application of the F-terms. Similarly, we can also start from $q_3$, which leads to the following subdiagram
\begin{equation}
	\label{eq:q32}
	\begin{tikzpicture}[baseline={(0,0.5)}]
		\node[draw,circle,minimum size=1mm] (circle1) at (0,0) {};
		\node[draw,circle,minimum size=1mm,fill]      (circle2) at (-1,1) {};
		\node[draw,circle,minimum size=1mm,fill]      (circle3) at (-1,-1) {};
		\node[]      (circlen) at (-2,1) {};
		\node[]      (circlenp1) at (-2,-1) {};

		\draw[-] (circle1) to[bend right=40]
		node[pos=0.5, yshift=6pt,above, inner sep=1pt] {$q_3$} (circle2);
		
		\draw[-] (circle1) to[bend left=40]
		node[pos=0.5, below,yshift=-6pt, inner sep=1pt] {$q_3$} (circle3);
		
		\draw[-,dashed] (circle2) to
		node[pos=0.5, below,yshift=-6pt, inner sep=1pt] {} (circlen);
		\draw[-,dashed] (circle3) to
		node[pos=0.5, below,yshift=-6pt, inner sep=1pt] {} (circlenp1);
	\end{tikzpicture}
\end{equation}
Because of the $\SO(6)$ F-terms, only one $q_1$ pair and only one $q_2$ pair, with their respective $\SO$ indices contracted, can be used. 
\begin{equation}
	\label{eq:q1q2}
	\begin{tikzpicture}[baseline={(0,0.5)}]
		\node[draw,circle,minimum size=1mm] (circle1) at (0,1) {};
		\node[draw,circle,minimum size=1mm,fill]      (circle2) at (1,2) {};
		\node[draw,circle,minimum size=1mm,fill]      (circle3) at (1,-0) {};
		\node[]      (circlen) at (2,2) {};
		\node[]      (circlenp1) at (2,-0) {};

		\draw[-] (circle1) to[bend left=40]
		node[pos=0.5, yshift=6pt,above, inner sep=1pt] {$q_1$} (circle2);
		
		\draw[-] (circle1) to[bend right=40]
		node[pos=0.5, below,yshift=-6pt, inner sep=1pt] {$q_1$} (circle3);
		
		\draw[-,dashed] (circle2) to
		node[pos=0.5, below,yshift=-6pt, inner sep=1pt] {} (circlen);
		\draw[-,dashed] (circle3) to
		node[pos=0.5, below,yshift=-6pt, inner sep=1pt] {} (circlenp1);
	\end{tikzpicture}\qquad,\qquad
	\begin{tikzpicture}[baseline={(0,0.5)}]
		\node[draw,circle,minimum size=1mm] (circle1) at (0,1) {};
		\node[draw,circle,minimum size=1mm,fill]      (circle2) at (1,2) {};
		\node[draw,circle,minimum size=1mm,fill]      (circle3) at (1,0) {};
		\node[]      (circlen) at (2,2) {};
		\node[]      (circlenp1) at (2,0) {};

		\draw[-] (circle1) to[bend left=40]
		node[pos=0.5, yshift=6pt,above, inner sep=1pt] {$q_2$} (circle2);
		
		\draw[-] (circle1) to[bend right=40]
		node[pos=0.5, below,yshift=-6pt, inner sep=1pt] {$q_2$} (circle3);
		
		\draw[-,dashed] (circle2) to
		node[pos=0.5, below,yshift=-6pt, inner sep=1pt] {} (circlen);
		\draw[-,dashed] (circle3) to
		node[pos=0.5, below,yshift=-6pt, inner sep=1pt] {} (circlenp1);
	\end{tikzpicture}
\end{equation}
Therefore, every candidate invariant is assembled from the five building blocks displayed in equations \eqref{eq:q36}, \eqref{eq:q34}, \eqref{eq:q32} and \eqref{eq:q1q2}. Every remaining singlet is built from chains containing at most six consecutive $q_3$ fields, together with alternating pairs of $q_1$'s and pairs of $q_2$'s. The possible sub-structures within a singlet are the following four:

	\begin{equation}
		\begin{tikzpicture}[baseline={(0,0)}]
			\node[] (circle0) at (-1,0) {};
			\node[draw,circle,minimum size=1mm,fill] (circle1) at (0,0) {};
			\node[draw,circle,minimum size=1mm]      (circle2) at (1,0) {};
			\node[draw,circle,minimum size=1mm,fill]      (circle3) at (2,0) {};
			\node[draw,circle,minimum size=1mm]      (circle4) at (3,0) {};
			\node[draw,circle,minimum size=1mm,fill]      (circle5) at (4,0) {};
			\node[draw,circle,minimum size=1mm]      (circle6) at (5,0) {};
			\node[draw,circle,minimum size=1mm,fill]      (circle7) at (6,0) {};
			\node[]      (circle8) at (7,0) {};
	
			\draw[-,dashed] (circle0) to
			node[pos=0.5, above,yshift=6pt, inner sep=1pt] {} (circle1);
			\draw[-] (circle1) to
			node[pos=0.5, above,yshift=6pt, inner sep=1pt] {$q_3$} (circle2);
			\draw[-] (circle2) to
			node[pos=0.5, above,yshift=6pt, inner sep=1pt] {$q_3$} (circle3);
			\draw[-] (circle3) to
			node[pos=0.5, above,yshift=6pt, inner sep=1pt] {$q_1$} (circle4);
			\draw[-] (circle4) to
			node[pos=0.5, above,yshift=6pt, inner sep=1pt] {$q_1$} (circle5);
			\draw[-] (circle5) to
			node[pos=0.5, above,yshift=6pt, inner sep=1pt] {$q_3$} (circle6);
			\draw[-] (circle6) to
			node[pos=0.5, above,yshift=6pt, inner sep=1pt] {$q_3$} (circle7);
			\draw[-,dashed] (circle7) to
			node[pos=0.5, above,yshift=6pt, inner sep=1pt] {} (circle8);
		\end{tikzpicture}
	\end{equation}

	\begin{equation}
	\begin{tikzpicture}[baseline={(0,0)}]
		\node[] (circle0) at (-1,0) {};
		\node[draw,circle,minimum size=1mm,fill] (circle1) at (0,0) {};
		\node[draw,circle,minimum size=1mm]      (circle2) at (1,0) {};
		\node[draw,circle,minimum size=1mm,fill]      (circle3) at (2,0) {};
		\node[draw,circle,minimum size=1mm]      (circle4) at (3,0) {};
		\node[draw,circle,minimum size=1mm,fill]      (circle5) at (4,0) {};
		\node[draw,circle,minimum size=1mm]      (circle6) at (5,0) {};
		\node[draw,circle,minimum size=1mm,fill]      (circle7) at (6,0) {};
		\node[]      (circle8) at (7,0) {};
		
		\draw[-,dashed] (circle0) to
		node[pos=0.5, above,yshift=6pt, inner sep=1pt] {} (circle1);
		\draw[-] (circle1) to
		node[pos=0.5, above,yshift=6pt, inner sep=1pt] {$q_3$} (circle2);
		\draw[-] (circle2) to
		node[pos=0.5, above,yshift=6pt, inner sep=1pt] {$q_3$} (circle3);
		\draw[-] (circle3) to
		node[pos=0.5, above,yshift=6pt, inner sep=1pt] {$q_2$} (circle4);
		\draw[-] (circle4) to
		node[pos=0.5, above,yshift=6pt, inner sep=1pt] {$q_2$} (circle5);
		\draw[-] (circle5) to
		node[pos=0.5, above,yshift=6pt, inner sep=1pt] {$q_3$} (circle6);
		\draw[-] (circle6) to
		node[pos=0.5, above,yshift=6pt, inner sep=1pt] {$q_3$} (circle7);
		\draw[-,dashed] (circle7) to
		node[pos=0.5, above,yshift=6pt, inner sep=1pt] {} (circle8);
	\end{tikzpicture}
\end{equation}

\begin{equation}
\label{eq:bluedashed}
\begin{tikzpicture}[baseline={(0,0)}]
    \node[] (circlem1) at (-2,0) {};
    \node[draw,circle,minimum size=1mm,fill] (circle0) at (-1,0) {};
    \node[draw,circle,minimum size=1mm] (circle1) at (0,0) {};
    \node[draw,circle,minimum size=1mm,fill] (circle2) at (1,0) {};
    \node[draw,circle,minimum size=1mm] (circle3) at (2,0) {};
    \node[draw,circle,minimum size=1mm,fill] (circle4) at (3,0) {};

    \node[draw,circle,minimum size=1mm,fill] (circle5) at (5,0) {};
    \node[draw,circle,minimum size=1mm] (circle6) at (6,0) {};
    \node[draw,circle,minimum size=1mm,fill] (circle7) at (7,0) {};
    \node[draw,circle,minimum size=1mm] (circle8) at (8,0) {};
    \node[draw,circle,minimum size=1mm,fill] (circle9) at (9,0) {};
    \node[] (circle10) at (10,0) {};
    
    \draw[-,dashed] (circlem1) to
    node[pos=0.5, above,yshift=6pt, inner sep=1pt] {} (circle0);
    
    \draw[-] (circle0) to
    node[pos=0.5, above,yshift=6pt, inner sep=1pt] {$q_3$} (circle1);
    
    \draw[-] (circle1) to
    node[pos=0.5, above,yshift=6pt, inner sep=1pt] {$q_3$} (circle2);
    
    \draw[-] (circle2) to
    node[pos=0.5, above,yshift=6pt, inner sep=1pt] {$q_{1,2}$} (circle3);
    
    \draw[-] (circle3) to
    node[pos=0.5, above,yshift=6pt, inner sep=1pt] {$q_{1,2}$} (circle4);

    \draw[-,dashed,blue,very thick] (circle4) --
    node[midway,fill=white,inner sep=1pt] {$\cdots$}
    (circle5);

    \draw[-] (circle5) to
    node[pos=0.5, above,yshift=6pt, inner sep=1pt] {$q_{1,2}$} (circle6);
    
    \draw[-] (circle6) to
    node[pos=0.5, above,yshift=6pt, inner sep=1pt] {$q_{1,2}$} (circle7);
    
    \draw[-] (circle7) to
    node[pos=0.5, above,yshift=6pt, inner sep=1pt] {$q_3$} (circle8);
    
    \draw[-] (circle8) to
    node[pos=0.5, above,yshift=6pt, inner sep=1pt] {$q_3$} (circle9);
    
    \draw[-,dashed] (circle9) to
    node[pos=0.5, above,yshift=6pt, inner sep=1pt] {$q_3$} (circle10);
\end{tikzpicture}
\end{equation}

	\begin{equation}
	\begin{tikzpicture}[baseline={(0,0)}]
		\node[] (circle1) at (0,0) {};
		\node[draw,circle,minimum size=1mm,fill]      (circle2) at (1,0) {};
		\node[draw,circle,minimum size=1mm]      (circle3) at (2,0) {};
		\node[draw,circle,minimum size=1mm,fill]      (circle4) at (3,0) {};
		\node[draw,circle,minimum size=1mm]      (circle5) at (4,0) {};
		\node[draw,circle,minimum size=1mm,fill]      (circle6) at (5,0) {};
		\node[]      (circle7) at (6,0) {};
        
		\draw[-,dashed] (circle1) to
		node[pos=0.5, above,yshift=6pt, inner sep=1pt] {$q_3$} (circle2);
		\draw[-] (circle2) to
		node[pos=0.5, above,yshift=6pt, inner sep=1pt] {$q_3$} (circle3);
		\draw[-] (circle3) to
		node[pos=0.5, above,yshift=6pt, inner sep=1pt] {$q_3$} (circle4);
		\draw[-] (circle4) to
		node[pos=0.5, above,yshift=6pt, inner sep=1pt] {$q_3$} (circle5);
		\draw[-] (circle5) to
		node[pos=0.5, above,yshift=6pt, inner sep=1pt] {$q_3$} (circle6);
		\draw[-,dashed] (circle6) to
		node[pos=0.5, above,yshift=6pt, inner sep=1pt] {$q_3$} (circle7);
	\end{tikzpicture},
\end{equation}
where in equation \eqref{eq:bluedashed}, the blue dashed line denotes any number of alternating $q_1$ and $q_2$ pairs.

 Applying the $\USp(8)$ F-term sequentially, starting from the pairs of $q_3$'s that are connected to $q_1$ or $q_2$, and taking into account the $\SO(6)$ F-terms, we see that all invariants reduce to those made of only alternating $q_1$ and $q_2$ pairs, or of only $q_3$'s:
\begin{equation}
	\begin{tikzpicture}[baseline={(0,0)}]
		\node[] (circlem1) at (-2,0) {};
		\node[draw,circle,minimum size=1mm,fill] (circle0) at (-1,0) {};
		\node[draw,circle,minimum size=1mm] (circle1) at (0,0) {};
		\node[draw,circle,minimum size=1mm,fill]      (circle2) at (1,0) {};
		\node[draw,circle,minimum size=1mm]      (circle3) at (2,0) {};
		\node[draw,circle,minimum size=1mm,fill]      (circle4) at (3,0) {};
		\node[draw,circle,minimum size=1mm]      (circle5) at (4,0) {};
		\node[draw,circle,minimum size=1mm,fill]      (circle6) at (5,0) {};
		\node[draw,circle,minimum size=1mm]      (circle7) at (6,0) {};
		\node[draw,circle,minimum size=1mm,fill]      (circle8) at (7,0) {};
		\node[]      (circle9) at (8,0) {};
		
		\draw[-,dashed] (circlem1) to
		node[pos=0.5, above,yshift=6pt, inner sep=1pt] {} (circle0);
		\draw[-] (circle0) to
		node[pos=0.5, above,yshift=6pt, inner sep=1pt] {$q_2$} (circle1);
		\draw[-] (circle1) to
		node[pos=0.5, above,yshift=6pt, inner sep=1pt] {$q_2$} (circle2);
		\draw[-] (circle2) to
		node[pos=0.5, above,yshift=6pt, inner sep=1pt] {$q_1$} (circle3);
		\draw[-] (circle3) to
		node[pos=0.5, above,yshift=6pt, inner sep=1pt] {$q_1$} (circle4);
		\draw[-] (circle4) to
		node[pos=0.5, above,yshift=6pt, inner sep=1pt] {$q_2$} (circle5);
		\draw[-] (circle5) to
		node[pos=0.5, above,yshift=6pt, inner sep=1pt] {$q_2$} (circle6);
		\draw[-] (circle6) to
		node[pos=0.5, above,yshift=6pt, inner sep=1pt] {$q_1$} (circle7);
		\draw[-] (circle7) to
		node[pos=0.5, above,yshift=6pt, inner sep=1pt] {$q_1$} (circle8);
		\draw[-,dashed] (circle8) to
		node[pos=0.5, above,yshift=6pt, inner sep=1pt] {} (circle9);
	\end{tikzpicture}
\end{equation}
	\begin{equation}
	\begin{tikzpicture}[baseline={(0,0)}]
		\node[] (circle1) at (0,0) {};
		\node[draw,circle,minimum size=1mm,fill]      (circle2) at (1,0) {};
		\node[draw,circle,minimum size=1mm]      (circle3) at (2,0) {};
		\node[draw,circle,minimum size=1mm,fill]      (circle4) at (3,0) {};
		\node[draw,circle,minimum size=1mm]      (circle5) at (4,0) {};
		\node[draw,circle,minimum size=1mm,fill]      (circle6) at (5,0) {};
		\node[]      (circle7) at (6,0) {};
		\draw[-,dashed] (circle1) to
		node[pos=0.5, above,yshift=6pt, inner sep=1pt] {$q_3$} (circle2);
		\draw[-] (circle2) to
		node[pos=0.5, above,yshift=6pt, inner sep=1pt] {$q_3$} (circle3);
		\draw[-] (circle3) to
		node[pos=0.5, above,yshift=6pt, inner sep=1pt] {$q_3$} (circle4);
		\draw[-] (circle4) to
		node[pos=0.5, above,yshift=6pt, inner sep=1pt] {$q_3$} (circle5);
		\draw[-] (circle5) to
		node[pos=0.5, above,yshift=6pt, inner sep=1pt] {$q_3$} (circle6);
		\draw[-,dashed] (circle6) to
		node[pos=0.5, above,yshift=6pt, inner sep=1pt] {$q_3$} (circle7);
	\end{tikzpicture}
\end{equation}
Next, we note that the F-term at the $\USp(8)$ node, equation \eqref{eq;usp8Fterm}, can be squared to yield the following diagrammatic relation (upon squaring, the diagrams with four $q_1$'s or four $q_2$'s are set to zero by the $\SO(6)$ F-terms):
\begin{equation}
	\label{eq:simpedFterm}
	\begin{tikzpicture}[baseline={(0,0)}]
		\node[draw,rectangle,minimum size=1mm,fill]      (circle2) at (1,0) {};
		\node[draw,circle,minimum size=1mm]      (circle3) at (2,0) {};
		\node[draw,circle,minimum size=1mm,fill]      (circle4) at (3,0) {};
		\node[draw,circle,minimum size=1mm]      (circle5) at (4,0) {};
		\node[draw,rectangle,minimum size=1mm,fill]      (circle6) at (5,0) {};

		\draw[-] (circle2) to
		node[pos=0.5, above,yshift=6pt, inner sep=1pt] {$q_1$} (circle3);
		\draw[-] (circle3) to
		node[pos=0.5, above,yshift=6pt, inner sep=1pt] {$q_1$} (circle4);
		\draw[-] (circle4) to
		node[pos=0.5, above,yshift=6pt, inner sep=1pt] {$q_2$} (circle5);
		\draw[-] (circle5) to
		node[pos=0.5, above,yshift=6pt, inner sep=1pt] {$q_2$} (circle6);
	\end{tikzpicture}~~
	=~~
	\begin{tikzpicture}[baseline={(0,0)}]
	\node[draw,rectangle,minimum size=1mm,fill]      (circle2) at (1,0) {};
	\node[draw,circle,minimum size=1mm]      (circle3) at (2,0) {};
	\node[draw,circle,minimum size=1mm,fill]      (circle4) at (3,0) {};
	\node[draw,circle,minimum size=1mm]      (circle5) at (4,0) {};
	\node[draw,rectangle,minimum size=1mm,fill]      (circle6) at (5,0) {};
    
	\draw[-] (circle2) to
	node[pos=0.5, above,yshift=6pt, inner sep=1pt] {$q_3$} (circle3);
	\draw[-] (circle3) to
	node[pos=0.5, above,yshift=6pt, inner sep=1pt] {$q_3$} (circle4);
	\draw[-] (circle4) to
	node[pos=0.5, above,yshift=6pt, inner sep=1pt] {$q_3$} (circle5);
	\draw[-] (circle5) to
	node[pos=0.5, above,yshift=6pt, inner sep=1pt] {$q_3$} (circle6);
	\end{tikzpicture}
\end{equation}
This identifies any operator with only alternating $q_1$ and $q_2$ pairs with an operator made of only $q_3$'s. Lastly, any gauge invariant made of only $q_3$'s can always be related to an operator made of only $q_5$'s by the F-terms along branch 3. The F-term at the terminal $\SO(4)$ node then sets it to zero.

 We must also rule out gauge invariants built using the totally antisymmetric epsilon tensor at the $\SO$ nodes. We first note that if the epsilon tensor of the same gauge node appears twice, an identity relates the product to a sum of products of the symmetric bilinear tensors, so we may restrict to operators that contain at most one epsilon tensor per gauge node.

Then, we consider the three $\SO$ nodes at the end of three branches, $\SO(4)$ and $\SO(6)$ in the example in Figure \ref{fig:usp_3valentexample}. Such operators can only take the form
\begin{equation}
\epsilon^{a_1 a_2\cdots a_n}\ (q_{I})^{\alpha_1}_{a_1}\,(q_{I})^{\alpha_2}_{a_2}\cdots(q_{I})^{\alpha_n}_{a_n}\,\cdots\,,
\end{equation}
where $I$ labels the half-hypers $q_1,q_2$ or $q_5$, $n=4$ or $6$ for $\SO(4)$ or $\SO(6)$, and $a_i$ and $\alpha_i$ are the fundamental indices for corresponding $\SO$ and $\USp$ nodes. The $\alpha_i$'s are totally antisymmetrized. Since $\USp(8)$ has rank $4$, any fully antisymmetric tensor with six fundamental indices can be reduced to tensors which contain $\Omega$-traces among the $q$'s. By the F-terms of the $\SO(6)$ nodes \eqref{eq:Fterms_SO6}, such operators vanish in the quotient. Same argument goes through for the $\SO(4)$ node attached to the $\USp(4)$ node.

We are only left with the bi-valent $\SO$ nodes in branch 3. In the specific example in Figure \ref{fig:usp_3valentexample}, there is only $\SO(8)$ node whose epsilon tensor is available and potential operators take the form
\begin{equation}
\epsilon^{a_1 a_2\cdots a_8}\ (q_{I_1})^{\alpha(I_1)}_{a_1}\,(q_{I_2})^{\alpha(I_2)}_{a_2}\cdots(q_{I_8})^{\alpha(I_8)}_{a_8}\,\dots\,,
\end{equation}
where $a_i$ are the fundamental indices for $\SO(8)$, $I_i$ labels $q_3$ or $q_4$ to the left or right of $\SO(8)$, and $\alpha(I_i)$ is the corresponding $\USp$ index. 

Suppose some of the $\USp$ indices $\alpha(I_i)$ are paired with half-hypers other than $q_3$ and $q_4$, \textit{i.e.}\ with $q_1, q_2$ or $q_5$. Then the $\SO$ indices of those $q_1, q_2$ or $q_5$ inherit the total antisymmetry, because the $\SO(8)$ indices $a_1,\dots, a_8$ are antisymmetrized. These antisymmetric $\SO$ indices cannot be saturated using only the symmetric bilinear tensor $\delta_{ab}$ of the corresponding $\SO$ node. A nonzero contraction would require another epsilon tensor there, which is unavailable by the restriction above. The only remaining option is that the $\USp$ indices $\alpha(I_i)$ are contracted among themselves, \textit{i.e.}
\begin{equation}
\epsilon^{a_1 a_2\cdots a_8}\,\Omega_{\alpha(I_1)\alpha(I_2)}\cdots\Omega_{\alpha(I_7)\alpha(I_8)}\,(q_{I_1})^{\alpha(I_1)}_{a_1}\,(q_{I_2})^{\alpha(I_2)}_{a_2}\cdots(q_{I_8})^{\alpha(I_8)}_{a_8}\,.
\end{equation}
Using the F-terms of $\SO(8)$ (equation \ref{eq:Fterms_SO8}), the $\USp$ contractions involving $q_3$ can again be traded for $q_4$, reducing the candidate to
\begin{equation}
\epsilon^{a_1\cdots a_8}\,\Omega_{i_1 i_2}\cdots\Omega_{i_7 i_8}\,(q_4)^{i_1}_{a_1}\cdots(q_4)^{i_8}_{a_8}\,.
\end{equation}
This operator vanishes because the eight $\USp(4)$ indices $i_1,\ldots,i_8$ take only four distinct values, but the antisymmetric contraction in the $\SO(8)$ indices effectively requires them to be distinct. The same argument extends to the higher-rank members of the family. When there are more bi-valent $\SO$ nodes in branch 3, we can make the same argument for each of them starting from the leftmost $\SO$ node in branch 3.

This concludes our enumeration of the potential gauge invariants of the family of Figure \ref{fig:usp_3valent}. As independent supporting evidence, we have computed the HL index of the two lowest-rank members of the family,\footnote{We use the duality $-\SO(5)-\USp(2)-\SO(3)\;\cong\;-\USp(4)-\SO(4)$ to write the shortest equivalent quiver. We write $A-B^n$ for an $A$ node with $n$ separate $B$ branches.}
\begin{align}
\USp(4)-\SO(4)^3 :\qquad
&\mathcal{I}_{\rm HL} \;=\; 1+6\,t^4+t^7\,,
\\
\SO(4)-\USp(4)-\SO(8)-\USp(8)-\bigl(\SO(6)\bigr)^2 :\qquad
&\mathcal{I}_{\rm HL} \;=\; 1 + 3\,t^6 - 8\,t^9 + 3\,t^{11} + t^{15}\,.
\end{align}
Both indices truncate, consistent with the absence of a Higgs branch in these theories.

\medskip
\noindent
{\bf Case C: At least one length-two branch is built from a special representation.}~
The candidate special branches without flavor symmetry that fit Figure~\ref{fig:possible3valent} are
\begin{enumerate}
\item[--] $-\SO(13)-\mathbf{\tfrac{1}{2}S}$ and $-\SO(11)-\mathbf{\tfrac{1}{2}S}$,
\item[--] $-\SO(12)-\mathbf{\tfrac{1}{2}S}$ and $-\SO(12)-\mathbf{\tfrac{1}{2}C}$,
\item[--] $-\USp(6)-\mathbf{\tfrac{1}{2}asym3}$,
\item[--] $-\USp(8)-G_2$.
\end{enumerate}
Apart from $-\USp(8)-G_2$, each branch may appear at most once in the quiver, otherwise the resulting theory carries a flavor symmetry. We eliminate these in turn.

\underline{\it $-\SO(13)-\mathbf{\tfrac{1}{2}S}$ and $-\SO(11)-\mathbf{\tfrac{1}{2}S}$ branches.} The second length-two branch would have to terminate at a $\USp(n)$ node attached to the $\SO(13)$ or $\SO(11)$ trunk, but neither $\SO(13)$--$\USp(n)$ nor $\SO(11)$--$\USp(n)$ can be conformal. These are ruled out.

\underline{\it $-\SO(12)-\mathbf{\tfrac{1}{2}S}$ and $-\SO(12)-\mathbf{\tfrac{1}{2}C}$ branches.} For a central $\SO(12)$ trunk, the ranks along branch~3 grow:
\begin{equation}
\begin{tikzpicture}[baseline={(0,-0.5ex)}]
		\node       (circle1)      at (0,0)   { $\SO(12)$};
		\node      (circle3)    at (1.7,0)    {$\USp(12)$};
        \node        (circle2)    at (3.4,0)    {$\SO(16)$};
        \node        (circle6)    at (5,0)    {$\cdots$};
		\node      (circle4)    at (-1.5,0)    {$\mathbf{\tfrac{1}{2}S}$};
		\node       (circle5)    at (0,1)    {$\mathbf{\tfrac{1}{2}C}$};
		\draw[-] (circle3.east) -- (circle2.west);
        \draw[-] (circle2.east) -- (circle6.west);
		\draw[-] (circle1.east) -- (circle3.west) ;
		\draw[-] (circle1.north) -- (circle5.south);
		\draw[-] (circle1.west) -- (circle4.east);
\end{tikzpicture}~\text{or}~
\begin{tikzpicture}[baseline={(0,-0.5ex)}]
		\node       (circle1)      at (0,0)   { $\SO(12)$};
		\node      (circle3)    at (1.7,0)    {$\USp(12)$};
        \node        (circle2)    at (3.4,0)    {$\SO(16)$};
        \node        (circle6)    at (5,0)    {$\cdots$};
		\node      (circle4)    at (-1.5,0)    {$\mathbf{\tfrac{1}{2}S/C}$};
		\node       (circle5)    at (0,1)    {$\USp(4)$};
		\draw[-] (circle3.east) -- (circle2.west);
        \draw[-] (circle2.east) -- (circle6.west);
		\draw[-] (circle1.east) -- (circle3.west) ;
		\draw[-] (circle1.north) -- (circle5.south);
		\draw[-] (circle1.west) -- (circle4.east);
\end{tikzpicture}
\end{equation}
so the chain cannot terminate without further branching. In either case, the resulting graph contains the subgraph of Figure \ref{fig:chain}, and the theory has a Higgs branch.

\underline{\it $-\USp(6)-\mathbf{\tfrac{1}{2}asym3}$ branch.} The unique balanced quiver built from it is
\begin{equation}
\tikz[]{
		\node (L) at (-1.8,-0.25) {$\mathbf{\tfrac{1}{2}asym3}$};
		\node (LU) at (-1.5,0.5) {$\SO(5)$};
		\node (C) at (0,0) {$\USp(6)$};
		\node (R) at (1.5,0) {$\SO(6)$};
		\node (RU) at (3,0) {$\USp(2)$}; 
        \node (RU2) at (4.5,0) {$\SO(2)$}; 
		\draw (C) -- (L);
		\draw (C) -- (LU);
		\draw (C) -- (R);
		\draw (R) -- (RU);
        \draw (RU) -- (RU2);
}
\end{equation}
which is not conformal.

\underline{\it $-\USp(8)-G_2$ branch.} There are two balanced quivers built from it:
\begin{enumerate}
\item[(1)]
\begin{equation}
\label{eq:g2g2usp8so6}
\begin{tikzpicture}[baseline={(0,-0.5ex)}]
		\node       (circle1)      at (0,0)   { $\mathrm{USp}(8)$};
		\node      (circle3)    at (1.6,0)    {$\mathrm{SO}(6)$};
		\node      (circle4)    at (-1.5,0)    {$\mathrm{G}_2$};
		\node       (circle5)    at (0,0.8)    {$\mathrm{G}_2$};
		\draw[-] (circle1.east) -- (circle3.west) ;
		\draw[-] (circle1.north) -- (circle5.south);
		\draw[-] (circle1.west) -- (circle4.east);
\end{tikzpicture}\qquad (n_h,\,n_v) = (80,\,79)\,,
\end{equation}
\item[(2)]
\begin{equation}
\begin{tikzpicture}[baseline={(0,-0.5ex)}]
		\node       (circle1)      at (0,0)   { $\mathrm{USp}(8)$};
		\node      (circle3)    at (1.6,0)    {$\mathrm{SO}(7)$};
		\node      (circle4)    at (-1.6,0)    {$\mathrm{SO}(6)$};
		\node       (circle5)    at (0,0.8)    {$\mathrm{G}_2$};
		\node      (circle6)    at (3.2,0)    {$\mathrm{USp}(2)$};
		\node       (circle7)    at (4.4,0)    {$1$};
		\draw[-] (circle1.east) -- (circle3.west) ;
		\draw[-] (circle1.north) -- (circle5.south);
		\draw[-] (circle1.west) -- (circle4.east);
		\draw[-] (circle3.east) -- (circle6.west);
		\draw[-] (circle6.east) -- (circle7.west);
\end{tikzpicture}\qquad (n_h,\,n_v) = (88,\,89)\,.
\end{equation}
\end{enumerate}
Theory (1) has $n_{\rm h}-n_{\rm v}=1>0$, which implies a non-trivial Higgs branch. Theory (2) has $n_{\rm h}-n_{\rm v}=-1$, and the criterion is inconclusive. We compute its HL index,
    \begin{equation}
        \small
        \begin{split}
            \mathcal{I}_{\text{HL}}=&\Big(1 + t + 2 t^2 + 3 t^3 + 5 t^4 + 3 t^5 + t^6 + t^7 - 2 t^8 - 8 t^9 - 3 t^{10}- 5 t^{11} - 2 t^{12} - 2 t^{13} + 3 t^{14} - 2 t^{15} \\
            &+ t^{16} + 2 t^{17} + t^{18} + t^{19} + t^{20}\Big)/\Big((1 - t) (1 + t)^2 (1 + t^2) (1 - t + t^2) (1 + t + t^2) (1 + t^4)\Big)~.
        \end{split}
        \end{equation}
The index does not truncate and theory (2) has a Higgs branch. This exhausts the case-by-case analysis. The only candidate interacting Higgsless Lagrangian SCFTs are the four entries (i)--(iv) of the result stated in Section~\ref{subsec:strategy}.

\section{Examples}
\label{sec:examples}
In the previous section we identified a list of candidate Higgsless Lagrangian SCFTs. 
The bulk of this section is devoted to detailed studies of two of them: the sporadic theories $\USp(4)$ with half-hypermultiplets in the $\mathbf{16}$ (item (i) of the result of Section~\ref{subsec:strategy}) and $\SU(3)\times\SU(2)$ with half-hypermultiplets in the $\mathbf{8}\times\mathbf{2}$ (item (ii)). 
In each case, we first confirm (by  group-theoretic arguments and/or by detailed calculations of the Higgs Hilbert series and of the Hall-Littlewood index) 
that the Higgs branch is indeed trivial.

The standard BRST prescription to determine the VOA \cite{Beem:2013sza} from the VOAs of free hypermultiplets and free vector multiplets becomes cumbersome as the rank of the theory increases. Already in our examples it is more efficient to use a bootstrap approach. In each case, by examining the Schur and Macdonald index of the theory, we make an ansatz for the possible strong generators and their chiral dimensions. 
We then write the most general OPEs with unfixed coefficients and require the Jacobi identities to be satisfied. We implement the OPEs and impose Jacobi identities in \texttt{Mathematica} using the package \texttt{OPEdefs} \cite{Thielemans:1994er}. The Jacobi identities uniquely fix these coefficients, up to normalization of the generators. It is worth noting that the Jacobi identities hold only thanks to nontrivial null states, which are set to zero in the simple quotient. (Recall that in the SCFT/VOA correspondence, we are always instructed to consider the simple quotient of the VOA, as a basic consequence of four-dimensional unitarity.)
Once the OPEs of the strong generators are fixed, we compute the vacuum character of the VOA up to order $q^{8}$ and find a match with the Schur index.  This serves as a non-trivial check since the Schur index with certain generators can only be reproduced if the right null states are taken into account. Using null states at the first few levels, we are able to show that the strong generators are nilpotent in ${\cal R}_\mathcal{V}$, confirming that these two novel VOAs are indeed
$C_2$-cofinite,
as expected from the Higgs branch conjecture. While we do not perform the BRST computation directly, it would provide a definitive identification of the bootstrapped VOAs with $\mathcal{V}[\mathcal{T}]$.

We will also observe that both VOAs enjoy an $\mathfrak{sl}_2$ outer automorphism. This is expected on general grounds from the results of \cite{Beem:2025guj}. Indeed, for a Lagrangian $\mathcal N=2$ superconformal gauge theory, the Schur VOA is realized as the relative semi-infinite BRST cohomology of a symplectic-boson matter VOA. Since the latter is graded-unitary, and the corresponding Hamiltonian action is good, Theorem 4.1 of \cite{Beem:2025guj} implies that the BRST cohomology is again graded-unitary. Moreover, by Proposition 4.6 of \cite{Beem:2025guj}, the $\U(1)_r$ grading on the cohomology extends to a canonical $\USp(2)$ action by vertex algebra automorphisms, or equivalently to a complexified $\mathfrak{sl}_2$ action. For semisimple gauge group, this statement is compatible with the iterated-cohomology results of Section 4.4 of \cite{Beem:2025guj}, which in fact indicate one such $\USp(2)$ structure for each simple gauge factor, possibly with some acting trivially.

Finally, we examine the modularity properties of these two VOAs. In each case we have at our disposal a closed form expression for the vacuum character, so it is a simple matter to compute its S-transform and see that it contains logarithmic pseudocharacters. This establishes that both VOAs are strongly finite but non-rational. For completeness, and as a preparation for future cases where closed form expressions may not be easily available, we also discuss modularity from the viewpoints of the modular linear differential equation and of the high-temperature expansion.

Before tackling the two interacting cases, let us warm
up with a brief review of free vector multiplets and their discrete gaugings. While very simple from a physics perspective, these are already  non-trivial examples of strongly finite logarithmic VOAs.
    
\subsection{Free examples: Free vector multiplets and their discrete gaugings}
	
	The simplest 4d $\mathcal{N}=2$ Lagrangian SCFT that does not have a Higgs branch is the theory of
     free $\mathcal{N}=2$ vector multiplets,
     with central charges
     \begin{equation}
		c_{\text{4d}}=\frac{1}{6},\; a_{\text{4d}}=\frac{5}{24}\,.
	\end{equation}
     A collection of $d$ vector multiplets is associated
     to the VOA generated by
  $d$ pairs of symplectic fermions
  $\lambda^\alpha$, with OPE
	\begin{equation}
    \label{eq:sympfermOPE}
		\lambda^{\alpha}(z)\lambda^{\beta}(0)\sim \frac{\epsilon^{\alpha\beta}}{(z-w)^2}~.
	\end{equation} 
    This VOA is denoted by
$SF_d$ in the mathematical literature (see e.g.~\cite{Kausch:2000fu,Creutzig:2013hma,Creutzig:2016fms}). The symplectic fermion VOA has an $\mathfrak{sl}_2$ automorphism.
	 As the $\lambda^\alpha$ are fermionic, they are obviously nilpotent in $\mathcal{R_V}$ and therefore $SF_d$ is $C_2$-cofinite.
     
	 Let us start with the case of a single free vector multiplet, with associated VOA $SF_1$. The Schur index (vacuum character) and the HL index can be written in the following closed form,
	\begin{align}
		\mathcal{I}_{\text{Schur}}(q)&=q^{-\frac{1}{12}}\chi_0(q)=(q;q)_{\infty}^2=q^{-\frac{1}{12}}\eta(q)^2~,\\
		\mathcal{I}_{\text{HL}}(q)&=1-t~.
	\end{align}
	Here the $q$-Pochhammer symbol is defined as $(x;q)_{\infty}=\prod^\infty_{n=0}(1-xq^n)$ while the Dedekind $\eta$ function is defined as $\eta(q)=q^{\frac{1}{24}}(q;q)_{\infty}$.
    The HL index is a finite series in $t$, as expected. The vacuum character solves second-order modular linear differential equation (MLDE)~\cite{Mathur:1988na}, $\mathcal{D}_{SF_1} \chi_0 = 0$, where\footnote{While presenting the MLDEs explicitly, we will use $D_q^{(k)}$ to denote derivatives composed of Ramanujan–Serre derivative $\partial_{(k)}$
\begin{equation}
	D_q^{(k)}:=\partial_{2k-2}\dots\partial_{2}\partial_{0},~~\partial_{(k)}:=q\partial_q+k\mathbb{E}_2(\tau)~
\end{equation}
and the following convention for Eisenstein series
\begin{equation}
	\mathbb{E}_{2k}(\tau):=-\frac{B_{2k}}{(2k)!}+\frac{2}{(2k-1)!}\sum_{n\geq 1}\frac{n^{2k-1}q^n}{1-q^n}~,
\end{equation}
where $B_{2k}$ is the $2k$-th Bernoulli number.} $\mathcal{D}_{SF_1}=D^{(2)}_q+5\mathbb{E}_4$. The indicial equation has  two degenerate solutions, $h_1=h_2=0$. The other solution to the MLDE is given by
	\begin{equation}
		\pchi_0^L=-\frac{\log(q)}{2\pi}\eta(q)^2~.
	\end{equation}
	In the following, we will always use $\pchi^{L^n}$ to denote the logarithmic solution where the order of $\log q$ is $n$. The appearance of a logarithmic solution suggests that $SF_1$ is a non-rational $C_2$-cofinite VOA. This is easily confirmed
    by showing that the logarithmic solution does appear in the modular transformation of the vacuum character. The modular
    $S$ and $T$ matrices in the basis $\{\chi_0,\pchi_0^L\}$ follow immediately from the modular transformation properties of the Dedekind eta function,
	\begin{equation}
		S=\begin{pmatrix}
			0 & 1\\
			1 & 0
		\end{pmatrix},\;\;\;
		T=\begin{pmatrix}
			e^{\pi i/6} & 0\\
			e^{-\pi i/3} & e^{\pi i/6}
		\end{pmatrix}~.
	\end{equation}
	 $SF_1$ has a single simple module, namely itself. The two solutions to the MLDE correspond to the ordinary character of this simple module and to a pseudocharacter.

	\subsubsection*{$\mathbb{Z}_2$ gauging of a single  vector multiplet: $SF_1^+=\mathcal{W}_2$}
	
	The  free vector multiplet SCFT has a discrete $\mathbb{Z}_2$ symmetry with the action
	\begin{align}
		\mathbb{Z}_2: \; \varphi\rightarrow-\varphi
	\end{align}
	where $\varphi\in \{A_\mu,\lambda_\alpha,\tilde{\lambda}_{\dot{\alpha}},\phi\}$ of the vector multiplet. Correspondingly, there is also a $\mathbb{Z}_2$ symmetry of $SF_1$ that acts with a minus sign on the symplectic fermions. The VOA associated with the $\mathbb{Z}_2$ discrete gauging  of a free vector multiplet is the even part of the symplectic fermions, $SF^+_1$. $SF^+_1$ coincides with the simplest triplet algebra $\mathcal{W}_2$ \cite{Gaberdiel:1996np,Gaberdiel:2001tr,Flohr:2005cm,Gaberdiel:2006pp}. This VOA is generated by the stress tensor $T$ and a bosonic $\mathfrak{sl}_2$-triplet $W_{\alpha\beta}$ with conformal weight $h=3$, which can be written in terms of the symplectic fermions $\lambda_\alpha$,
	\begin{align}
		T=-\frac{1}{2}\epsilon^{\alpha\beta}\lambda_\alpha\lambda_\beta, \;\; W_{\alpha\beta}=\lambda_{(\alpha}\partial\lambda_{\beta)}.
	\end{align}
	The Schur index/vacuum character of $SF_1^+$ is
	\begin{align}
		q^{\frac{1}{12}}\mathcal{I}_{\mathrm{Schur}}(q)=\chi_0(q)=\frac{1}{2}\left(\frac{\eta(q^2)^2}{\eta(q)^2}+\eta(q)^2\right).
	\end{align}
The vacuum character solves following MLDE of order 5,
	\begin{equation}\label{eq:MLDE_W_2}
		\begin{split}
			\mathcal{D}_{SF_1^+}=&D^{(5)}_q-\frac{815}{4}\mathbb{E}_4D^{(3)}_q-\frac{6825}{8}\mathbb{E}_6D^{(2)}_q+\frac{3025}{4}\mathbb{E}^2_4D^{(1)}_q-\frac{25025}{8}\mathbb{E}_4\mathbb{E}_6. 
		\end{split}
	\end{equation}
	All solutions to the MLDE are collected in Table \ref{tab:W_2_MDE_solutions} with their corresponding weights.
	\begin{table}[t]
		\centering
		\begin{tabular}{|c|c| c|}
			\hline
			$h$ & & Solution to MLDE \\
			\hline
			$0$ & $\chi_0$ & $\frac{1}{2}\left(\frac{\eta(q^2)^2}{\eta(q)^2}+\eta(q)^2\right)$\\
			$0$& $\pchi^L_{0}$ & $ \frac{1}{2\pi i}\log(q)\eta(q)^2=\frac{1}{2\pi i}\log(q)(\chi_0-\chi_1)$\\
			$1$ &  $\chi_{1}$ & $\frac{1}{2}\left(\frac{\eta(q^2)^2}{\eta(q)^2}-\eta(q)^2\right)$\\
			$-\frac{1}{8}$ & $\chi_{-1/8}$ & $\frac{1}{2}\left(\frac{\eta(q)^4}{\eta(q^2)^2\eta(q^{1/2})^2}+\frac{\eta(q^{1/2})^2}{\eta(q)^2}\right)$\\
			$\frac{3}{8}$& $\chi_{3/8}$  & $\frac{1}{2}\left(\frac{\eta(q)^4}{\eta(q^2)^2\eta(q^{1/2})^2}-\frac{\eta(q^{1/2})^2}{\eta(q)^2}\right)$\\
			\hline
		\end{tabular}
		\caption{Closed form solutions of the MLDE (equation \eqref{eq:MLDE_W_2}) for characters of the VOA $SF_1^+$ corresponding to $\mathbb{Z}_2$ quotient of the free vector multiplet.}
		\label{tab:W_2_MDE_solutions}
	\end{table}
	The span of $\{\chi_0,\pchi_0^L,\chi_1,\chi_{-1/8},\chi_{3/8}\}$ is closed under $SL(2,\mathbb{Z})$. The $S$ and $T$ matrices in this basis are given by
	\begin{equation}
		S=\left(
		\begin{array}{ccccc}
			0 & -\frac{i}{2}  & 0 & \frac{1}{4} & -\frac{1}{4} \\
			i &0 & -i &0 & 0 \\
			0 & \frac{i}{2}  & 0 & \frac{1}{4} & -\frac{1}{4} \\
			1 & 0 & 1 & \frac{1}{2} & \frac{1}{2} \\
			-1 & 0 & -1 & \frac{1}{2} & \frac{1}{2} \\
		\end{array}
		\right)~,~
		T=\left(
		\begin{array}{ccccc}
			e^{\frac{i \pi }{6}} & 0 & 0 & 0 & 0 \\
			e^{\frac{i \pi }{6}} & e^{\frac{i \pi }{6}} & e^{\frac{5i \pi }{6}} & 0 & 0 \\
			0 & 0 & e^{\frac{i \pi }{6}} & 0 & 0 \\
			0 & 0 & 0 & e^{-\frac{i \pi }{12}} & 0 \\
			0 & 0 & 0 & 0 & e^{\frac{11i\pi }{12} } \\
		\end{array}\right).
	\end{equation}
	The four non-logarithmic solutions to the MLDE in \eqref{eq:MLDE_W_2} correspond to the characters of the four simple modules of $SF_1^+$. As expected~\cite{miyamoto2015c}, the logarithmic solution $\pchi^L_0$ can be written as a linear combination of characters of simples with coefficient a polynomial in $\log (q)$ \cite{Miyamoto:2002ar}. 
	
	\subsubsection*{General $d$: $SF_d$, $SF_d^+$}
	
	For $d$ free vector multiplets, the associated VOA $SF_d$ is just the $d$-fold tensor product of $SF_1$. Just as in the case of $d=1$, taking the $\mathbb{Z}_2$-orbifold theory, its VOA $SF_d^+$ is the $\mathbb{Z}_2$-even part of $SF_d$. It also has four simple modules and their characters are generalizations of those for $SF_1^+$ with conformal weights $0,1,-d/8,(4-d)/8$. More details can be found in \cite{Creutzig:2016fms}. Although we have only discussed $\mathbb{Z}_2$ orbifolds, the automorphism group of symplectic fermions is larger, and  other $C_2$-cofinite VOAs  can be obtained by taking more general orbifolds~\cite{Creutzig:2014xea}.

	\subsection{Interacting  examples}

We now graduate to  two interacting Lagrangian examples.

	\subsubsection{$\mathrm{USp}(4)$ with half-hypers in $\mathbf{16}$}
	\label{sec:usp416b2}
	
	The next simplest example is the $\USp(4)$ gauge theory with half-hypermultiplets in the $\mathbf{16}$ representation of $\USp(4)$. This theory has no flavor symmetry and $n_{\text{h}}=8,n_{\text{v}}=10,n_{\text{h}}-n_{\text{v}}=-2<0$. Thus, from our previous discussion, it is a candidate Higgsless theory.
	
	\subsubsection*{Absence of Higgs branch}

 To illustrate how one may proceed in general, we start by computing the cheapest invariant, the Hall-Littlewood index -- as discussed in Section \ref{sec:scindex}, the truncation of the HL index is an indicator for the theory to be Higgsless. The HL index is given by the matrix integral
	\begin{equation}
		\begin{split}
		\mathcal{I}_{\mathrm{HL}}(t)=&\oint  \frac{dx_1 dx_2}{2\pi i x_12\pi i x_2} \Delta_{\USp(4)}(x_1,x_2)\,\,\frac{\prod_{\alpha}(1-\mathbf{x}^\alpha\ t)}{\prod_{\rho}(1-\mathbf{x}^\rho \ t^{1/2})}~,
		\end{split}
	\end{equation}
	where $\Delta_{\USp(4)}(x_1,x_2)$ is the Haar measure, $\alpha$ are the roots and $\rho$ are the weights of $\mathbf{16}$ of $\USp(4)$. Computing the contour integral explicitly yields
    \begin{equation}\label{eq:usp4_hl}
    \begin{split}
    \mathcal{I}_{\mathrm{HL}}(t)=&1-t^2+t^3,\\
    =&\frac{1 - t^4 + t^5 - t^6 + t^7 - t^8}{(1 + t^2) (1 - t^3)}~.
    \end{split}
	\end{equation}
    In the second line, we have rewritten the finite HL index in terms of a fraction where the denominator contains the generators and the numerator encodes the relations. In this case, there is one dimension-two fermionic HL chiral ring generator and one dimension-three bosonic HL chiral ring generator. 
    The finite HL index implies that there are no Higgs branch generators under the two assumptions discussed in Section \ref{sec:scindex}. 

    In this case we can actually do better and give a rigorous argument that the theory does not have a Higgs branch, by computing its Higgs branch Hilbert series. Let us denote the half-hypermultiplets in the $\mathbf{16}$ of $\USp(4)$ by $q_{[ab]c}$ where $a,b,c$ are the vector indices of $\USp(4)$ and the first two indices are anti-symmetrized. In addition, $q_{[ab]c}$ satisfy the following two sets of constraints
	\begin{align}
		q_{[ab]c}\Omega^{ab}&=0\; \; \; (\Omega\mathrm{-tracelessness})~,\\
		q_{[ab]c}+q_{[bc]a}+q_{[ca]b}&=0\; \; \; \mathrm{(cyclic~identity)}~,
	\end{align}
	where $\Omega$ is the invariant tensor of $\USp(4)$, $\Omega^{ab}=\big(\begin{smallmatrix}0 & 1\\ -1 & 0\end{smallmatrix}\big)\otimes \mathds{1}_{2}$. In total, there are $\frac{4\times3}{2}\times4-4-4=16$ components in $q_{[ab]c}$. The F-term relations are
	\begin{align}
		q_{[ab]c}q_{[ij]k}\Omega^{bj}\Omega^{ck}=0, \; \; (ai)\ \mathrm{symmetrized}.
	\end{align}
	There are $\frac{4\times 5}{2}=10$ relations. Then, we find the Hilbert series for the quotient ring $\mathbb{C}[q_{[ab]c}]/\langle \text{F-terms}\rangle$ using \textsf{Macaulay2} and perform the contour integral to project onto gauge-invariant operators. We obtain precisely $\mathcal{I}_{\text{Higgs}}(t)=1$. This shows that this theory does not have a Higgs branch.

	\subsubsection*{Strong generators}

To identify the low-lying operators in the VOA, we examine the series expansion of the Schur and Macdonald indices at low orders in the fugacities $q$ and $T$ (see Appendix~\ref{app:sci} for a summary of our index conventions):
	\begin{equation}
    \small
		\begin{split}
			\mathcal{I}_{\text{Schur}}(q)&=1-q^2+2 q^3+2 q^4+4 q^5+q^6+2 q^7+5 q^8+4 q^9+12 q^{10}+14 q^{11}+16 q^{12}+O(q^{13})\\
			&=\mathrm{PE}\left[\frac{-q^2+3 q^3+4 q^5-6 q^6+4 q^7-11 q^8+3 q^9-11 q^{10}+23 q^{11}-7 q^{12}+O(q^{13})}{1-q}\right]~,\\
			\mathcal{I}_{\text{Mac}}(q,T)&=1 - T^2 q^2 + (T + T^3) q^3 + (T + T^3) q^4 + (T + T^2 + T^3 + T^4) q^5+O(q^6)\\
			&=\mathrm{PE}\left[\frac{-T^2 q^2+\left(T+T^2+T^3\right) q^3+\left(T^2+T^3+T^4+T^5\right) q^5+O\left(q^6\right)}{1-q}\right]~.
		\end{split}
	\end{equation}
    Here PE denotes the plethystic exponential. The argument of PE, also known as the plethystic logarithm, captures  generators and nulls of a VOA. At low orders, a bosonic generator or fermionic relation of dimension $h$ and R-charges $R$ and $r$ contributes to the Macdonald index as $q^h T^{R+r}$, while a fermionic generator or bosonic relation contributes as $-q^h T^{R+r}$. At higher orders, this interpretation becomes less direct as there may exist relation among relations (syzygies).

    We summarize our proposal for the strong generators in Table~\ref{tab:generators_usp4}. Let us briefly illustrate our reasoning.
    We know on general grounds that a (single) stress tensor must be present, and must in fact be a strong generator (as  in this case there are no affine currents for a Sugawara construction). The stress tensor has $h=2$, $R=1$ and $r=0$,
    contributing $Tq^2$ to the plethystic log of the Macdonald index. As we find no such monomial,
    the only scenario compatible with 4d superconformal representation theory 
     is the existence of a fermionic Schur operator\footnote{It helps to recall that a Schur operator is bosonic if $2r$ is even and fermionic if $2r$ is odd.} from a $\bar {\cal D}_{1, (0, 0)}$ multiplet, which
     has $h=2$, $R=3/2$ and $r=-1/2$ and thus contributes $- T q^2$,  canceling the stress tensor contribution. The conjugate multiplet ${\cal D}_{1, (0, 0)}$, which must be present by CPT symmetry, contributes
     $-T^2 q^2$, fully explaining the index at $h=2$.
 At $h=3$, we straightforwardly read off the existence of three additional bosonic generators with $R=2$ and $r=\pm1,0$, which we  interpret as arising from suitable Schur multiplets. The simplest guess is that 
this exhausts 
 the complete list of strong generators.

We also indicate in Table~\ref{tab:generators_usp4}  the schematic form of the generators in terms of their elementary constituents: 
$q$ denotes the symplectic bosons in the $\mathbf{16}$ of $\USp(4)$, corresponding to the half-hypermultiplets, while $\lambda,\tilde{\lambda}$ denotes the symplectic fermions in the adjoint of $\USp(4)$, corresponding to the free vector multiplet. For brevity, we suppress $\USp(4)$ indices, and any word written as a product of letters $q,\lambda,\tilde{\lambda}$ and $\partial$ is to be understood to denote the appropriate $\USp(4)$ contractions to form a gauge singlet. 
    
The  $r$-charge assignments in Table~\ref{tab:generators_usp4} exhibit the $\mathfrak{sl}_2$ automorphism, guaranteed on general grounds by Proposition 4.6 of \cite{Beem:2025guj} as recalled at the beginning of this section, which descends from the $\mathfrak{sl}_2$ automorphism of the symplectic fermions in the adjoint of $\USp(4)$. We can arrange the generators into a singlet of dimension 2 (the stress tensor $T$), a fermionic doublet $\Lambda_a$ of dimension 2 and a bosonic triplet $W_{(ab)}$ of dimension 3. Here $a,b,c,d$ are fundamental indices of $\mathfrak{sl}(2)$ and $W_{(ab)}$ is symmetric. Collecting the symplectic fermions as a doublet as in equation \eqref{eq:sympfermOPE} ($\lambda_+:=\lambda,~\lambda_-:=\tilde{\lambda}$), the schematic form of the generators in Table~\ref{tab:generators_usp4} can be written in an explicitly $\mathfrak{sl}_2$ covariant form as
\begin{align}
    T&\simeq q\partial q+\lambda_a\lambda_b\epsilon^{ab}~,\\
    \Lambda_a&\simeq qq\lambda_a~,\\
    W_{ab}&\simeq qq\lambda_a\lambda_b~.
\end{align}

\begin{table}[t]
\centering
\renewcommand{\arraystretch}{1.5}
\begin{tabular}{|c|c|c|c|c|c|c|}
    \hline
    Name & Multiplet & Index contribution& \makecell{Schematic form\\ of operators} & $h$ & $SU(2)_R$ &$U(1)_r$\\
    \hline
    $T$ & $\hat{\mathcal{C}}_{0(0,0)}$ & $\frac{Tq^2}{1-q}$ & $q\partial q+\lambda\tilde{\lambda}$ & $2$ & $1$ &$0$\\
    $\Lambda_+$ & $\mathcal{D}_{1(0,0)}$ & $\frac{-T^2q^2}{1-q}$ &$qq\tilde{\lambda}$ & $2$ & $\frac{3}{2}$ &$\frac{1}{2}$\\
    $\Lambda_-$ & $\bar{\mathcal{D}}_{1(0,0)}$ & $\frac{-Tq^2}{1-q}$ & $qq\lambda$ & $2$ & $\frac{3}{2}$ &$-\frac{1}{2}$\\
    $W_{++}$ & $\mathcal{D}_{\frac{3}{2}(0,\frac{1}{2})}$ & $\frac{T^3q^3}{1-q}$ & $qq\tilde{\lambda}\tilde{\lambda}$ & $3$ & $2$ &$1$\\
    $W_{+-}$ & $\hat{\mathcal{C}}_{1(0,0)}$ &  $\frac{T^2q^3}{1-q}$ & $qq\lambda\tilde{\lambda}$ & $3$ & $2$ &$0$\\
    $W_{--}$ & $\bar{\mathcal{D}}_{\frac{3}{2}(\frac{1}{2},0)}$ &  $\frac{Tq^3}{1-q}$ & $qq\lambda \lambda$ & $3$ & $2$ &$-1$\\
    \hline
\end{tabular}
\caption{VOA generators for the $\USp(4)$ theory with half-hypermultiplets in $\mathbf{16}$. The first two columns list the name and the four-dimensional multiplet associated with each generator. The third column lists the contribution of each multiplet to the Macdonald index. The fourth column lists the schematic form of each generator in terms of the elementary fields in the Lagrangian theory, where $q$ is the generator of free half-hypermultiplet and $\lambda,\tilde{\lambda}$ are the generators of free vector multiplet. The last three columns list the quantum numbers of the generators. The HL chiral ring generators are $\Lambda_+$ and $W_{++}$ (they correspond to the denominator in (\ref{eq:usp4_hl})) while the HL anti-chiral ring generators are $\Lambda_-$ and $W_{--}$.}
\label{tab:generators_usp4}
\end{table}

\subsubsection*{Associated strongly finite VOA}

We now proceed to bootstrap the vertex algebra. Its central charge is $c_{\rm 2d}=-28$, and the OPE of the stress tensor with itself takes the canonical form; the other strong generators are Virasoro primaries of the appropriate dimensions. We then write the most general ansatz for the 
singular OPEs of the strong generators, and fix coefficients imposing Jacobi identities. As already remarked, null relations are crucial for the Jacobi identities to hold. The null relations required are collected in Appendix \ref{app:nulls}.  We find the following OPEs:
    \begin{equation}
    \footnotesize
	\begin{split}
	    \Lambda_a(z) \Lambda_b(w)&\sim\frac{-7\epsilon_{ab}}{(z-w)^4}+\frac{\epsilon_{ab}T(w)}{(z-w)^2}+\frac{\frac{1}{2}\epsilon_{ab}T'(w)+W_{ab}(w)}{(z-w)}~,\\
		W_{a b}(z) W_{cd}(w)&\sim\epsilon_{(a(c}\epsilon_{b)d)}\left(\frac{-14}{(z-w)^6}+\frac{3T(w)}{(z-w)^4}+\frac{
        \frac{3}{2}T'(w)}{(z-w)^3}+\frac{\frac{3}{4}T''(w)-\frac{1}{2}T^2(w)}{(z-w)^2}+\frac{\frac{1}{4}T'''(w)-\frac{1}{2}T'T(w)}{(z-w)}\right)\\
		& + \left(\frac{\epsilon_{(a(c}W_{b)d)}(w)}{(z-w)^3}+\frac{\epsilon_{(a(c}W_{b)d)}'(w)}{(z-w)^2}+\frac{\frac{1}{2}\epsilon_{(a(c}W_{b)d)}''(w)-\epsilon_{(a(c}T W_{b)d)}(w)}{(z-w)}\right)\\
		& - \left(\frac{\epsilon_{(a(c}\Lambda_{b)}\Lambda_{d)}(w)}{(z-w)^2}+\frac{\epsilon_{(a(c}\Lambda_{b)}\Lambda_{d)}'(w)}{(z-w)}\right)~,\\
		W_{ab}(z) &\Lambda_c(w)\sim\frac{4\epsilon_{(ac}\Lambda_{b)}(w)}{(z-w)^3}+\frac{3\epsilon_{(ac}\Lambda_{b)}'(w)}{(z-w)^2}+\frac{\frac{3}{2}\epsilon_{(ac}\Lambda_{b)}''(w)-\epsilon_{(ac}T\Lambda_{b)}(w)}{(z-w)}~.
	\end{split}
	\end{equation}

    To show that this VOA is $C_2$-cofinite, we compute all the null states in the VOA up to dimension $h=7$, which provides us with enough relations in $\mathcal{R}_\mathcal{V}$ to show that all strong generators are nilpotent. These relations are collected in Table \ref{tab:relusp4}. 
    \begin{table}[t]
    \renewcommand{\arraystretch}{1.5}
		\centering
		\begin{tabular}{|c|c|}
			\hline
			$h$ & Relations in $\mathcal{R}_\mathcal{V}$\\
			\hline
			$5$ & $\left(W_{ab}\Lambda_c\right)|_\mathbf{4}$\\
			$6$ & $ W_{++}^2,~
			W_{--}^2,~
			W_{+-}W_{++},~
			W_{+-}W_{--},~ 
			2W_{+-}^2+W_{++}W_{--},~
			\frac{5}{6}T^3+4T\Lambda_+\Lambda_-+8W_{+-}^2 $\\
			$7$ & $W_{ab}\Lambda_+\Lambda_-,~T^2W_{--},~T^2W_{++},~TW_{--}\Lambda_-,~TW_{++}\Lambda_+$\\
			\hline
		\end{tabular}
		\caption{Relations in $\mathcal{R}_\mathcal{V}$ corresponding to $\USp(4)$ with half-hypermultiplets in $\mathbf{16}$. These polynomial expressions vanish in $\mathcal{R}_\mathcal{V}$. The relations are shown up to $h=7$, obtained from computing the null states in VOA. Here $(\dots)|_{\mathbf{4}}$ denotes projection to the quartet of $\mathfrak{sl}_2$.}
		\label{tab:relusp4}
	\end{table}
	Nilpotency of $\Lambda_a$ is automatic, as they are fermionic generators. Next, $W_{++},W_{--}$ are nilpotent directly because of the first two relations at $h=6$, $W_{++}^2=0$ and $W_{--}^2=0$. Finally, nilpotency of $W_{+-}$ and $T$ can be shown using the existing relations in $\mathcal{R}_\mathcal{V}$. Specifically, from the relation $2W_{+-}^2+W_{++}W_{--}=0$, we obtain
    \begin{equation}
        W^4_{+-}\sim W_{++}^2W_{--}^2 =0 ~.
    \end{equation}
    Since both $W_{++}^2$ and $W_{--}^2$ vanish in $\mathcal{R_V}$, $W_{+-}$ is nilpotent. Similarly, from $\frac{5}{6}T^3+4T\Lambda_+\Lambda_-+8W_{+-}^2=0$, we obtain
    \begin{equation}
        T^6\sim 4T^2W_{+-}(W_{+-}\Lambda_-\Lambda_+)+4W_{+-}^4 =0~.
    \end{equation}
    Since $W_{+-}\Lambda_-\Lambda_+$ and $W_{+-}^4$ vanish in $\mathcal{R_V}$, $T$ is also nilpotent. We have thus shown that the associated VOA is $C_2$-cofinite, confirming the prediction of the Higgs branch conjecture for this theory.

\subsubsection*{Modularity and non-rationality}

We now proceed to show that the VOA associated with this theory is non-rational by identifying a pseudocharacter  in the $S$-transformation of the vacuum character.

    The Schur index (normalized by $q^{-c_{\rm 2d}/24}$) is annihilated by the following order 7 modular linear differential operator,
	\begin{equation}
		\begin{split}
			\mathcal{D}_{\mathrm{\USp(4)}}=&D^{(7)}_q-\frac{2344}{5}\mathbb{E}_4D^{(5)}_q-\frac{30072}{5}\mathbb{E}_6D^{(4)}_q+34000\mathbb{E}^2_4D^{(3)}_q-17248\mathbb{E}_4\mathbb{E}_6D^{(2)}_q\\&-(232848\mathbb{E}_6^2+126720\mathbb{E}_4^3)D^{(1)}_q-1552320\mathbb{E}^2_4\mathbb{E}_6~.
		\end{split}
	\end{equation}
As usual, we have arrived at this  differential operator by making an ansatz of increasing order and requiring that it annihilates the $q$ series expansion (obtained from the matrix integral) up to a tractable power of $q$. We have then checked that the vacuum solution of the MLDE retains integral coefficients to much higher order. The 7 independent solutions to the MLDE are collected in Table~\ref{tab:usp4MLDEsolsclosedform}, where we use $\chi_h$ to denote a non-logarithmic solution with weight $h$ and $\pchi^{L^n}_h$ to denote a logarithmic solution that has weight $h$ and is proportional to $(\log q)^n$. 
     
     The presence of logs in the $q$-expansions of two of the non-vacuum solutions
     is a {\it prima facie} indication of non-rationality. In fact, in this concrete case we have at our disposal  closed form expressions for the 6 non-spurious solutions\footnote{The closed form expression for the Schur index was suggested to us by Shlomo S. Razamat and was also used in \cite{ArabiArdehali:2023bpq}.} and can perform explicitly
     their $SL(2, \mathbb{Z})$ transformations.
     \begin{table}[t]
    \centering
    \renewcommand{\arraystretch}{1.5}
    \begin{tabular}{|c|c| c|}
    \hline
    $h$ & & Solution to MLDE \\
    \hline
    $0$ & $\chi_0$ & $\frac{\eta(q^5)-\eta(q)^5}{5\eta(q)}=q^{4/24}\frac{(q^5;q^5)-(q;q)^5}{5(q;q)}$\\
    $-\frac{4}{5}$& $\chi_{-4/5}$ & $q^{11/30} H(q)^2=q^{11/30}\left(\frac{1}{(q^2;q^5)(q^3;q^5)}\right)^2=q^{11/30}\sum_{n,m=0}^\infty\frac{q^{n^2+m^2+n+m}}{(q;q)_n(q;q)_m}$\\
    $-1$ &  $\chi_{-1}$ & $q^{1/6}G(q)H(q)=\frac{\eta(q^5)}{\eta(q)}=q^{1/6}\frac{(q^5;q^5)}{(q;q)}=q^{1/6}\sum_{n,m=0}^\infty\frac{q^{n^2+n+m^2}}{(q;q)_n(q;q)_m}$\\
     $-\frac{6}{5}$& $\chi_{-6/5}$  & $q^{-1/30}G(q)^2=q^{-1/30}\left(\frac{1}{(q;q^5)(q^4;q^5)}\right)^2=q^{-1/30}\sum_{n,m=0}^\infty\frac{q^{n^2+m^2}}{(q;q)_n(q;q)_m}$\\
    $-1$ &  $\pchi^L_{-1}$ & $\frac{1}{2\pi i}\log (q)\eta(q)^4=q^{1/6}\tau(q;q)^4$\\
    $-1$ &$\pchi^{L^2}_{-1}$ & $\frac{-1}{4\pi^2}\log (q)^2\eta(q)^4=q^{1/6}\tau^2(q;q)^4$\\
    \hline
    $\frac{1}{3}$ &  & $1-\frac{306 q}{6517}-\frac{146013786 q^2}{81169235}+\frac{257136354564 q^3}{117621129535}+\frac{220359482784459 q^4}{118410329371880}+O[q]^6$\\
    \hline
    \end{tabular}
    \caption{Closed form solution of the MLDE for $\USp(4)$ with half-hypermultiplets in $\mathbf{16}$. The last solution corresponding to $h=\frac{1}{3}$ is a spurious solution. $H(q)$ and $G(q)$ are the Rogers-Ramanujan functions.}
    \label{tab:usp4MLDEsolsclosedform}
	\end{table}
    In the basis $\{\chi_0,\chi_{-4/5},\chi_{-1},\chi_{-6/5},\pchi^{L}_{-1},\pchi^{L^2}_{-1}\}$, the full $S$ and $T$ matrices are 
	\begin{equation}
		S=\left(
		\begin{array}{cccccc}
			0 & -\frac{1}{5 \sqrt{5}} & -\frac{1}{5 \sqrt{5}} & \frac{1}{5 \sqrt{5}} & 0 & \frac{1}{5} \\
			0 & \frac{1}{10} \left(\sqrt{5}+5\right) & -\frac{2}{\sqrt{5}} & \frac{1}{10} \left(5-\sqrt{5}\right) & 0 & 0 \\
			0 & -\frac{1}{\sqrt{5}} & -\frac{1}{\sqrt{5}} & \frac{1}{\sqrt{5}} & 0 & 0 \\
			0 & \frac{1}{10} \left(5-\sqrt{5}\right) & \frac{2}{\sqrt{5}} & \frac{1}{10} \left(\sqrt{5}+5\right) & 0 & 0 \\
			0 & 0 & 0 & 0 & 1 & 0 \\
			5 & 0 & -1 & 0 & 0 & 0 \\
		\end{array}
		\right)~.
	\end{equation}
    
	\begin{equation}
		T= \left(
		\begin{array}{cccccc}
			e^{\frac{i \pi }{3}} & 0 & 0 & 0 & 0 & 0 \\
			0 & e^{\frac{11 i \pi }{15}} & 0 & 0 & 0 & 0 \\
			0 & 0 & e^{\frac{i \pi }{3}} & 0 & 0 & 0 \\
			0 & 0 & 0 & e^{-\frac{1}{15} (i \pi )} & 0 & 0 \\
			-5 e^{\frac{i \pi }{3}} & 0 & e^{\frac{i \pi }{3}} & 0 & e^{\frac{i \pi }{3}} & 0 \\
			-5 e^{\frac{i \pi }{3}} & 0 & e^{\frac{i \pi }{3}} & 0 & 2 e^{\frac{i \pi }{3}} & e^{\frac{i \pi }{3}} \\
		\end{array}
		\right)~;
	\end{equation}
    In particular, $S(\chi_0)=-\frac{1}{5\sqrt{5}}(\chi_{-4/5}+\chi_{-1}-\chi_{-6/5})+\frac{1}{5}\pchi_{-1}^{L^2}$,  so one of the logarithmic solution appears in the $S$-transformation of $\chi_0$. This provides a conclusive argument that the associated VOA is non-rational. 
Note also, in agreement with the general structure of pseudocharacters described in \cite{Miyamoto:2002ar},  that the logarithmic solutions can be written as a linear combination of characters of simple modules with coefficient an order-$n$ monomial in $\log q$, $\pchi_{-1}^{L^n}=(\frac{1}{2\pi i}\log q)^n\eta(q)^4=(\frac{1}{2\pi i}\log q)^n(-5\chi_{0}+\chi_{-1})$.

The high-temperature asymptotics of this theory have been discussed at length in \cite{ArabiArdehali:2023bpq} from  the viewpoint of the matrix integral of the Schur index. Let us briefly review that analysis.
One parametrizes $\mathbf{x}$ in equation \eqref{eq:rains-theta} as $\mathbf{x}=(x_1,x_2)$. The positive roots for  $\USp(4)$ and the positive weights for $\mathbf{16}$ representation are
\begin{equation}
    \begin{split}
    &\alpha_+= \{(2, 0),(0, 2),(1, 1),(1, -1)\}\\
    &\rho_+=\{(2, 1),(2, -1),(1, 2),(1, -2),(1, 0),(0, 1)\}~.
    \end{split}
\end{equation}    
$L(\mathbf{x})$ has a minimum with $L_*=-\frac{1}{5}$, which corresponds to $h_{\mathrm{min}}=-\frac{6}{5}$. Therefore the leading term in the high-temperature expansion is independent of $\log \tilde{q}$ and comes from $\chi_{-6/5}$. The next-to-leading asymptotics is given by a degenerate extremum of the higher Rains function $L(x_1,x_2)+\left\{x_2\right\}$. It has a two-dimensional critical surface on which $L(x_1,x_2)+\left\{x_2\right\}=0$, corresponding to weight $h_*=-1$.  Therefore, it contributes a term of the form $\tilde{q}^{h_*-\frac{c_{\rm 2d}}{24}}\cdot P_2(\log \tilde{q})$ to the high-temperature expansion, where $P_2(x)$ is a degree-two polynomial in $x$. This contribution comes from $\pchi_{-1}^{L^2}$. This is consistent with the $S$-transformation of the closed form of the Schur index.

\subsubsection{$\mathrm{SU}(3)\times \mathrm{SU}(2)$ with half-hypers in  $\mathbf{8}\times\mathbf{2}$}\label{sec:su3su282}

The second interacting theory that we discuss is $\SU(3)\times \SU(2)$ gauge theory with half-hypers in the $\mathbf{8}\times\mathbf{2}$. This theory does not have any flavor symmetry and has $n_{\text{h}}=8,n_{\text{v}}=11$, such that $n_{\text{h}}-n_{\text{v}}=-3<0$. As in the previous example, we will show that this theory is Higgsless and  that its associated VOA is strongly finite and non-rational.

\subsubsection*{Absence of Higgs branch}

There is a quick group-theoretic argument that strongly constrains the Higgs chiral ring of this theory. Let us consider the half-hypermultiplets $q_m$, in the representation $\mathbf{8}\times \mathbf{2}$ of $\mathrm{SU}(3)\times \mathrm{SU}(2)$. Each $q_m$ is a ${\mathrm{3\times3}}$ traceless matrix and $m$ is the fundamental index for $\SU(2)$. Then, the F-term equations for the two simple gauge groups are
\begin{align}
    \mathrm{Tr}(q_{(m}q_{n)})=0,\; \; \SU(2) \text{ F-term equations}~,\\
    \epsilon^{mn}[q_m,q_n]=0,\; \; \SU(3) \text{ F-term equations}~.
\end{align}
The $\SU(3)$ F-term states that $q_1$ and $q_2$ commute. Any gauge invariant is a polynomial in the traces
\begin{align}
    T_{m_1\cdots m_k}\;:=\;\mathrm{Tr}(q_{m_1} q_{m_2}\ldots q_{m_k})~,
\end{align}
with all $\SU(2)$ indices contracted by $\epsilon$ tensors, possibly across different trace factors. Whenever the two indices of an $\epsilon$ land in the {\it same} trace factor, commutativity and cyclicity allow us to bring the corresponding $q$’s next to each other, and $\epsilon^{ab}q_aq_b=\tfrac{1}{2}\epsilon^{ab}[q_a,q_b]=0$. Since the $q_a$ commute, each tensor $T_{a_1\cdots a_k}$ is totally symmetric; moreover $T_{ab}=0$, its symmetric part being the $\SU(2)$ F-term and its antisymmetric part the trace of a commutator. The remaining candidates are cross-contractions of the higher symmetric tensors $T_{a_1\cdots a_k}$, $k\geq 3$. Since the $q_a$ commute, they can generically be diagonalized simultaneously, with eigenvalues $\lambda^{(i)}_a$, $i=1,2,3$. The constraints $\sum_i \lambda^{(i)}_a=0$ and $T_{ab}=\sum_i\lambda^{(i)}_a\lambda^{(i)}_b=0$ force the two eigenvalue vectors to be proportional, $\lambda^{(i)}_2=c\,\lambda^{(i)}_1$: in the two-dimensional space of traceless eigenvalue vectors there are only two independent solutions of $\lambda\cdot\lambda=0$, and two such vectors that are also orthogonal to each other must be parallel. Hence on the F-term locus $q_2=c\,q_1$, every trace factorizes as $T_{a_1\cdots a_k}=\mathrm{Tr}(q_1^{\,k})\,v_{a_1}\cdots v_{a_k}$ with $v=(1,c)$, and every $\epsilon$ contraction produces a factor $\epsilon^{ab}v_av_b=0$. All gauge invariants therefore vanish on the F-term locus. The ring-level statement is confirmed by a direct computation. Using \textsf{Macaulay2} to obtain the Hilbert series for $\mathbb{C}[q_a]/\langle \text{F-terms} \rangle$ and projecting onto gauge invariant operators, we find that the Higgs branch Hilbert series is $\mathcal{I}_{\mathrm{Higgs}}=1$. We also computed the HL index and it truncates as expected. Specifically, it takes the following form
\begin{equation}
    \begin{split}
    \mathcal{I}_{\mathrm{HL}}(t)=&\oint \frac{dx_1 dx_2}{2\pi i x_12\pi i x_2} \Delta_{\SU(3)}(x_1,x_2)\oint \frac{dy}{2\pi i y} \Delta_{\SU(2)}(y)\,\frac{\prod_{\alpha_1}(1-\mathbf{x}^{\alpha_1}\ t)\prod_{\alpha_2}(1-\mathbf{y}^{\alpha_2}\ t)}{\prod_{\rho}(1- \tilde{\mathbf{x}}^\rho\ t^{1/2})}~,
    \end{split}
\end{equation}
where $\Delta_{\SU(2)},\Delta_{\SU(3)}$ are the respective Haar measure, $\alpha_1,\alpha_2$ are the roots of $\SU(3)$ and $\SU(2)$, $\rho$ are the weights of $\mathbf{8}\times \mathbf{2}$ of $\SU(3)\times \SU(2)$, and $\tilde{\mathbf{x}}=(x_1,x_2,y)$, $\mathbf{x}=(x_1,x_2)$. Carrying out the contour integral yields the full HL index,
\begin{equation}\label{eq:su2su3_hl}
    \mathcal{I}_{\text{HL}}(t)=1+2t^3=\frac{1-3 t ^{6}+2 t ^{9}}{\left(1-t ^3\right)^2}~.
\end{equation}

\subsubsection*{Strong generators}

The denominator  of the HL index indicates the presence of two dimension-three bosonic HL chiral ring generators. To identify the other strong generators, we compute the first few terms of the $q$ expansions of the Schur and Macdonald indices,
\begin{equation}
\small
    \begin{split}
        \mathcal{I}_{\text{Schur}}(q)&=1+q^2+8 q^3+9 q^4+16 q^5+26 q^6+54 q^7+99 q^8+150 q^9+252 q^{10}+O(q^{11})\\
        &=\mathrm{PE}\left[\frac{q^2+7 q^3-27 q^6-7 q^7-q^8+91 q^9+125 q^{10}+O\left(q^{11}\right)}{1-q}\right]~,\\
        \mathcal{I}_{\text{Mac}}(q,T)=&1+q^2 T+q^3 \left(3 T+3 T^2+2 T^3\right)+q^4 \left(3 T+4 T^2+2 T^3\right)\\&+q^5 \left(3 T+6 T^2+5 T^3+2 T^4\right)+q^6
        \left(3 T+9 T^2+9 T^3+5 T^4\right)+O(q^7)\\
        =&\mathrm{PE}\left[\frac{T q^2 + (2 T + 3 T^2 + 2 T^3) q^3 + (-3 T^2 - 6 T^3 - 9 T^4 - 6 T^5 - 
            3 T^6) q^6+O(q^7)}{1-q}\right]~.
    \end{split}
\end{equation}
The Macdonald index suggests that there is only one bosonic generator of the VOA at $h=2$, which is the stress tensor. At $h=3$, there are {\it two} sets of bosonic operators with $R=2$ and $r=\pm 1,0$ and an additional bosonic operator with $R=2,r=0$.

Our ansatz is then that  the associated VOA is strongly generated by the stress tensor $T$, two dimension-three bosonic $\mathfrak{sl}(2)$-triplets $U_{i=+,0,-},V^{i=+,0,-}$, and a dimension-three bosonic $\mathfrak{sl}(2)$-singlet $W$. 

It is illuminating to view this theory as obtained by gauging an $\SU(3)$ subgroup of the $\SO(8)$ flavor symmetry of $\SU(2)$ SQCD with $N_f=4$. The $\SU(3)$ is embedded such that the adjoint of $\SO(8)$ has the following branching
\begin{equation}    
\mathbf{28}\to\mathbf{8}\oplus\mathbf{10}\oplus\overline{\mathbf{10}}~.
\end{equation}
The embedding index is $I_{\SU(3)\hookrightarrow\SO(8)}=3$, so that the induced level matches the conformal gauging condition $k_{\text{2d}}=3\times(-2)=-6=-2h^\vee$. From the VOA perspective this corresponds to applying the BRST prescription to the affine Kac-Moody VOA $\hat{\mathfrak{so}}(8)_{-2}$ (the VOA of $\SU(2)$ $N_f=4$ SQCD), gauging the $\hat{\mathfrak{su}}(3)_{-6}$ subalgebra.
In this presentation the strong generators take the schematic forms collected in Table \ref{tab:generators_su3su2_so8}. There, $J_{\mathbf{8}}$, $J_{\mathbf{10}}$ and $J_{\overline{\mathbf{10}}}$ denote the components of the $\mathfrak{so}(8)$ currents in the branching above ($\SU(3)$ indices are suppressed). 
$\lambda^a$ denotes the symplectic fermions in the adjoint of $\SU(3)$, with $\lambda^+:=\lambda~,\lambda^-:=\tilde{\lambda}$ to make the $\mathfrak{sl}_2$ covariance manifest.
In these notations, the generators take the schematic form,
 \begin{align}  W&\simeq J_{\mathbf{8}}\lambda^a\lambda^b\epsilon_{ab}~,\\    U^i&\simeq J_{\mathbf{10}}\lambda^a\lambda^b\sigma_{ab}^i~,\\ V_i&\simeq J_{\mathbf{\overline{10}}}\lambda^a\lambda^b\sigma_{ab}^i~.
\end{align}
The $\mathfrak{sl}_2$ automorphism is again guaranteed by Proposition 4.6 of \cite{Beem:2025guj}, as recalled at the beginning of this section; here one applies it to the $\mathfrak{su}(3)$ gauging, the graded-unitary structure of the symplectic bosons having descended to $\hat{\mathfrak{so}}(8)_{-2}$ along the $\mathfrak{su}(2)$ gauging by Theorem 4.1 of the same paper. Note that in this presentation the triplet structure is carried entirely by the $\SU(3)$ symplectic fermions; the $\SU(2)$ gauginos do not appear in the generators. The $\SU(3)$ tensor products are consistent with this description and with the OPEs found below in equations \eqref{eq:OPEsu3su2}: $\mathbf{10}\otimes\mathbf{10}\supset\overline{\mathbf{10}}$, $\overline{\mathbf{10}}\otimes\overline{\mathbf{10}}\supset\mathbf{10}$ and $\mathbf{10}\otimes\overline{\mathbf{10}}\supset \mathbf{1}+\mathbf{8}$, in agreement with the $U_iU_j\sim V^k$, $V^iV^j\sim U_k$ and $U_iV^j\sim\delta_i^j(1+W+\cdots)$ structure of the OPE algebra.

\begin{table}[t]
\centering
\renewcommand{\arraystretch}{1.5}
\centering
    \begin{tabular}{|c|c|c|c|c|c|c|}
    \hline
    Name & Multiplet &  \makecell{Index \\contribution} & \makecell{Schematic form \\ of operators} & $h$ & $SU(2)_R$ &$U(1)_r$\\
    \hline
    $T$ & $\hat{\mathcal{C}}_{0(0,0)}$ &  $\frac{Tq^2}{1-q}$ & $T_{\mathrm{sug}}+\lambda^a\lambda^b\epsilon_{ab}$ & $2$ & $1$ &$0$\\
    $U_+$,$V^+$ & $2\times\mathcal{D}_{\frac{3}{2}(0,\frac{1}{2})}$ & $\frac{2T^3q^3}{1-q}$ &  $J_{\mathbf{10}}\tilde{\lambda}\tilde{\lambda},\ J_{\overline{\mathbf{10}}}\tilde{\lambda}\tilde{\lambda}$ & $3$ & $2$ &$1$\\
    $U_0$,$V^0$,$W$ & $3\times\hat{\mathcal{C}}_{1(0,0)}$ & $\frac{3T^2q^3}{1-q}$ & $J_{\mathbf{10}}\lambda\tilde{\lambda},\ J_{\overline{\mathbf{10}}}\lambda\tilde{\lambda},\ J_{\mathbf{8}}\lambda\tilde{\lambda}$ & $3$ & $2$ &$0$\\
    $U_-$,$V^-$ & $2\times\bar{\mathcal{D}}_{\frac{3}{2}(\frac{1}{2},0)}$ & $\frac{2Tq^3}{1-q}$ & $J_{\mathbf{10}}\lambda \lambda,\ J_{\overline{\mathbf{10}}}\lambda \lambda$ & $3$ & $2$ &$-1$\\
    \hline
\end{tabular}
\caption{
VOA generators for gauging $\mathfrak{su}(3)\subset \mathfrak{so}(8)_{-2}$.  The first two columns list the name and the four-dimensional multiplet associated with each generator with its multiplicity. The third column lists the contribution of each multiplet to the Macdonald index. The fourth column lists the schematic form of each generator in terms of the elementary fields in the Lagrangian theory. $\lambda^a$ is the symplectic fermion from the $\SU(3)$ free vector multiplet where $a$ is the $\mathfrak{sl}_2$ index. The $\mathfrak{so}(8)$ current $J_{\mathbf{28}}$ decomposes as $J_{\mathbf{28}}=J_{\mathbf{8}}+J_{\mathbf{10}}+J_{\overline{\mathbf{{10}}}}$ of $\mathfrak{su}(3)$. $T_{\mathrm{sug}}\simeq\left(J_{\mathbf{8}}J_{\mathbf{8}}\right)|_{\mathbf{1}}$ is the Sugawara stress tensor for $\mathfrak{so}(8)_{-2}$. The HL chiral ring generators are $U_+$ and $V^+$ (they are captured by the denominator in \eqref{eq:su2su3_hl}), while the HL anti-chiral ring generators are $U_-$ and $V^-$.}
\label{tab:generators_su3su2_so8}
\end{table}

\subsubsection*{Associated strongly finite VOA}

The VOA has central charge $c_{\rm 2d}=-30$ and the OPEs of the stress tensor with itself and other operators take the canonical form. 
Writing the most general ansatz and imposing Jacobi identities, we find that the
OPEs among the rest of the strong generators take the form
\begin{equation}
\label{eq:OPEsu3su2}
\footnotesize
\begin{split}
    U_i(z) U_j(w)&\sim  \epsilon_{ijk}  \left(\frac{2V^k(w)}{(z-w)^3}+\frac{(V^k)'(w)}{(z-w)^2}+\frac{\frac{1}{2}(V^k)''(w)-T V^k(w)}{(z-w)}\right)~,\\
    V^i(z) V^j(w)&\sim \epsilon^{ijk}  \left(\frac{-2U_k(w)}{(z-w)^3}+\frac{-U_k'(w)}{(z-w)^2}+\frac{-\frac{1}{2}U_k''(w)+T U_k(w)}{(z-w)}\right)~,\\
    U_i(z) V^j(w)&\sim \delta_{i}^{j} \left(\frac{10}{(z-w)^6}+\frac{-2T(w)}{(z-w)^4}+\frac{-T'(w)}{(z-w)^3}+\frac{-\frac{3}{8}T''(w)+\frac{1}{4}T^2(w)}{(z-w)^2}+\frac{\frac{1}{4}T'T(w)-\frac{1}{12}T'''(w)}{(z-w)}\right)\\&+\delta^j_i\left(\frac{W(w)}{(z-w)^3}+\frac{\frac{1}{2}W'(w)}{(z-w)^2}+\frac{\frac{1}{4}W''(w)-\frac{1}{2}T W(w)}{(z-w)}\right)~, \\
    W(z) U_j(w)&\sim   \left(\frac{2U_j(w)}{(z-w)^3}+\frac{U_j'(w)}{(z-w)^2}+\frac{\frac{1}{2}U_j''(w)-T U_j(w)}{(z-w)}\right)~,\\
    W(z) V^j(w)&\sim   \left(\frac{-2V^j(w)}{(z-w)^3}+\frac{-(V^j)'(w)}{(z-w)^2}+\frac{-\frac{1}{2}(V^j)''(w)+T V^j(w)}{(z-w)}\right)~,\\
    W(z)W(w)&\sim \left(\frac{20}{(z-w)^6}+\frac{-4T(w)}{(z-w)^4}+\frac{-2T'(w)}{(z-w)^3}+\frac{-\frac{3}{4}T''(w)+\frac{1}{2}T^2(w)}{(z-w)^2}+\frac{\frac{1}{2}T'T(w)-\frac{1}{6}T'''(w)}{(z-w)}\right)~.
\end{split}
\end{equation}
The null states needed for the Jacobi identities to hold are collected in Appendix \ref{app:nulls}. Note that    $T$ and $W$ form a $W_3$ subalgebra.~\cite{Zamolodchikov:1985wn,Bouwknegt:1992wg, Blumenhagen:1990jv}\footnote{Our normalization for the $W$ generator differs from the conventional one: $W_{\rm here} = i \sqrt{2} \, W_{\rm standard}$.} We also observe a $\mathbb{Z}_3$ automorphism under which $T$ and $W$ are neutral and $V^i \to \omega V^i$,  $U_i \to \omega^2 U_i$, with $\omega$ a cube root of unity.

The null states in the VOA lead to relations in $\mathcal{R}_\mathcal{V}$, which allows one to show that this VOA is $C_2$-cofinite. In Table \ref{tab:relsu3su2}, we enumerated all such relations up to dimension $h=8$. $U_i,V^i$ are nilpotent directly because of the first two relations at $h=6$, $U_i U_j=0$ and $V^i V^j=0$. The relation $2U_0 V^0-W^2=0$ at $h=6$ implies that $W$ is nilpotent, $W^4= 4U_0^2(V^0)^2=0$. Similarly, the relation $T^4+8TU_+V^+=0$ at $h=8$ implies that $T$ is nilpotent, $T^8=64 (TU_+^2)(T(V^+)^2)=0$. Thus, the associated VOA is $C_2$-cofinite.\footnote{Note that the $W_3$ subalgebra  is by itself {\it not} $C_2$-cofinite. See~\cite{Beem:2026lkq} for a discussion of 
how graded unitarity~\cite{ArabiArdehali:2025fad} restricts the allowed values of the central charge for 
$W_3$ algebras that arise in the SCFT/VOA correspondence, under the assumption that the {\it whole} VOA is $W_3$. }
\begin{table}[t]
    \centering
    \renewcommand{\arraystretch}{1.5}
    \begin{tabular}{|c|c|}
        \hline
        $h$ & Relations in $\mathcal{R}_\mathcal{V}$ \\
        \hline
        $6$ & $ U_i U_j,~
        V^i V^j,~
        U_i V^{j\neq i},~
        W U_i,~
        W V^i,~
        2U_i V^i-W^2$\\
        $7$ & $T^2 U_i,~
        T^2 V^i ,~
        T^2 W$ \\
        $8$ & $T^4+8TU_iV^i,\  T^4+8T W^2,\ TU_iU_j,\ TV^iV^j,\ TWU_i,\ TWV^i,\ TU_iV^{j\neq i}$ \\
        \hline
    \end{tabular}
    \caption{Relations in $\mathcal{R}_\mathcal{V}$ corresponding to $\SU(3)\times \SU(2)$ with half-hypermultiplets in $\mathbf{8\times 2}$. These polynomial expressions vanish in $\mathcal{R}_\mathcal{V}$. The relations are shown up to $h=8$ obtained by computing the null states in the VOA. Repeated $i$ indices are {\it not} summed: e.g.\ $2U_iV^i-W^2$ is a relation for each $i$ separately.}
    \label{tab:relsu3su2}
\end{table}

\subsubsection*{Modularity and non-rationality}	

In this subsection, we discuss the high-temperature asymptotics of the Schur index or the vacuum character of the associated VOA and show that the VOA is non-rational. As in Section \ref{sec:usp416b2}, this example admits a closed-form expression for the Schur index\footnote{The closed form expressions for this example were worked out with Harshal Kulkarni.}, allowing us to write modular transformation matrices explicitly and then compare the result with the asymptotics obtained from the Rains function analysis.

We first discuss the MLDE approach. We find that the Schur index of this theory solves the following MLDE of order 7,
\begin{equation}
    \begin{split}
        \mathcal{D}_{\text{\SU(3)}\times\text{\SU(2)}}=&D^{(7)}_q-385\mathbb{E}_4D^{(5)}_q-6020\mathbb{E}_6D^{(4)}_q+16275\mathbb{E}^2_4D^{(3)}_q-207900\mathbb{E}_4\mathbb{E}_6D^{(2)}_q\\&-(1440600\mathbb{E}_6^2+592875\mathbb{E}_4^3)D^{(1)}_q-8505000\mathbb{E}^2_4\mathbb{E}_6~.
    \end{split}
\end{equation}
The solutions to the MLDE are collected in Table \ref{tab:su3su2MLDEsols} with their corresponding weights. There is also a spurious solution with non-integer coefficient and weight $h=\frac{1}{12}$. The logarithmic solution is a linear combination of characters of simple modules with coefficient as a third-order polynomial in $\log q$, $\pchi_{-1}^{L^n}=(\frac{1}{2\pi i}\log q)^n(-6\chi_{0}+\chi_{-1})$. Since the closed form expressions are known, we can perform the $S$ and $T$ transformations explicitly and check that the $S$-transformation of the vacuum character contains the logarithmic solution $\pchi^{L^3}_{-1}$. The full $S$ and $T$ matrices in the basis $\{\chi_0,\chi_{-1},\chi_{-4/3},\pchi^L_{-1},\pchi^{L^2}_{-1},\pchi^{L^3}_{-1}\}$ are

\begin{equation}
 S=   \left(
\begin{array}{cccccc}
-\frac{1}{3 \sqrt{3}} & -\frac{1}{9 \sqrt{3}} & \frac{1}{27      \sqrt{3}} & 0 & 0 & -\frac{i}{9} \\
 -\frac{2}{\sqrt{3}} & -\frac{2}{3 \sqrt{3}} & \frac{2}{9 \sqrt{3}} & 0 & 0 & \frac{i}{3} \\
 6 \sqrt{3} & 2 \sqrt{3} & \frac{1}{\sqrt{3}} & 0 & 0 & 0 \\
 0 & 0 & 0 & 0 & -i & 0 \\
     0 & 0 & 0 & i & 0 & 0 \\
 6 i & -i & 0 & 0 & 0 & 0 \\
    \end{array}
    \right)
\end{equation}

\begin{equation}
    T=\left(
\begin{array}{cccccc}
 e^{\frac{\pi  i}{2}} & 0 & 0 & 0 & 0 & 0 \\
 0 & e^{\frac{\pi  i}{2}} & 0 & 0 & 0 & 0 \\
 0 & 0 & e^{\frac{1}{6} (-\pi ) i} & 0 & 0 & 0 \\
 -6 e^{\frac{\pi  i}{2}} & e^{\frac{\pi  i}{2}} & 0 & e^{\frac{\pi  i}{2}} & 0 & 0 \\
 -6 e^{\frac{\pi  i}{2}} & e^{\frac{\pi  i}{2}} & 0 & 2 e^{\frac{\pi  i}{2}} & e^{\frac{\pi  i}{2}} & 0 \\
 -6 e^{\frac{\pi  i}{2}} & e^{\frac{\pi  i}{2}} & 0 & 3 e^{\frac{\pi  i}{2}} & 3 e^{\frac{\pi  i}{2}} & e^{\frac{\pi  i}{2}} \\
\end{array}
\right)
\end{equation}
Since $S(\chi_0)=-\frac{1}{27\sqrt{3}}(9\chi_{0}+3\chi_{-1}-\chi_{-4/3})-\frac{i}{9}\pchi_{-1}^{L^3}$, the leading character has $h=-\frac{4}{3}$ and the logarithmic solution $\pchi_{-1}^{L^3}$ appears in the $S$-transformation of $\chi_0$. Hence, the associated VOA is non-rational.

Let us reproduce the asymptotics from the Rains functions approach. For this theory, we parametrize $\mathbf{x}$ in equation \eqref{eq:rains-theta}  as $\mathbf{x}=(x_1,x_2,y)$. The set of positive roots of $\SU(3)\times\SU(2)$ and the positive weights of $\mathbf{8}\times\mathbf{2}$ representation are 
\begin{equation}
\begin{split}
    &\alpha_+\in \{(0, 0, 2), (1, -1, 0), (2, 1, 0), (1, 2, 0)\},\\
    &\rho_+\in \{(1, -1, 1), (2, 1, 1), (1, 2, 1), (0, 0, 1), (1, -1, -1), (2, 1, -1), (1, 2, -1), (0, 0, -1)\}.
\end{split}
\end{equation}
By minimizing numerically, we find that the Rains function has minima at the critical points $\mathbf{x}=(0,\pm\frac{1}{3},\pm\frac{1}{3})$, with minimum value $-\frac{1}{3}$, corresponding to the smallest weight $h_{*}=-\frac{4}{3}$. Since the critical set is zero-dimensional, the leading term in the high-temperature expansion is free of $\log\tilde{q}$; it is accounted for by the character $\chi_{-4/3}$. The next-to-leading term has weight $h_*=-1$. We find a two-dimensional critical surface of the ordinary Rains function with $L_*=0$, and a three-dimensional critical surface of the higher Rains function $L(\mathbf{x})+\left\{y\right\}$ on which $L(\mathbf{x})+\left\{y\right\}=0$. Both contribute the same exponent, so the subleading term takes the form $\tilde{q}^{h_*-\frac{c_{\rm 2d}}{24}}\cdot P_3(\log \tilde{q})$, with the degree of the polynomial determined by the largest critical dimension. This matches the $S$-transformation of the closed form, in which the leading character has $h=-\frac{4}{3}$ and the subleading terms include the $(\log \tilde{q})^3$ contribution of $\pchi_{-1}^{L^3}$.

  \begin{table}[t]
    \centering
    \renewcommand{\arraystretch}{1.5}
    \begin{tabular}{|c|c| c|}
        \hline
        $h$  &  & Solution to MLDE \\
        \hline
        $0$ & $\chi_0$ & $\frac{1}{9}\left(\left(\frac{\eta(q^3)}{\eta(q)}\right)^3-\eta(q)^6\right)$\\
        $-1$ &$\chi_{-1}$ & $\frac{1}{3}\left(2\left(\frac{\eta(q^3)}{\eta(q)}\right)^3+\eta(q)^6\right)$\\
        $-\frac{4}{3}$& $\chi_{-4/3}$ & $3\left(\frac{\eta(q^3)}{\eta(q)}\right)^3+\left(\frac{\eta(q^{1/3})}{\eta(q)}\right)^3$\\
        $-1$ & $\pchi^{L}_{-1}$ &$\frac{1}{2\pi i}\log (q)\eta(q)^6$\\
        $-1$ &$\pchi^{L^2}_{-1}$ & $\frac{-1}{4\pi^2}\log (q)^2 \eta(q)^6$\\
        $-1$ &$\pchi^{L^3}_{-1}$ & $\frac{-1}{8\pi^3 i}\log (q)^3 \eta(q)^6$\\
        \hline
        $\frac{1}{12}$ &  &$1-\frac{1131616 q}{11328125}+\frac{580146787497296 q^2}{870459933203125}+\frac{871303427682414263168 q^3}{110768637879897265625}+O\left(q^{4}\right)$
        \\
        \hline
    \end{tabular}
    
    \caption{Solutions of the MLDE for $\mathrm{SU}(3)\times \mathrm{SU}(2)$ with half-hypermultiplets in $\mathbf{8}\times\mathbf{2}$. The last solution corresponding to $h=\frac{1}{12}$ is a spurious solution.}
    \label{tab:su3su2MLDEsols}
\end{table}

\section{Discussion}
\label{sec:discussion}

Our results invite further work in several directions. The  most immediate is to construct the VOAs associated to the remaining candidates, namely the infinite quiver family  and the third sporadic theorym, and to show that they are strongly finite and logarithmic. We close by highlighting a few questions of a different nature:

\begin{itemize}
\item   The two VOAs constructed in Section~\ref{sec:examples} come with closed-form vacuum characters, explicit $S$ and $T$ matrices, and pseudocharacters. It would be very interesting to determine their full representation categories, such as the simple modules, projective covers, pseudotrace functions, and to test against these new examples the general expectations for log-modular tensor categories and Verlinde-type formulas for strongly finite VOAs~\cite{Creutzig_2017,McRae:2021yyb}. This would directly address the scarcity-of-examples problem that motivated this work.

\item  All known Higgsless SCFTs with {\it rational} VOAs, namely the Argyres-Douglas theories, are isolated, with empty conformal manifold. The Lagrangian Higgsless theories classified here have exactly marginal gauge couplings, and it is precisely the gaugino sector that generates the logarithms. It is tempting to speculate that, within the Higgsless world, non-rationality of $\mathcal{V}[\mathcal{T}]$ is tied to the existence of a conformal manifold. A sharp test is provided by the putative rank-one SCFT $IV^*_{Q=1}$ appearing in the classification of rank-one Seiberg-Witten geometries~\cite{Argyres:2015ffa,Argyres:2015gha,Argyres:2016xmc,Argyres:2016xua,Argyres:2020nrr}: it is isolated and Higgsless, yet its candidate VOAs, the doublet $\mathcal{A}(4)$ and triplet $\mathcal{W}(4)$ algebras, are logarithmic~\cite{Deb:2025cqr}. Settling the existence of this theory would either refute the speculation or remove the putative counterexample.

\item  For SCFTs with a non-trivial Higgs branch, the associated VOAs admit free-field realizations organized by the effective field theory at generic points of the Higgs branch~\cite{Beem:2019tfp,Beem:2019snk,Beem:2021jnm,Beem:2024fom}. That construction is unavailable, by definition, for every theory considered in this paper. Is there an analogous organizing principle for strongly finite VOAs  that would render the Higgsless examples equally computable?

\item  The list of Higgsless SCFTs does not conform to any obvious pattern. It would be interesting to find a unifying construction, such as a string-theory/orientifold or six-dimensional realization of the SO/USp family, that explains both the existence of the infinite sequence and the sparseness of the sporadic cases. Relatedly, the $\USp(2)$ outer automorphism established via~\cite{Beem:2025guj}, which acts very visibly on the small generator sets of our two VOAs, awaits a clear four-dimensional interpretation. The examples constructed here provide a concrete laboratory for this question.
\end{itemize}

\acknowledgments
	We would like to thank Christopher Beem, Federico Bonetti, Thomas Creutzig, Harshal Kulkarni, Shlomo S. Razamat, Brandon C. Rayhaun, Wenbin Yan and Zhenghao Zhong for useful discussions. We are grateful to Arash A. Ardehali for useful discussions on the high temperature asymptotics of the Schur index. We would especially like to thank Martin Ro\v{c}ek and Matteo Sacchi for detailed discussions on various aspects of this project. This work is supported in part by NSF grant PHY-2513893 and by the
	Simons Foundation grant 681267 (Simons Investigator Award).
\appendix

  \section{Superconformal indices}
    \label{app:sci}
   The superconformal index of a 4d $\mathcal{N}=2$ SCFTs is given by \cite{Romelsberger:2005eg,Kinney:2005ej,Dolan:2008qi}
	\begin{equation}
		\mathcal{I}(p,q,t)=\text{Tr}\,(-1)^F \left(\frac{p q}t\right)^{-r}p^{j_2-j_1}\,q^{j_1+j_2}\,t^{R}\,\prod_{i=1}^{\text{rank\,} G_F}u_i^{F_i}~,
	\end{equation}
	where $p,q$ and $t$ are superconformal fugacities and $u_i$ are the flavour fugacities for the flavor group $G_F$. The superconformal index captures protected data of the theory and its various limits count operators satisfying different shortening conditions \cite{Gadde:2011uv}. The limits relevant to this paper are
	
	\textbf{Hall-Littlewood limit:} $q\to 0$ and $p\to 0$,
	\begin{equation}
		\mathcal{I}_{\text{HL}}(t)=\text{Tr}(-1)^Ft^{R+r}~,\;\;\;\;~E-2R-r=0~,~j_1=0~.
	\end{equation}
	
	\textbf{Schur limit:} $t\to q$ and $p$ arbitrary,
	\begin{equation}
		\mathcal{I}_{\text{Schur}}(q)=\text{Tr}(-1)^F q^{E-R}~,\;\;\;\;~ E-2R=j_1+j_2,~r=j_2-j_1.
	\end{equation}
	
	\textbf{Macdonald limit:} $p\to0$ and $q,t$ arbitrary,
	\begin{equation}
		\label{eq:macindex}
		\mathcal{I}_{\text{Mac}}(q,t)=\text{Tr}(-1)^F q^{E-2R-r} t^{R+r}~,\;\;\;~ E+2j_1-2R-r=0.
	\end{equation}
    In the Macdonald limit, it is often convenient to write $t=q T$.
    
	For a Lagrangian theory, the superconformal index can be written as a matrix integral with contributions from the vector multiplet and hypermultiplets. The Macdonald limit of the index takes the form
	\begin{equation}
		\label{eq:integralexpression}
		\mathcal{I}_{\text{Mac}}(q,t)=\oint [d\mathbf{a}]\mathcal{I}_{\text{vec}}(\mathbf{a},t,q)\mathcal{I}_{\text{hyp}}(\mathbf{a},t,q)~,
	\end{equation}
	where $\oint [d\mathbf{a}]$ denotes the integral over gauge fugacities along with the the Haar measure and $\mathcal{I}_{\text{hyp}}$ and $\mathcal{I}_{\text{vec}}$ denote the indices for the vector and hypermultiplet, which can be written as the \textit{plethystic exponential}\footnote{~$
		\text{PE}\left[f(x_1,x_2,\dots,x_n)\right]:=e^{\sum_{k=1}^\infty \frac{1}{k}f(x_1^k,x_2^k,\dots,x_n^k)}$} of single particle indices as follows
	\begin{equation}
		\mathcal{I}_{\text{hyp}}(\mathbf{a},t,q)=\text{PE}\left[\frac{t^{\frac{1}{2}}}{1-q}\chi^{\mathfrak{g}}_{\mathcal{R}}(\mathbf{a})\right]~,~
		\mathcal{I}_{\text{vec}}(\mathbf{a},t,q)=\text{PE}\left[\frac{-t-q}{1-q}\chi^{\mathfrak{g}}_{\text{adj}}(\mathbf{a})\right]~.
	\end{equation}
	$\chi^{\mathfrak{g}}_\mathcal{R}(\mathbf{a})$ denotes the character of representation $\mathcal{R}$ of the half-hypermultiplets in the gauge algebra $\mathfrak{g}$. The Schur and Hall-Littlewood limit can be taken starting from the Macdonald limit, as described above.

\section{High temperature limit of the Schur index}
\label{app:htl}
In this appendix, following the asymptotic analysis of \cite{ArabiArdehali:2023bpq}, we explain how logarithmic terms arise in the asymptotic expansion of the modular transform of the Schur index of a Lagrangian theory. The asymptotic expansion is governed by the ordinary Rains function and by a family of higher Rains functions. When any one of these functions is constant on a positive-dimensional locus that contributes to the asymptotics, we shall call this locus a critical surface. The contribution from a $n$-dimensional critical surface naturally produces a term proportional to $(\log \widetilde q)^n$, provided the corresponding coefficient is non-zero.

Consider a Lagrangian theory with gauge group $G$ and half-hypermultiplets in a pseudoreal representation $\Rrep$.  The Schur index can be written as
\begin{equation}
	\mathcal{I}_{\rm Schur}(q)=
	\frac{(q;q)_{\infty}^{2r_G}}{\theta(q^{1/2};q)^{n_{\rho_0}/2}}
	\int \frac{d^{r_G}x}{|W|}\,
	\frac{\prod_{\alpha_+}\theta(z^{\pm\alpha_+};q)}
	{\prod_{\rho_+}\theta(q^{1/2}z^{\rho_+};q)}~.
\end{equation}
Here $\theta(p;q)\coloneqq (p;q)_\infty(q/p;q)_\infty$, $r_G$ is the rank of $G$, $|W|$ is the order of the Weyl group, $\alpha_+$ are the positive roots, $\rho_+$ are positive weights of $\Rrep$, $n_{\rho_0}$ is the number of zero weights, $x=(x_1,\ldots,x_{r_G})$ are the Cartan holonomy variables, and $z_i=e^{2\pi i x_i}$ are the corresponding gauge fugacities.

After an $S$-transformation, the asymptotic expansion is controlled by the Schur-limit Rains function and by its higher analogues \cite{ArabiArdehali:2015ybk,ArabiArdehali:2023bpq}.  The ordinary Rains function is
\begin{equation}
	\label{eq:rains-theta}
	L(\mathbf{x})
	=
	\frac{1}{2}\sum_{\rho_+}\vartheta(\rho_+\cdot \mathbf{x})
	-\sum_{\alpha_+}\vartheta(\alpha_+\cdot \mathbf{x}) ,
	\qquad
	\vartheta(x)=\{x\}\bigl(1-\{x\}\bigr),
\end{equation}
where $\{x\}=x-\lfloor x\rfloor$.  Locally, in a chamber where no root or weight hyperplane is crossed, this becomes a piecewise-linear function.  Indeed, using the four-dimensional ABJ anomaly cancellation condition \cite{ArabiArdehali:2015ybk}, the quadratic terms cancel and one obtains
\begin{equation}
	\label{eq:rains-piecewise-linear}
	L(\mathbf{x})
	=
	\frac{1}{2}\sum_{\rho_+}\left|\rho_+\cdot\mathbf{x}\right|
	-
	\sum_{\alpha_+}\left|\alpha_+\cdot\mathbf{x}\right| .
\end{equation}
The higher Rains functions are obtained from this expression by adding non-negative integral combinations of the elementary linear terms that arise in the expansion of the theta functions.  Thus they have the schematic form
\begin{equation}
	\label{eq:higher-rains-simple}
	L^{(n)}(\mathbf{x})
	=
	L(\mathbf{x})+
	\sum_j m_j\ell_j(\mathbf{x}),
	\qquad m_j\in\mathbb{Z}_{\geq 0},
\end{equation}
with
\begin{equation}
	\ell_j(\mathbf{x})\in
	\left\{
	\{\alpha_+\cdot\mathbf{x}\},
	1-\{\alpha_+\cdot\mathbf{x}\},
	\{\rho_+\cdot\mathbf{x}\},
	1-\{\rho_+\cdot\mathbf{x}\}
	\right\} .
\end{equation}
The superscript $(n)$ is only a label distinguishing the different choices of the non-negative integers $m_j$. 

We shall call a positive-dimensional locus a critical surface if the relevant Rains function is constant on it and this locus possibly contributes to the corresponding asymptotic sector.  If the critical surface has dimension $d$, the contribution of that sector has the form
\begin{equation}
	\label{eq:log-polynomial-general}
	\exp\!\left[-\frac{2\pi i}{\tau}\bigl(2(a-c)+L_*\bigr)\right]
	P_d(\log \widetilde q),
\end{equation}
where $L_*$ is the value of the relevant Rains function on the critical surface, and $P_d$ is a polynomial of degree at most $d$.  For example, a one-dimensional critical surface can possibly contribute a $\log\widetilde q$ term.  This conclusion assumes, of course, that the coefficient of the logarithm does not vanish.

We now explain why such critical surfaces are expected.
Let us consider first a theory built from full hypermultiplets.  If the hypermultiplet representation is $R\oplus \overline R$, then the factor $1/2$ in equation \eqref{eq:rains-piecewise-linear} is cancelled by the pairing of weights in $R$ and $\overline R$.  Along a one-dimensional subspace defined by
\begin{equation}
x_1=x_2=\ldots =x_m,
\end{equation}
$\frac{1}{2}\sum_{\rho_+}|\rho_+\cdot \mathbf{x}|$ in $L(\mathbf{x})$ can be replaced by $\sum_{\rho_+'}|\rho_+'\cdot \mathbf{x}|$ where $\rho_+'$ are the positive weights of $R$. On this subspace, $L$ is therefore piecewise linear with integer slopes. Since the elementary monomials $\{\alpha_+\cdot \mathbf{x}\}$, $1-\{\alpha_+\cdot \mathbf{x}\}$, $\{\rho_+\cdot \mathbf{x}\}$, $1-\{\rho_+\cdot \mathbf{x}\}$ available in \eqref{eq:higher-rains-simple} are linear with unit slopes of both signs, a suitable choice of the non-negative integers $m_j$ produces a higher Rains function that is exactly constant on a segment of this one-dimensional subspace, indicating a logarithmic term.

For theories built only from half-hypermultiplets which cannot be paired into full hypermultiplets, the same idea requires a slightly different implementation. The argument for full hypermultiplets used the pair $R\oplus \overline R$ to obtain an integer coefficient for $\mathbf{x}$ variables. 
To recover a similar structure, we restrict to a convenient rank-two subgroup, which we take to be $SU(3)\subseteq G$. Under this subgroup, the pseudoreal representation $\mathcal R$ decomposes as
\begin{equation}
	\Rrep_{SU(3)\subseteq G}
	=
	\bigoplus_i R_i
	\oplus
	\bigoplus_a \left(\mathsf{R}_a\oplus \overline {\mathsf{R}}_a\right),
\end{equation}
where the $R_i$ are real $\SU(3)$ representations and the complex representations $\mathsf{R}_a$ occur together with their conjugates.  We then restrict to the one-dimensional subspace defined by
\begin{equation}
	y_1=y_2=y~
\end{equation}
in the Cartan of $\SU(3)$. Here $y_i = f_i(\mathbf{x})$ and $f_i$ is a linear
function determined by the branching.  On this line, the contributions from the complex pairs $\mathsf{R}_a\oplus \overline {\mathsf{R}}_a$ behave as in the full-hypermultiplet case.  For the real $\SU(3)$ representations, the weights $\lambda$ are invariant under complex conjugation and under $\lambda\mapsto -\lambda$.  Therefore the positive weights can be chosen compatibly with the symmetry exchanging the two Dynkin labels. After restriction to $y_1=y_2=y$, the remaining contribution is again a piecewise-linear function of $y$ with integer slope.  Since the higher Rains functions allow the addition of elementary root and weight terms with non-negative integer coefficients, one can always cancel this slope. For instance, the term $\{\alpha_+\cdot \mathbf{y}\}$ with $\alpha_+=(2,-1)$ (or its companion $1-\{\alpha_+\cdot \mathbf{y}\}$, with slope of the opposite sign), added with a suitable non-negative integer coefficient, cancels the remaining coefficient of $|y|$ in a higher Rains function. 

There are a few low-rank cases where the above $\SU(3)$-subgroup argument does not directly apply, namely theories in which every simple gauge factor is of type $\SU(2)$ or $\USp(4)$, so that no simple factor contains an $\SU(3)$ with the required properties. Consulting the list of allowed hypermultiplets of conformal theories (Tables 2, 3 and 4 of \cite{Bhardwaj:2013qia}), one checks that the conformal theories built solely from such factors and from half-hypermultiplets are the $A_1$ class $\mathcal{S}$ theories, built from $\SU(2)$ tri-fundamentals, and the $\USp(4)$ gauge theory with half-hypermultiplets in the $\mathbf{16}$. In both cases one verifies directly that a higher Rains function is constant on a locus of positive dimension; for the $\USp(4)$ theory the modular transform is computed explicitly in the main text and indeed displays logarithmic behavior.

Therefore, the Rains-function asymptotics predicts at least a single power of $\log \widetilde q$ in the modular transform of the Schur index. More generally, higher-dimensional contributing loci would produce higher powers of $\log \widetilde q$. Thus logarithmic terms are expected quite generally, unless the corresponding coefficients vanish accidentally. This is consistent with the empirical observation that known vertex operator algebras associated to four-dimensional $\mathcal N=2$ Lagrangian SCFTs are non-rational.

\section{Glossary of terms used in the Bhardwaj-Tachikawa classification}
\label{app:gloss}
Here we collect the terms that are frequently used in Section \ref{sec:classrec}, which are relevant for the classification of Lagrangian theories. These terms are defined in the Section \ref{sec:classrec}, we collect them here for easy reference.
\begin{enumerate}
    \item $n$-gons: hypermultiplets in representations of n simple groups.
   \item $k$-valent node: A node to which $k$ branches are attached.
    \item Main subgraph: part of the graph where the hypermultiplets are in bifundamental representation of the adjacent nodes. 
    \item Current at an edge: Consider a hypermultiplet in the bi-fundamental of $X(m)\times Y(n)$, where $X,Y=\SO,\USp,\SU$. The current at the edge is defined to be $m+2-n$ for half-hypermultiplet in $\text{fund}\times\text{fund}$ of $\USp(m)\times \SO(n)$ and $m-n$ otherwise. 
    \item Trunk: part of the main subgraph where the current is zero 
    \item Branch: Part of the graph which is not a trunk.
    \item Branch current: let $h^\vee_0$ be the dual Coxeter number of the zeroth node of the branch, and let $\beta$ be the beta-function contribution to the zeroth node from the first node. The branch current is the difference $h^\vee_0-\beta$, multiplied by $1$, $1$ or $2$ according to whether the zeroth node is of type $\SU$, $\SO$ or $\USp$, respectively. A branch is called \textit{small} if the branch current is non-negative, and \textit{large} otherwise. 
\end{enumerate}

\section{List of relevant branches}\label{app:branches}

\begin{itemize}
    \item The only large branch: \begin{tikzpicture}[baseline={(0,-0.5ex)},squarenode/.style={rectangle, draw=black, minimum size=7mm}]
         \node (r0) at (0.5,0) { $\cdots$};
          \node (r1) at (2,0) { $\USp(8)$};
          \node[squarenode] (r2) at (3.8,0) { $\mathbf{\frac{1}{2}asym3}$};

          \draw[-,shorten <=-2pt,shorten >=- 2pt] (r0) -- (r1);
          \draw[-,shorten <=-2pt,shorten >=0pt] (r1) -- (r2);
        \end{tikzpicture}
    \item \begin{tikzpicture}[baseline={(0,-0.5ex)}]
         \node (r0) at (0.3,0) { $\cdots$};
          \node (r1) at (1.8,0) { $\SO(m_2)$};
          \node (r2) at (3.6,0) {$\USp(m_1)$};
          \node (r3) at (5.4,0) { $\SO(m_0)$};

          \draw[-,shorten <=-2pt,shorten >=- 2pt] (r0) -- (r1);
          \draw[-,shorten <=-2pt,shorten >=- 2pt] (r1) -- (r2);
          \draw[-,shorten <=-2pt,shorten >=- 2pt] (r2) -- (r3);
          \end{tikzpicture}
    \item \begin{tikzpicture}[baseline={(0,-0.5ex)}]
         \node (r0) at (0.3,0) { $\cdots$};
          \node (r1) at (1.8,0) { $\USp(m_2)$};
          \node (r2) at (3.6,0) {$\SO(m_1)$};
          \node (r3) at (5.4,0) { $\USp(m_0)$};

          \draw[-,shorten <=-2pt,shorten >=- 2pt] (r0) -- (r1);
          \draw[-,shorten <=-2pt,shorten >=- 2pt] (r1) -- (r2);
          \draw[-,shorten <=-2pt,shorten >=- 2pt] (r2) -- (r3);
          \end{tikzpicture}
    \item Branches with special representations:
    \begin{itemize} 
        \item     \begin{tikzpicture}[
        baseline={(0,-0.5ex)},
        squarenode/.style={rectangle, draw=black,  minimum size=7mm},
        ]
        \node     (circle1)      at (0.5,0)   { $\USp(6)$};
        \node     (circle2)    at (2.5,0)    {$G_2$};
        \node     (circle3)    at (4.5,0)    {$\USp(2)$};
        \node (A) at (-1,0) {$\cdots$};

        \draw[-] (A.east) -- (circle1.west);
        \draw[-] (circle1.east) -- (circle2.west) node[pos=0.8, above] {$\mathrm{\mathbf{7}}$}; 
        \draw[-] (circle2.east) -- (circle3.west)  node[pos=0.2, above] {$\mathrm{\mathbf{7}}$};
        
        \end{tikzpicture}
        \item 
        \begin{tikzpicture}[
        baseline={(0,-0.5ex)},
        squarenode/.style={rectangle, draw=black,  minimum size=7mm},
        ]

        \node     (circle1)      at (0,0)   { $\USp(8)$};
        \node      (circle2)    at (2,0)    {$G_2$};
        \node (A) at (-1.5,0) {$\cdots$};

        \draw[-] (A.east) -- (circle1.west);
        \draw[-] (circle1.east) -- (circle2.west) node[pos=0.8, above] {$\mathrm{\mathbf{7}}$};
        
        \end{tikzpicture}
        \item 
        \begin{tikzpicture}[
        baseline={(0,-0.5ex)},
        squarenode/.style={rectangle, draw=black,  minimum size=7mm},
        ]

        \node     (circle1)      at (0,0)   { $\USp(6)$};
        \node[squarenode]        (square1)    at (2,0)    {$\mathrm{\mathbf{\frac{1}{2}asym3}}$};
        \node (A) at (-1.5,0) {$\cdots$};

        \draw[-] (A.east) -- (circle1.west);
        \draw[-] (circle1.east) -- (square1.west);
        
        \end{tikzpicture}
        \item 
        \begin{tikzpicture}[
        baseline={(0,-0.5ex)},
        squarenode/.style={rectangle, draw=black,  minimum size=7mm},
        ]

        \node      (circle1)      at (0,0)   { $\SO(11)$};
        \node[squarenode]     (square1)    at (2,0)    {$\mathrm{\mathbf{\frac{1}{2}S}}$};
        \node (A) at (-1.5,0) {$\cdots$};

        \draw[-] (A.east) -- (circle1.west);
        \draw[-] (circle1.east) -- (square1.west);
        
        \end{tikzpicture}
        \item 
        \begin{tikzpicture}[
        baseline={(0,-0.5ex)},
        squarenode/.style={rectangle, draw=black,  minimum size=7mm},
        ]

        \node     (circle1)  at (0,0)    { $\SO(12)$};
        \node[squarenode]        (square1)    at (2,0) {$\mathbf{\frac{1}{2}S}$};
        \node (A) at (-1.5,0) {$\cdots$};

        \draw[-] (circle1.east) -- (square1.west);
        \draw[-] (A.east) -- (circle1.west);
        
        \end{tikzpicture}
        \item 
        \begin{tikzpicture}[
        baseline={(0,-0.5ex)},
        squarenode/.style={rectangle, draw=black,  minimum size=7mm},
        ]

        \node     (circle1)  at (0,0)    { $\SO(12)$};
        \node[squarenode]        (square1)    at (2,0) {$\mathbf{\frac{1}{2}C}$};
        \node (A) at (-1.5,0) {$\cdots$};

        \draw[-] (circle1.east) -- (square1.west);
        \draw[-] (A.east) -- (circle1.west);
        
        \end{tikzpicture}
        \item 
        \begin{tikzpicture}[
        baseline={(0,-0.5ex)},
        squarenode/.style={rectangle, draw=black,  minimum size=7mm},
        ]
  
        \node      (circle1)      at (0,0)   { $\SO(13)$};
        \node[squarenode]        (square1)    at (2,0)    {$\mathrm{\mathbf{\frac{1}{2}S}}$};
        \node (A) at (-1.5,0) {$\cdots$};
 
        \draw[-] (A.east) -- (circle1.west);
        \draw[-] (circle1.east) -- (square1.west);
        
        \end{tikzpicture}
    \end{itemize}

    \item Branches with further branching:
    \begin{itemize}
        \item \begin{tikzpicture}[
        baseline={(0,-0.5ex)},
        squarenode/.style={rectangle, draw=black,  minimum size=7mm},
        ]
        \node    (circle1)      at (0,0)   { $\SO(8)$};
        \node      (circle2)    at (2,0)    {$\USp(2)$};
        \node     (circle3)    at (2,-0.8)    {$\USp(2)$};
        \node     (circle4)    at (2,0.8)    {$\USp(2)$};
        \node      (circle5)    at (-2,0)    {$\USp(6)$};
        \node (A) at (-3.5,0) {$\cdots$};

        \draw[-] (A.east) -- (circle5.west);
        \draw[-] (circle1.east) -- (circle2.west);
        \draw[-] (circle1.east) -- (circle3.west);
        \draw[-] (circle1.east) -- (circle4.west);
        \draw[-] (circle1.west) -- (circle5.east);

        \end{tikzpicture}
        \item 
        \begin{tikzpicture}[
        baseline={(0,-0.5ex)},
        squarenode/.style={rectangle, draw=black,  minimum size=7mm},
        ]

        \node    (circle1)      at (0,0)   { $\USp(6)$};
        \node    (circle2)    at (2,0)    {$\SO(8)$};
        \node      (circle3)    at (4,0.5)    {$\USp(2)$};
        \node     (circle4)    at (4,-0.5)    {$\USp(4)$};
        \node     (circle5)    at (6,-0.5)    {$\SO(4)$};
        \node (A) at (-1.5,0) {$\cdots$};

        \draw[-] (A.east) -- (circle1.west);
        \draw[-] (circle1.east) -- (circle2.west); 
        \draw[-] (circle2.east) -- (circle3.west); 
        \draw[-] (circle2.east) -- (circle4.west);
        \draw[-] (circle4.east) -- (circle5.west);
        
        \end{tikzpicture}
        \item 
        \begin{tikzpicture}[
        baseline={(0,-0.5ex)},
        squarenode/.style={rectangle, draw=black,  minimum size=7mm},
        ]

        \node     (circle1)      at (0,0)   { $\USp(4k+2)$};
        \node     (circle2)    at (3,0)    {$\SO(4k+3)$};
        \node     (circle3)    at (6,0.5)    {$\USp(2k)$};
        \node      (circle4)    at (6,-0.5)    {$\USp(2k)$};
        \node (A) at (-2,0) {$\cdots$};

        \draw[-] (A.east) -- (circle1.west);
        \draw[-] (circle1.east) -- (circle2.west); 
        \draw[-] (circle2.east) -- (circle3.west); 
        \draw[-] (circle2.east) -- (circle4.west) ;
        
        \end{tikzpicture}
        \item 
        \begin{tikzpicture}[
        baseline={(0,-0.5ex)},
        squarenode/.style={rectangle, draw=black,  minimum size=7mm},
        ]

        \node     (circle1)      at (0,0)   { $\USp(4k)$};
        \node     (circle2)    at (2.5,0)    {$\SO(4k)$};
        \node     (circle3)    at (5,0.5)    {$\USp(2k-2)$};
        \node     (circle4)    at (5,-0.5)    {$\USp(2k-2)$};
        \node (A) at (-1.5,0) {$\cdots$};

        \draw[-] (A.east) -- (circle1.west);
        \draw[-] (circle1.east) -- (circle2.west); 
        \draw[-] (circle2.east) -- (circle3.west); 
        \draw[-] (circle2.east) -- (circle4.west) ;
        
        \end{tikzpicture}
    \end{itemize}

    \end{itemize}

\subsection*{Dualities among branches}
\label{app:branch_dualities}

A number of the branches listed above are related to one another by group-theoretic isomorphisms or class~$\mathcal{S}$ dualities, and only one representative of each duality class need be considered in the analysis of Section~\ref{sec:classrec}.

\paragraph{$\SO(7)$-spinor branches via $\SO(8)$ triality.}
The decorated branch involving a half-hypermultiplet in $\mathbf{S}\times\mathbf{fund}$ of $\SO(7)\times\USp(2m)$ with $1\leq m\leq 5$, where $\mathbf{S}$ is the spinor of $\SO(7)$, is equivalent to the union of two simpler branches: a $2$-gon half-hyper in the bi-fundamental of $\SO(7)\times\USp(2m)$ together with a single $1$-gon half-hyper in the fundamental of $\USp(2m)$. To see this, embed $\SO(7)\subset\SO(8)$ as the stabilizer of one of the directions of the $\SO(8)$ flavor symmetry of the $\USp(2m)$ node. The $\SO(8)$ vector and spinor representations branch as $\mathbf{8}_v\rightarrow\mathbf{1}\oplus\mathbf{7}$ and $\mathbf{8}_s\rightarrow\mathbf{8}_s$, which decomposes the $\SO(7)\times\USp(2m)$ representation content into the two simpler branches.

\paragraph{$\SO(5)\cong\USp(4)$ duality.}
Two decorated branches involve the $\SO(5)\cong\USp(4)$ isomorphism:
\begin{enumerate}
\item[--] $-\bigl(\SO(5)\cong\USp(4)\bigr)-\SO(4)$,
\item[--] $-\bigl(\USp(4)\cong\SO(5)\bigr)-\USp(2)-\SO(3)$,
\end{enumerate}
where the notation $\bigl(\SO(5)\cong\USp(4)\bigr)$ means that the half-hyper to the left of the node is in the fundamental of $\SO(5)$, while that to the right is in the fundamental of $\USp(4)$ (equivalently, the spinor of $\SO(5)$). From the class~$\mathcal{S}$ perspective, the branches $-\SO(5)-\USp(2)-\SO(3)$ and $-\USp(4)-\SO(4)$ are dual: $\SO(4)$ SQCD is described by a torus with two $\USp(4)$-flavored punctures, which is dual to $\mathcal{N}=4$ $\SU(2)$ SYM gauged with a trifundamental by $\SU(2)$, and $\mathcal{N}=4$ $\SU(2)$ SYM is in turn dual to $\SO(3)$ SQCD. All four branches lie in a single duality class, and a single representative suffices for the analysis.

\section{Hall-Littlewood index for some low-rank quivers}
\label{app:hlindquiv}
\subsection*{Loops}
Here we collect Hall-Littlewood indices of some theories in order to eliminate theories with a Higgs branch.
\begin{itemize}
    \item 
    \begin{tikzpicture}[baseline={(0,2ex)}]

        \node        (circle1)    at (0,0)    { $\SO(4)$};
        \node       (circle2)    at (2,0)    {$\USp(2)$};

        \node       (circle5)    at (2,1)    {$\SO(4)$};
        \node        (circle6)   at (0,1)    {$\USp(2)$};
    
        \draw[-] (circle1.east) -- (circle2.west);
        \draw[-] (circle2.north) -- (circle5.south);
        \draw[-] (circle6.south) -- (circle1.north);
        \draw[-] (circle6.east) -- (circle5.west);
    \end{tikzpicture}
    \begin{equation}
    \begin{split}
        \mathcal{I}_{\mathrm{HL}}=&
   \frac{(1-t)(1 + t - t^2 - 3 t^3 + t^4 + t^5)}{1 - t^2}.
    \end{split}
    \end{equation}
    \item 
    \begin{tikzpicture}[baseline={(0,2ex)}]

        \node        (circle1)    at (0,0)    { $\SO(4)$};
        \node       (circle2)    at (2,0)    {$\USp(2)$};
        \node       (circle3)    at (4,0)    {$\SO(4)$};
        \node       (circle4)    at (4,1)    {$\USp(2)$};

        \node       (circle5)    at (2,1)    {$\SO(4)$};
        \node        (circle6)   at (0,1)    {$\USp(2)$};
    
        \draw[-] (circle1.east) -- (circle2.west);
        \draw[-] (circle2.east) -- (circle3.west);
        \draw[-] (circle3.north) -- (circle4.south);
        \draw[-] (circle6.south) -- (circle1.north);
        \draw[-] (circle6.east) -- (circle5.west);
        \draw[-] (circle5.east) -- (circle4.west);
    \end{tikzpicture}
    \begin{equation}
    \begin{split}
        \mathcal{I}_{\mathrm{HL}}=&
   \frac{(1 - t)^3 (1 + 3 t + 3 t^2 + t^3 - 3 t^4 - 3 t^5 - t^6)}{1 - t^3}.
    \end{split}
    \end{equation}
    \item 
    \begin{tikzpicture}[baseline={(0,2ex)}]

        \node        (circle1)    at (0,0)    { $\SO(6)$};
        \node       (circle2)    at (2,0)    {$\USp(4)$};

        \node       (circle5)    at (2,1)    {$\SO(6)$};
        \node        (circle6)   at (0,1)    {$\USp(4)$};
    
        \draw[-] (circle1.east) -- (circle2.west);
        \draw[-] (circle2.north) -- (circle5.south);
        \draw[-] (circle6.south) -- (circle1.north);
        \draw[-] (circle6.east) -- (circle5.west);
    \end{tikzpicture}
    \begin{equation}
    \begin{split}
        \mathcal{I}_{\mathrm{HL}}=&
   \frac{(1 - t)^2 (1 - t^2)(1 + t + t^2 - t^4 - t^5)^2}{(1 - t^2)^2 (1 - t^4)}.
    \end{split}
    \end{equation}
\end{itemize}

\subsection*{HL index for graphs including $\mathcal{O}_{\mathrm{chain}}$}\label{app:O_chain}

\begin{itemize}
    \item 
    \begin{tikzpicture}[baseline={(0,2ex)}]

    \node        (circle1)  at (-3.2,0)      { $\SO(3)$};
    \node      (circle2)    at (-1.6,0)   {$\USp(2)$};
    \node        (square1)    at (-1.6,0.9)   {$1$};
    \node     (circle3)       at (0,0) {$\SO(4)$};
    \node       (circle4)       at (1.6,0) {$\USp(2)$};
    \node       (square2)    at (1.6,0.9)  {$1$};
    \node      (circle5)       at (3.2,0) {$\SO(3)$};

    \draw[-] (circle1.east) -- (circle2.west);
    \draw[-] (circle2.east) -- (circle3.west);
    \draw[-] (square1.south) -- (circle2.north);
    \draw[-] (square2.south) -- (circle4.north);

    \draw[-] (circle3.east) -- (circle4.west);
    \draw[-] (circle4.east) -- (circle5.west);

    \end{tikzpicture}:

    \begin{equation}
    \begin{split}
        \mathcal{I}_{\mathrm{HL}}=&\frac{(1 - t)^2 (1 + 2 t + t^2 - 2 t^3 - t^4)}{1 - t^2};
    \end{split}
    \end{equation}
\item 
    \begin{tikzpicture}[baseline={(0,2ex)}]

    \node        (circle1)  at (-3.2,0)      { $\SO(3)$};
    \node      (circle2)    at (-1.6,0)   {$\USp(2)$};
    \node        (square1)    at (-1.6,0.9)   {$1$};
    \node     (circle3)       at (0,0) {$\SO(4)$};
    \node       (circle4)       at (1.6,0) {$\USp(2)$};
    \node     (circle5)       at (3.2,0) {$\SO(4)$};
    \node       (circle6)       at (4.8,0) {$\USp(2)$};
    \node       (square2)    at (4.8,0.9)  {$1$};
    \node      (circle7)       at (6.4,0) {$\SO(3)$};

    \draw[-] (circle1.east) -- (circle2.west);
    \draw[-] (circle2.east) -- (circle3.west);
    \draw[-] (circle3.east) -- (circle4.west);
    \draw[-] (circle4.east) -- (circle5.west);
    \draw[-] (circle5.east) -- (circle6.west);
    \draw[-] (circle6.east) -- (circle7.west);
    \draw[-] (square1.south) -- (circle2.north);
    \draw[-] (square2.south) -- (circle6.north);

    \end{tikzpicture}:
    
    \begin{equation}
    \begin{split}
        \mathcal{I}_{\mathrm{HL}}=&\frac{(1 - t)^3 (1 + 3 t + 3 t^2 + t^3 - 3 t^4 - 3 t^5 - t^6)}{1 - t^3};
    \end{split}
    \end{equation}

\item 
    \begin{tikzpicture}[baseline={(0,2ex)}]

    \node        (circle1)  at (-3,0)      { $\SO(3)$};
    \node      (circle2)    at (-1.5,0)   {$\USp(2)$};
    \node        (square1)    at (-1.5,0.9)   {$1$};
    \node     (circle3)       at (0,0) {$\SO(4)$};
    \node       (circle4)       at (1.5,0) {$\USp(2)$};
    \node     (circle5)       at (3,0) {$\SO(4)$};
    \node       (circle6)       at (4.5,0) {$\USp(2)$};
    \node     (circle7)       at (6,0) {$\SO(4)$};
    \node       (circle8)       at (7.5,0) {$\USp(2)$};
    \node       (square2)    at (7.5,0.9)  {$1$};
    \node      (circle9)       at (9,0) {$\SO(3)$};

    \draw[-] (circle1.east) -- (circle2.west);
    \draw[-] (circle2.east) -- (circle3.west);
    \draw[-] (circle3.east) -- (circle4.west);
    \draw[-] (circle4.east) -- (circle5.west);
    \draw[-] (circle5.east) -- (circle6.west);
    \draw[-] (circle6.east) -- (circle7.west);
    \draw[-] (circle7.east) -- (circle8.west);
    \draw[-] (circle8.east) -- (circle9.west);
    \draw[-] (square1.south) -- (circle2.north);
    \draw[-] (square2.south) -- (circle8.north);

    \end{tikzpicture}:
    
    \begin{equation}
    \begin{split}
        \mathcal{I}_{\mathrm{HL}}=&\frac{(1 - t)^4 (1 + 4 t + 6 t^2 + 4 t^3 + t^4 - 4 t^5 - 6 t^6 - 4 t^7 - t^8)}{1 - t^4}.
    \end{split}
    \end{equation}

\end{itemize}

\subsection*{HL index for graphs including $\mathcal{O}_{n\geq4\text{-valent}}$}\label{app:O_4valent}
\begin{itemize}
    \item 
    \begin{tikzpicture}[baseline={(0,2ex)}]

        \node         (circle1)      at (0,0)   { $\USp(6)$};
        \node         (circle2)    at (-1,1)    {$\SO(5)$};
        \node         (circle3)    at (1.7,0)    {$\SO(5)$};
        \node         (circle4)    at (-2,0)    {$\SO(5)$};

        \node         (circle8)    at (1,1)    {$1$};

        \draw[-] (circle1.north) -- (circle2.south);
        \draw[-] (circle1.east) -- (circle3.west) ;
        \draw[-] (circle1.west) -- (circle4.east);

        \draw[-] (circle1.north) -- (circle8.south);
        \end{tikzpicture}:
    \begin{equation}
    \begin{split}
        \mathcal{I}_{\mathrm{HL}}=&
   \frac{1 - t + t^2 + 6 t^5 + 2 t^9 - t^{10}}{(1 - t) (1 - t^2)};
    \end{split}
    \end{equation}
    \item 
    \begin{tikzpicture}[baseline={(0,0ex)}]

        \node      (circle2)    at (-2.5,1)   {$1-\USp(2)$};
        \node      (circle9)    at (-2.5,0)   {$1-\USp(2)$};
        \node      (circle11)    at (-2.5,-1)   {$1-\USp(2)$};
        \node     (circle3)       at (0,0) {$SO(7)$};
        \node       (circle4)       at (1.7,0) {$\USp(4)$};
        \node     (circle5)       at (3.4,0) {$\SO(5)$};
        \node       (circle6)       at (5.1,0) {$\USp(2)$};
        \node     (circle7)       at (6.8,0) {$\SO(3)$};

        \draw[-] (circle2.east) -- (circle3.west);
   
        \draw[-] (circle9.east) -- (circle3.west);

        \draw[-] (circle11.east) -- (circle3.west);
        \draw[-] (circle3.east) -- (circle4.west);
        \draw[-] (circle4.east) -- (circle5.west);
        \draw[-] (circle5.east) -- (circle6.west);
        \draw[-] (circle6.east) -- (circle7.west);

        \end{tikzpicture}:
    \begin{equation}
    \begin{split}
        \mathcal{I}_{\mathrm{HL}}=&(1 - t) (1 + t + 3 t^2 + 6 t^3 + 2 t^4 - t^5 - 3 t^6 - 6 t^7 \\
        &- t^9 + t^{11})/(1 - t^2)^3 (1 - t^3).
    \end{split}
    \end{equation}
\end{itemize}

\subsection*{HL index for graphs including $\mathcal{O}_{\text{3-valent}}$}\label{app:O_3valent}
\begin{itemize}
    \item 
    \begin{tikzpicture}[baseline={(0,2ex)}]

        \node        (circle1)  at (-3.6,0)      { $\SO(4)$};
        \node      (circle2)    at (-1.8,0)   {$\USp(4)$};
        \node        (square1)    at (0,0.9)   {$\USp(4)$};
        \node     (circle3)       at (0,0) {$\SO(8)$};
        \node       (circle4)       at (1.8,0) {$\USp(4)$};
        \node     (circle5)       at (3.6,0) {$\SO(4)$};

        \node       (square2)    at (0,1.8)  {$\SO(4)$};

        \draw[-] (circle1.east) -- (circle2.west);
        \draw[-] (circle2.east) -- (circle3.west);
        \draw[-] (circle3.east) -- (circle4.west);
        \draw[-] (circle4.east) -- (circle5.west);

        \draw[-] (square1.south) -- (circle3.north);
        \draw[-] (square2.south) -- (square1.north);

        \end{tikzpicture}:
    \begin{equation}
    \begin{split}
        \mathcal{I}_{\mathrm{HL}}=&(1-t)^2 (1+2 t+3 t^2+3 t^3-2 t^4-7 t^5-t^6+5 t^7\\
        &+21 t^8+17 t^9+8 t^{10}-t^{11}-6 t^{12}-t^{13}+3 t^{14}\\
        &+2t^{15}+t^{16})/( \left(1-t^3\right) \left(1-t^4\right))
    \end{split}
    \end{equation}

\item 
     \begin{tikzpicture}[baseline={(0,-0.5ex)}]
     \node        (circle1)  at (-2,0)      { $\SO(6)$};
     \node        (circle2)  at (0,0)      { $\USp(8)$};
     \node        (circle3)  at (3,0.5)      { $\SO(7)-\USp(2)-1$};
     \node        (circle4)  at (3,-0.5)      { $\SO(7)-\USp(2)-1$};

     \draw[-] (circle1.east) -- (circle2.west);
     \draw[-] (circle2.east) -- (circle3.west);
     \draw[-] (circle2.east) -- (circle4.west);
    \end{tikzpicture}:

    \begin{equation}
    \begin{split}
        \mathcal{I}_{\mathrm{HL}}=&((1 + t^2) (1 - t^2 + t^4 + 6 t^7 + 6 t^8 - 6 t^9 \\
        &+ t^{11} + t^{13} - t^{15}))/((1 - t^4) (1 - t^3)).
    \end{split}
    \end{equation} 
\end{itemize}

\subsection*{HL index for graphs with a single $G_2$ node trunk}\label{app:G_2}

\begin{enumerate}
    \item    
        \begin{tikzpicture}[baseline={(0,2ex)}]

        \node        (circle7)    at (-3.2,0)    {$1$};
        \node         (circle1)      at (0,0)   { $G_2$};
        \node         (circle2)    at (-1,1)    {$\USp(2)$};
        \node         (circle3)    at (1.7,0)    {$\USp(2)$};
        \node         (circle4)    at (-2,0)    {$\USp(2)$};
        \node         (circle5)    at (-2.6,1)    {$1$};
        \node       (circle6)    at (3,0)    {$1$};

        \node         (circle8)    at (1,1)    {$\USp(2)$};
        \node         (circle9)    at (2.6,1)    {$1$};

        \draw[-] (circle1.north) -- (circle2.south);
        \draw[-] (circle1.east) -- (circle3.west) ;
        \draw[-] (circle2.west) -- (circle5.east) ;
        \draw[-] (circle1.west) -- (circle4.east);
         \draw[-] (circle3.east) -- (circle6.west);

        \draw[-] (circle7.east) -- (circle4.west);
        \draw[-] (circle1.north) -- (circle8.south);
        \draw[-] (circle8.east) -- (circle9.west);
        \end{tikzpicture}:
        \begin{equation}
        \begin{split}
            \mathcal{I}_{\mathrm{HL}}=&(1 - 5 t + 30 t^2 - 46 t^3 + 147 t^4 - 59 t^5 + 236 t^6 \\
            &- 5 t^7 + 236 t^8 - 59 t^9 + 147 t^{10} - 46 t^{11} + 30 t^{12} \\
            &- 5 t^{13} + t^{14})/((1 - t)^5 (1 - t^2)^6 (1 - t^3))
        \end{split}
        \end{equation}
    \item 
        \begin{tikzpicture}[baseline={(0,-0.5ex)}]

        \node        (circle1)  at (-3,0)      { $\SO(4)$};
        \node      (circle2)    at (-1.3,0)   {$\USp(4)$};
        \node        (square1)    at (-1.3,0.8)   {$1$};
        \node     (circle3)       at (0,0) {$G_2$};
        \node       (circle4)       at (1.4,0.8) {$\USp(2)$};
        \node     (circle5)       at (2.8,0.8) {$1$};
        \node    (circle6)  at (1.4,-0.8) {$\USp(2)$};
        \node     (square2) at (2.8,-0.8) {1};
   
        \draw[-] (circle1.east) -- (circle2.west);
        \draw[-] (circle2.east) -- (circle3.west);
        \draw[-] (circle3.east) -- (circle4.west);
        \draw[-] (circle4.east) -- (circle5.west);

        \draw[-] (square1.south) -- (circle2.north);

        \draw[-] (circle3.east) -- (circle6.west);
        \draw[-] (circle6.east) -- (square2.west);
        
        \end{tikzpicture}:
        \begin{equation}
        \begin{split}
            \mathcal{I}_{\mathrm{HL}}=&
            (1 - 3 t + 13 t^2 - 8 t^3 + 21 t^4 + 16 t^5 + 6 t^6 + 15 t^7 \\
            &+ 8 t^8 - 3 t^9 + 3 t^{10} - t^{11})/((1 - t)^3 (1 - t^2)^5 (1 - t^3))
        \end{split}
        \end{equation}

    \item 
    \begin{tikzpicture}[baseline={(0,-0.5ex)}]

        \node        (circle1)  at (-3,0.5)      { $1$};
        \node      (circle2)    at (-1.3,0.5)   {$\USp(2)$};
        \node        (circle8)  at (-3,-0.5)      { $1$};
        \node      (circle9)    at (-1.3,-0.5)   {$\USp(2)$};
        \node     (circle3)       at (0,0) {$G_2$};
        \node       (circle4)       at (1.4,0) {$\USp(4)$};
        \node     (circle5)       at (3,0) {$\SO(5)$};
        \node       (circle6)       at (4.6,0) {$\USp(2)$};
        \node     (circle7)       at (6.2,0) {$\SO(3)$};

        \draw[-] (circle1.east) -- (circle2.west);
        \draw[-] (circle2.east) -- (circle3.west);
        \draw[-] (circle8.east) -- (circle9.west);
        \draw[-] (circle9.east) -- (circle3.west);
        \draw[-] (circle3.east) -- (circle4.west);
        \draw[-] (circle4.east) -- (circle5.west);
        \draw[-] (circle5.east) -- (circle6.west);
        \draw[-] (circle6.east) -- (circle7.west);

        \end{tikzpicture}:
        \begin{equation}
        \begin{split}
            \mathcal{I}_{\mathrm{HL}}=&(1 - t + 5 t^2 + 4 t^3 + t^4 + 8 t^5 - 3 t^7 + 2 t^8 \\
            &- 6 t^9 + 2 t^{10} - t^{11} )/((1 - t) (1 - t^2)^4 (1 - t^3)).
        \end{split}
        \end{equation}
    
    \item 
        \begin{tikzpicture}[baseline={(0,2ex)}]

        \node        (circle1)  at (-3,0)      { $\SO(4)$};
        \node      (circle2)    at (-1.3,0)   {$\USp(4)$};
        \node        (square1)    at (-1.3,0.8)   {$1$};
        \node     (circle3)       at (0,0) {$G_2$};
        \node       (circle4)       at (1.4,0) {$\USp(4)$};
        \node     (circle5)       at (3,0) {$\SO(5)$};
        \node       (circle6)       at (4.6,0) {$\USp(2)$};
        \node     (circle7)       at (6.2,0) {$\SO(3)$};

        \draw[-] (circle1.east) -- (circle2.west);
        \draw[-] (circle2.east) -- (circle3.west);
        \draw[-] (circle3.east) -- (circle4.west);
        \draw[-] (circle4.east) -- (circle5.west);
        \draw[-] (circle5.east) -- (circle6.west);
        \draw[-] (circle6.east) -- (circle7.west);

        \draw[-] (square1.south) -- (circle2.north);

        \end{tikzpicture}:
        \begin{equation}
        \begin{split}
            \mathcal{I}_{\mathrm{HL}}=&((1 - t) (1 + t + 3 t^2 + 6 t^3 + 2 t^4 - t^5 - 3 t^6 \\
            &- 6 t^7 - t^9 + t^{11}))/((1 - t^2)^3 (1 - t^3)).
        \end{split}
        \end{equation}
    \item $\SO(3)-\USp(2)-\SO(5)-\USp(4)-G_2-\USp(4)-\SO(5)-\USp(2)-\SO(3)$:
        \begin{equation}
        \begin{split}
            \mathcal{I}_{\mathrm{HL}}=&((1 - t)^2(1 + 2 t + 4 t^2 + 7 t^3 + 4 t^4 - t^5 - 4 t^6 - 9 t^7 - 5 t^8 -5 t^9  \\
            &- 2 t^{10} + 3 t^{11} + 3 t^{12} + 2 t^{13} + t^{14}))/((1 - t^2) (1 - t^6)).
        \end{split}
        \end{equation}
    \item 
        \begin{tikzpicture}[baseline={(0,2ex)}]

        \node        (circle1)  at (-3,0)      { $\SO(4)$};
        \node      (circle2)    at (-1.3,0)   {$\USp(4)$};
        \node        (square1)    at (-1.3,0.8)   {$1$};
        \node     (circle3)       at (0,0) {$G_2$};
        \node       (circle4)       at (1.4,0) {$\USp(4)$};
        \node     (circle5)       at (3,0) {$\SO(4)$};

        \node       (square2)    at (1.4,0.8)  {$1$};

        \draw[-] (circle1.east) -- (circle2.west);
        \draw[-] (circle2.east) -- (circle3.west);
        \draw[-] (circle3.east) -- (circle4.west);
        \draw[-] (circle4.east) -- (circle5.west);

        \draw[-] (square1.south) -- (circle2.north);
        \draw[-] (square2.south) -- (circle4.north);

        \end{tikzpicture}:
        \begin{equation}
        \begin{split}
            \mathcal{I}_{\mathrm{HL}}=&(1 - t + 5 t^2 + 4 t^3 + t^4 + 8 t^5 - 3 t^7 + 2 t^8 \\
            &- 6 t^9 + 2 t^{10} - t^{11})/((1 - t) (1 - t^2)^4 (1 - t^3)).
        \end{split}
        \end{equation}
\end{enumerate}
We note in passing that theories 3 and 6 have the same ${\cal I}_{\rm HL}$. They also have the same $n_v$ and $n_h$ and their Coulomb branches are generated by the same set of invariants. This suggests that they may be related by an exact duality. It would be interesting to investigate this further.

\section{Null relations used for the Jacobi identities}\label{app:nulls}

The null states needed for the Jacobi identities to hold are shown in Table \ref{tab:nulls_Jacobi_usp4} for the VOA associated to the $\mathrm{USp}(4)$ gauge theory with a half-hyper in $\mathbf{16}$ and in Table \ref{tab:nulls_Jacobi_su3su2} for the VOA associated to the $\mathrm{SU}(3)\times \mathrm{SU}(2)$ theory with a half-hyper in the $\mathbf{8}\times\mathbf{2}$.

\begin{table}[htpb]
    \renewcommand{\arraystretch}{1.5}
		\centering
		\resizebox{\textwidth}{!}{\begin{tabular}{|c|c|}
			\hline
			$h$ & Null states appear in the Jacobi identities\\
			\hline
			$5$ & \makecell{$2 W_{+-}\Lambda_- + W_{--}\Lambda_+$\\ $ 2W_{+-}\Lambda_++W_{++}\Lambda_- $ \\$W_{++}\Lambda_+$ \\$W_{--}\Lambda_-$} \\
            \hline
			$6$ &  \makecell{$2 W_{+-} W_{--}+ T W_{--}' + T' W_{--}+ \Lambda_-'' \Lambda_-- \frac{1}{3} W_{--}^{(3)}$\\ $2 W_{+-}^2 + W_{++} W_{--}-\frac{1}{2}\Lambda_+ \Lambda_-'' + T W_{+-}' + T' W_{+-}+\frac{1}{2}\Lambda_+'' \Lambda_--\frac{1}{6}W_{+-}^{(3)}+\frac{1}{24}T^{(4)}$\\ $2 W_{++} W_{+-}+ T W_{++}' + T' W_{++}+ \Lambda_+'' \Lambda_+-\frac{1}{3}W_{++}^{(3)}$\\ $W_{--}\Lambda_+' + 2 W_{+-}\Lambda_-'+ W_{--}'\Lambda_++ 2 W_{+-}'\Lambda_-$\\ $2 W_{+-}\Lambda_+'+ W_{++}\Lambda_-'+ 2 W_{+-}'\Lambda_+ + W_{++}'\Lambda_-$\\ $W_{++}\Lambda_+' + W_{++}'\Lambda_+$\\ $W_{--}\Lambda_-' + W_{--}'\Lambda_-$\\ $ W_{--}W_{--}$ \\ $W_{++}W_{++}$}\\
            \hline
			$7$ & \makecell{$W_{--}' W_{--}$\\ $  3 T W_{++}'' + 6 W_{++} W_{+-}' + 6 T' W_{++}' + 6 W_{++}' W_{+-} + 3 T'' W_{++} + 3 \Lambda_+'' \Lambda_+' + 3 \Lambda_+^{(3)} \Lambda_+ - W_{++}^{(4)} $\\ $  3 T W_{--}'' + 6 W_{+-} W_{--}' + 6 T' W_{--}' + 6 W_{+-}' W_{--} + 3 T'' W_{--}   + 3 \Lambda_-'' \Lambda_-' + 3 \Lambda_-^{(3)} \Lambda_- - W_{--}^{(4)}  $\\ $ T W_{+-}'' + W_{++} W_{--}' -\frac{1}{3}\Lambda_+ \Lambda_-^{(3)} + 2 T' W_{+-}' + W_{++}' W_{--} + 4 W_{+-}' W_{+-} + T'' W_{+-} + \frac{1}{2} T'' T' + \Lambda_+'' \Lambda_-' + \frac{1}{6} T^{(3)} T + \frac{2}{3} \Lambda_+^{(3)} \Lambda_- -\frac{1}{4} W_{+-}^{(4)} -\frac{1}{60} T^{(5)}$}\\
			\hline
		\end{tabular}}
		\caption{Null states appear in the Jacobi identities for the VOA associated the $\mathrm{USp}(4)$ gauge theory with a half-hyper in $\mathbf{16}$. They show up at levels $h=5,6,7$.}
		\label{tab:nulls_Jacobi_usp4}
	\end{table}

\begin{table}[htpb]
    \renewcommand{\arraystretch}{1.5}
		\centering
		\resizebox{\textwidth}{!}{
        \begin{tabular}{|c|c|}
			\hline
			$h$ & Null states appear in the Jacobi identities\\
            \hline
			$6$ &  \makecell{$T W' - 2W^2 + 4U_0V^{0} + T'W - \frac{1}{3}W^{(3)}$\\ $-T (V^{+})' + 2W V^{+} - T'V^{+} + \frac{1}{3}(V^{+})^{(3)}$\\ $-T (V^{0})' + 2W V^{0} - T'V^{0} + \frac{1}{3}(V^{0})^{(3)}$\\ $T (V^{-})' + 2U_0U_+ + T'V^{-} - \frac{1}{3}(V^{-})^{(3)}$\\ $-T U_+' - 2W U_+ - T'U_+ + \frac{1}{3}U_+^{(3)}$\\ $-T U_0' - 2W U_0 - T'U_0 + \frac{1}{3}U_0^{(3)}$\\ $T U_-' - 2V^{0}V^{+} + T'U_- - \frac{1}{3}U_-^{(3)}$\\ $ T U_-' + WU_- - V^{0}V^{+} + T'U_- - \frac{1}{3}U_-^{(3)}$\\ $-T (V^{0})' + WV^{0} + U_-U_+ - T'V^{0} +\frac{1}{3}(V^{0})^{(3)}$\\ $T (V^{+})' - WV^{+} + U_-U_0 + T'V^{+}-\frac{1}{3}(V^{+})^{(3)}$\\ $-T (V^{-})' + WV^{-} - U_0U_+ - T'V^{-}+\frac{1}{3}(V^{-})^{(3)}$\\ $T U_0' + WU_0 + V^{-}V^{+} + T'U_0 - \frac{1}{3}U_0^{(3)}$\\ $- T U_+' - WU_+ + V^{-}V^{0} - T'U_++\frac{1}{3}U_+^{(3)}$\\ $U_iV^{j\neq i}, \qquad U_-V^{-}-U_0V^{0},
            \qquad -U_0V^{0}+U_+V^{+}, \qquad  (V^{i})^2,\qquad U_i^2$}\\
            \hline
			$7$ & \makecell{$3T^2W+TW''+12U_0(V^0)'+4T'W'-6W'W-4U_-'V^-+8U_0'V^0-4U_+'V^++\frac32T''W-\frac32T''T'-\frac12T^{(3)}T-\frac23W^{(4)}+\frac18T^{(5)}$ \\$-3T^2V^++2T(V^+)''+2T'(V^+)'-2W'V^++4U_-'U_0+\frac32T''V^+-\frac13(V^+)^{(4)}$ \\$\frac32T^2V^+-\frac32T(V^+)''+W(V^+)'-2T'(V^+)'+2W'V^+-2U_-'U_0-\frac54T''V^++\frac13(V^+)^{(4)}$ \\$-3T^2V^0+2T(V^0)''+2T'(V^0)'-2W'V^0-4U_-'U_+ +\frac32T''V^0-\frac13(V^0)^{(4)}$ \\$\frac32T^2V^0-\frac32T(V^0)''+W(V^0)'-2T'(V^0)'+2W'V^0+2U_-'U_+-\frac54T''V^0+\frac13(V^0)^{(4)}$ \\$-3T^2V^-+2W(V^-)'-2U_0U_+'-2T'(V^-)'+2U_0'U_+-\frac12T''V^-+\frac13(V^-)^{(4)}$ \\$ \frac32T^2V^-+\frac12T(V^-)''-W(V^-)' +2U_0U_+'+2T'(V^-)'+\frac34T''V^--\frac13(V^-)^{(4)}$ \\$-3T^2U_++2TU_+''+2T'U_+'+2W'U_+-4(V^-)'V^0
            +\frac32T''U_+-\frac13U_+^{(4)}$ \\$
            \frac32T^2U_+-\frac32TU_+''-WU_+'-2T'U_+'-2W'U_+
            +2(V^-)'V^0-\frac54T''U_++\frac13U_+^{(4)}$ \\$
            -3T^2U_0+2TU_0''+2T'U_0'+2W'U_0+4(V^-)'V^+ +\frac32T''U_0 -\frac13U_0^{(4)}$ \\$ \frac32T^2U_0-\frac32TU_0'' -WU_0'-2T'U_0'-2W'U_0 -2(V^-)'V^+ -\frac54T''U_0 +\frac13U_0^{(4)}$ \\$ -3T^2U_--2WU_-' +2V^0(V^+)' -2T'U_-'-2(V^0)'V^+ -\frac12T''U_- +\frac13U_-^{(4)}$ \\$ \frac32T^2U_-+\frac12TU_-'' +WU_-'-2V^0(V^+)' +2T'U_-' +\frac34T''U_- -\frac13U_-^{(4)}$ \\$ -T(V^0)''+W(V^0)' +U_-U_+'-2T'(V^0)'+W'V^0+U_-'U_+ -T''V^0+\frac13(V^0)^{(4)}$ \\$T(V^+)''-W(V^+)'+U_-U_0'+2T'(V^+)'-W'V^++U_-'U_0+T''V^+-\frac13(V^+)^{(4)}$ \\$TU_-''+WU_-'-V^0(V^+)'+2T'U_-'+W'U_--(V^0)'V^++T''U_--\frac13U_-^{(4)}$ \\$ -T(V^-)''+W(V^-)'-U_0U_+'-2T'(V^-)'+W'V^- -U_0'U_+ -T''V^-+\frac13(V^-)^{(4)}$ \\$TU_0''+WU_0'+V^-(V^+)'+2T'U_0'+W'U_0+(V^-)'V^++T''U_0-\frac13U_0^{(4)}$ \\$-TU_+''-WU_+'+V^-(V^0)'-2T'U_+'-W'U_++(V^-)'V^0 -T''U_++\frac13U_+^{(4)}$\\$U_-(V^-)' - U_0(V^0)'+U_-'V^- - U_0'V^0,\qquad-U_0(V^0)'+U_+(V^+)' -U_0'V^0+U_+'V^+, \qquad U_i(V^{j\neq i})'+U_i'V^{j\neq i},\qquad (V^i)'V^i, \qquad U_i'U_i,\qquad T^{(5)}$}\\
			\hline
		\end{tabular}
        }
		\caption{Null states appear in the Jacobi identities for the VOA associated to the $\mathrm{SU}(3)\times \mathrm{SU}(2)$ theory with a half-hyper in the $\mathbf{8}\times\mathbf{2}$. They show up at levels $h=6,7$.}
		\label{tab:nulls_Jacobi_su3su2}
	\end{table}

\newpage
\bibliographystyle{JHEP}
\bibliography{bibliref.bib}

\end{document}